\documentclass[fleqn,usenatbib]{mnras}

\usepackage{newtxtext,newtxmath}

\usepackage[T1]{fontenc}

\DeclareRobustCommand{\VAN}[3]{#2}
\let\VANthebibliography\thebibliography
\def\thebibliography{\DeclareRobustCommand{\VAN}[3]{##3}\VANthebibliography}

\usepackage{graphicx}	
\usepackage{amsmath}	

\usepackage{longtable}   

\title[SN Ia Magnitude Step from the local DM Environment]{Type Ia Supernova Magnitude Step from the local Dark Matter Environment}

\author[Steigerwald et al.]{
Heinrich Steigerwald,$^{1}$\thanks{E-mail: 
heinrich@steigerwald.name}
Davi Rodrigues,$^{1,3}$
Stefano Profumo,$^{2}$
Valerio Marra$^{1,3,4,5}$
\\
$^{1}$Núcleo de Astrofísica e Cosmologia, Universidade Federal do Espírito Santo, 29075-910, Vitória, ES, Brazil\\
$^{2}$Department of Physics and Santa Cruz Institute for Particle Physics, 1156 High St, University of California, Santa Cruz, CA 95064, USA\\
$^{3}$Departamento de Física, Universidade Federal do Espírito Santo, 29075-910, Vitória, ES, Brazil\\
$^{4}$INAF -- Osservatorio Astronomico di Trieste, via Tiepolo 11, 34131, Trieste, Italy\\
$^{5}$IFPU -- Institute for Fundamental Physics of the Universe, via Beirut 2, 34151, Trieste, Italy
}

\date{Accepted XXX. Received YYY; in original form ZZZ}

\pubyear{2021}

\begin{document}
\label{firstpage}
\pagerange{\pageref{firstpage}--\pageref{lastpage}}
\maketitle

\begin{abstract}
Residuals in the Hubble diagram at optical wavelengths and host galaxy stellar mass are observed to correlate in Type Ia supernovae (SNe Ia) (`magnitude step’). Among possible progenitor channels for the associated explosions, those based on dark matter (DM) have attracted significant attention, including our recent proposal that `normal' SNe Ia from bare detonations in sub-Chandrasekhar white dwarf stars are triggered by the passage of asteroid-mass primordial black holes (PBHs):  the magnitude step could then originate from a brightness dependence on stellar properties, on DM properties, or both. Here, we present a method to estimate the local DM density and velocity dispersion of the environment of SN Ia progenitors. We find a luminosity step of $0.52\pm 0.11\,$mag corresponding to bins of high vs low DM density in a sample of 222 low-redshift events from the Open Supernova Catalog. We investigate whether the magnitude step can be attributed to local DM properties alone, assuming asteroid-mass PBHs. Given the inverse correlation between SN Ia brightness and PBH mass, an intriguing explanation is a spatially-inhomogeneous PBH mass function. If so, a strong mass segregation in the DM density-dependent PBH mass scale is needed to explain the magnitude step. While mass segregation is observed in dense clusters, it is unlikely to be realized on galactic scales. Therefore, if DM consists of asteroid-mass PBHs, the magnitude step is more likely to exist, and dominantly to be attributed to local stellar properties.
\end{abstract}




\section{Introduction}

It is difficult to overstate the importance of thermonuclear supernovae (SNe Ia) for modern cosmology, that led to the discovery of the enigmatic cosmic acceleration \citep{1998AJ....116.1009R,1999ApJ...517..565P} and, more recently, to a growing tension between early and late Universe expansion rate measurements \cite[e.g.][for a status report]{2019NatAs...3..891V}.
Their praised standard candle nature is also regularly questioned with reference to the fact that the origin of SNe Ia itself continues a mystery, despite tremendous collective efforts over at least three decades \citep[e.g.][]{2011NatCo...2..350H}.

It is commonly assumed that the ignition is triggered in binary systems where some kind of mass transfer leads to the explosion of a heavy carbon-oxygen white dwarf (WD) star \citep[see, e.g.,][for a review]{2018PhR...736....1L}. Over the years, the various sub-scenarios of this basic setting have all suffered severe restrictions from a growing list of non-detections: the lack of H-emission in SN Ia spectra \citep{2007ApJ...670.1275L}, the upper limit on X-ray signatures from accreting WDs \citep{2010Natur.463..924G}, the non-detection of a pre-explosion companion in HST imaging \citep{2011Natur.480..348L}, the lack of early UV flash signatures of ejecta interaction with a companion \citep{2010ApJ...708.1025K}, the lack of ejected high velocity He-donor stars \citep{2021A&A...654A.103L}, among others.

This notorious lack of triggering evidence has led to look beyond the standard setting, and preferentially for a trigger that has the property of being invisible, such as, for example, dark matter (DM).
Initial studies focused on deflagrations (consistent with certain \textit{peculiar} SNe Ia) and include particle DM admixed WDs \citep{2013PhRvD..87l3506L,2015ApJ...812..110L,2021ApJ...914..138C},  heating from asymmetric DM collapse in WD stars (\citealt{Bramante:2015cua,2019PhRvD.100d3020A,2019PhRvD.100c5008J}; see, however, \citealt{SPMR21}), and dynamical friction heating from asteroid mass primordial black hole (PBH)-WD encounters \citep{Graham:2015apa}. For most cases, the result of these studies was to derive astrophysical constraints on particular DM models, based on the non-observation of large fractions of peculiar SN Ia-like transients.
However, in our recent reanalysis of the latter setting  \citep{2021PhRvL.127a1101S} with focus on detonation triggering (consistent with \textit{normal} SNe Ia), we have found that DM in the form of $\mathcal{O}(10^{23}$g$)$ PBHs can reproduce the observed SN Ia rate and accessorily their median brightness, apart from being consistent with all constraints found in the literature \citep{2020ARNPS..70..355C,2020PhRvD.101f3005S}. This evokes the possibility that a large fraction, if not all normal SNe Ia are triggered by DM. 

As one of the many unexplained observational aspects of SNe Ia, sub-luminous events (or fast decliners) occur more frequently in heavier host galaxies \citep{2010MNRAS.406..782S,2010ApJ...722..566L,2010ApJ...715..743K,2011ApJ...740...92G,2011ApJ...743..172D,2013ApJ...764..191H,2014MNRAS.445.1898C,2021ApJ...909...28R,2021arXiv210902456B}.
The correlation is known as the magnitude step, or `mass step', and is often used to standardize SNe Ia luminosity, mostly parametrized as a step-function. 
This functional form is debatable, as it assumes implicitly two progenitor channels for SNe Ia to account for the difficulty of a single binary system channel to explain all events \citep{2019NewAR..8701535S}.

A large number of underlying environmental properties have been investigated leading to more or less strong correlations. 
The procedure is generally to divide a sample of SNe Ia into bins of high and low values of a certain property, say $X$, and then to compute the difference of the mean magnitudes at maximum luminosity in the $B$-band, $\Delta \langle M_{B,\rm max}\rangle$ (the magnitude step is then informally called $X$-step).
To exemplify this procedure with recent works, we cite the derivation of a  global host stellar mass-step  \citep[$0.116\pm 0.030\,$mag, ][]{2021arXiv210902456B}, a metalicity-step \citep{2014MNRAS.445.1898C}, a local stellar mass-step \citep[$0.059\pm 0.024$\,mag, ][]{2020A&A...644A.176R}, a host morphology-step \citep[$0.058\pm 0.019$~mag, ][]{2020MNRAS.499.5121P}, and local specific star formation rate step \citep[$0.127\pm 0.032$~mag,][]{2021arXiv210902456B}. 
It is unclear if external factors like dust can \citep{2021ApJ...909...26B} or cannot \cite{2020arXiv200613803P,2021MNRAS.508.4310T} account for the observed luminosity differences.
Even though stellar age is the favored explanation \citep[e.g.][]{2014MNRAS.445.1898C}, the true underlying reason is still matter of debate. 

Given that in hierarchical structure formation,  stellar age is a reasonable tracer of the underlying DM density, it is a tempting possibility that the observed magnitude step was in fact a consequence of the DM environment, especially if some form of DM triggers the ignition. For this purpose, we propose, in a first part, a method to estimate the local DM density and velocity dispersion to which were exposed the progenitors of observed SNe Ia. We apply this method to publicly available SN Ia data from \texttt{The Open Supernova Catalog} (\texttt{TOSC}) and derive a `local DM density-step' which we find sufficiently pronounced to justify a second part. 

In a second part, we investigate under which requirements, the environmental dependence of SN Ia properties can be entirely attributed to the local properties of DM, when constituted by asteroid mass PBHs. The starting point is the property that the median SN Ia brightness is correlated with the PBH mass scale \citep{2021PhRvL.127a1101S}. Thus we hypothesis a spatially inhomogenous mass function for PBHs, which is loosely motivated by the fact that whenever a range of masses is present, gravitational interactions lead to mass segregation in which the most massive bodies have the most spatially compact distribution \citep{1969ApJ...158L.139S}. Of course, the present day concentration of PBHs on galactic scales is far too low to produce a sizable effect and any mass segregation, if existent, must be the left-over setting of early Universe dense clustering.
With these assumptions we ignore knowingly environmental dependent stellar properties, notably the fact that mass and composition of heavy WDs evolve with time due merger, spin-down, off-center carbon ignition and core transmutation, that actually do lead to environmental dependence of SN Ia properties (we investigate these effects in a dedicated study, \citealt{SMRP21}). Therefore, this second part should be considered as an `amplitude-of-effect' study rather than a realistic scenario.

The paper is organized as follows: In Sec.~\ref{sec:2} we present our method to determine the local DM environment of SNe Ia that we apply to observational data and infer a magnitude step. In Sec.~\ref{sec:3} we investigate the effects of PBH mass segregation on the environmental dependence of SN Ia brightnesses. In Sec.~\ref{sec:4} we discuss our methods and present a summary.


\section{Determining the local DM environment}\label{sec:2}

\subsection{Method}
The DM distribution in galaxies is usually satisfactorily described by a spherically symmetric density profile, $\rho_{\rm DM}(r)$, with normalization parameters that can be related to their total luminosity in stars \citep{Bell:2003cj,Moster:2012fv}. Determining the local DM density in the vicinity of SNe Ia can therefore be decomposed into three steps: (1) host galaxy identification, (2) determination of the host's total luminosity, and (3) determination of the three-dimensional radial position, $r$, of the SN Ia inside the host galaxy. 
The first step is trivial in most cases, but confirmation from comparison between individually measured redshifts (of host galaxy and SN) usually clears all doubts (we require $|z_{\rm SN}-z_{\rm gal}|< 0.005$). Also the second step is trivial, but the third merits some more detailed explanations. Let us decompose the radial position in its line-of-sight and perpendicular (orthogonal-to-sight) components
\begin{align}
r = \sqrt{r_{\!\parallel}^{\;2} + r_{\!\perp}^{\;2}}\,.
\end{align}
The perpendicular part is given by $r_{\!\perp} \simeq \theta \, d$, where $\theta$ is the angular separation between the SN and the center of its host galaxy on the sky, and $d$ the distance to the host galaxy. We focus on near Universe SNe Ia where the complications involved with curved space are avoided (we cut redshift $z_{\rm SN}<0.08$ to put a number).

\begin{figure}
    \centering
    \includegraphics[width=\columnwidth]{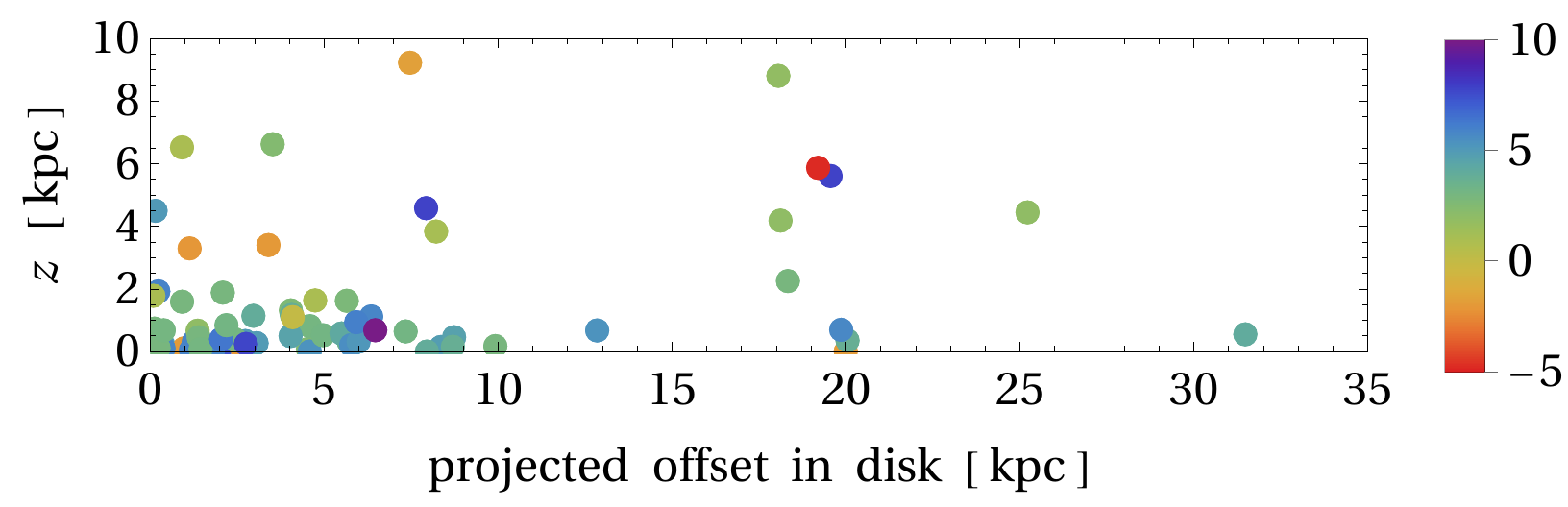}
    \caption{Vertical offset $z = r_{\perp} \sin \phi$ above the disk of 69 SNe Ia for the \texttt{OSC} that exploded in edge-on galaxies as a function of projected offset in disk $x = r_{\:\!\!\perp}\cos \phi$. The center of the host galaxy is located at the lower left corner and the disk extends along the $x$-axis.  Colors indicate Hubble type of host (elliptical $\in [-5,-3]$, lenticular $\in [-3,0]$, and spiral $\in [0,10]$)}
    \label{fig:vertical-height}
\end{figure}

The line-of-sight component $r_{\!\parallel}$ is far more tricky to determine. For example, the redshift of the SN spectrum contains no information about $r_{\!\parallel}$, 
but an estimator can be constructed since SNe Ia most likely occur in a disc. This assumption is corroborated by the characteristic delay time of SNe Ia, $200-500\,$Myr \citep{2009ApJ...707...74R}, which is inconsistent with thick disc or stellar halo membership. Only a small fraction of all SNe Ia (less than 10\%) has a delay time of the order of $10\,$Gyr and can possibly originate from outside the thin disc, see Table~\ref{tab:wd-population}. 
\begin{table}
    \centering
    \begin{tabular}{lccrr}
    \hline
    Population  &  $n_{\odot}$ [pc$^{-3}$]  &  $N$ & fraction  & age \\
    \hline 
    Bulge & & & $<$1\% & \\
    Thin disc   & $2.9\times 10^{-3}$ & $2.0\times 10^9$ & 17\%  & ASF \\
    Thick disc  & $1.7\times 10^{-3}$ & $3.9\times 10^9$ & 34\% & $10\,$Gyr  \\
    Halo        & $2.7\times 10^{-4}$ & $5.6\times 10^9$ & 49\% & $10.9\pm 0.4\,$Gyr \\
    \hline
    \end{tabular}
    \caption{Contributions to the WD population of the Milky Way (\protect\cite{2009JPhCS.172a2004N}, table 2, see also \protect\cite{2019MNRAS.482..965K}), $n_{\odot}$ is the local number density in the solar neighborhood, and ASF = active star formation.}
    \label{tab:wd-population}
\end{table}
To further confirm this trend, we have conducted a survey of the vertical offsets $z$ of archival SNe Ia that occured in near edge-on galaxies. With `near edge-on' we mean galaxies with inclination angle between the axis of symmetry and the line of sight between $86^{\circ}$ and $90^{\circ}$. Scrutinizing \texttt{TOSC}, we find that from 69 events, the median vertical offset $z$ from the disc plane is as low as $0.6\,$kpc and that 80\% have $z<2\,$kpc  in all types of galaxies combined (see Fig.~\ref{fig:vertical-height}). This is a conservative estimation, since selection effects due to dust extinction tend to obscure and hide events hidden in the disk, especially in edge-on galaxies where dust layer obtruding the line-of-sight is maximal.

\begin{figure}
    \centering
    \includegraphics[width=\columnwidth]{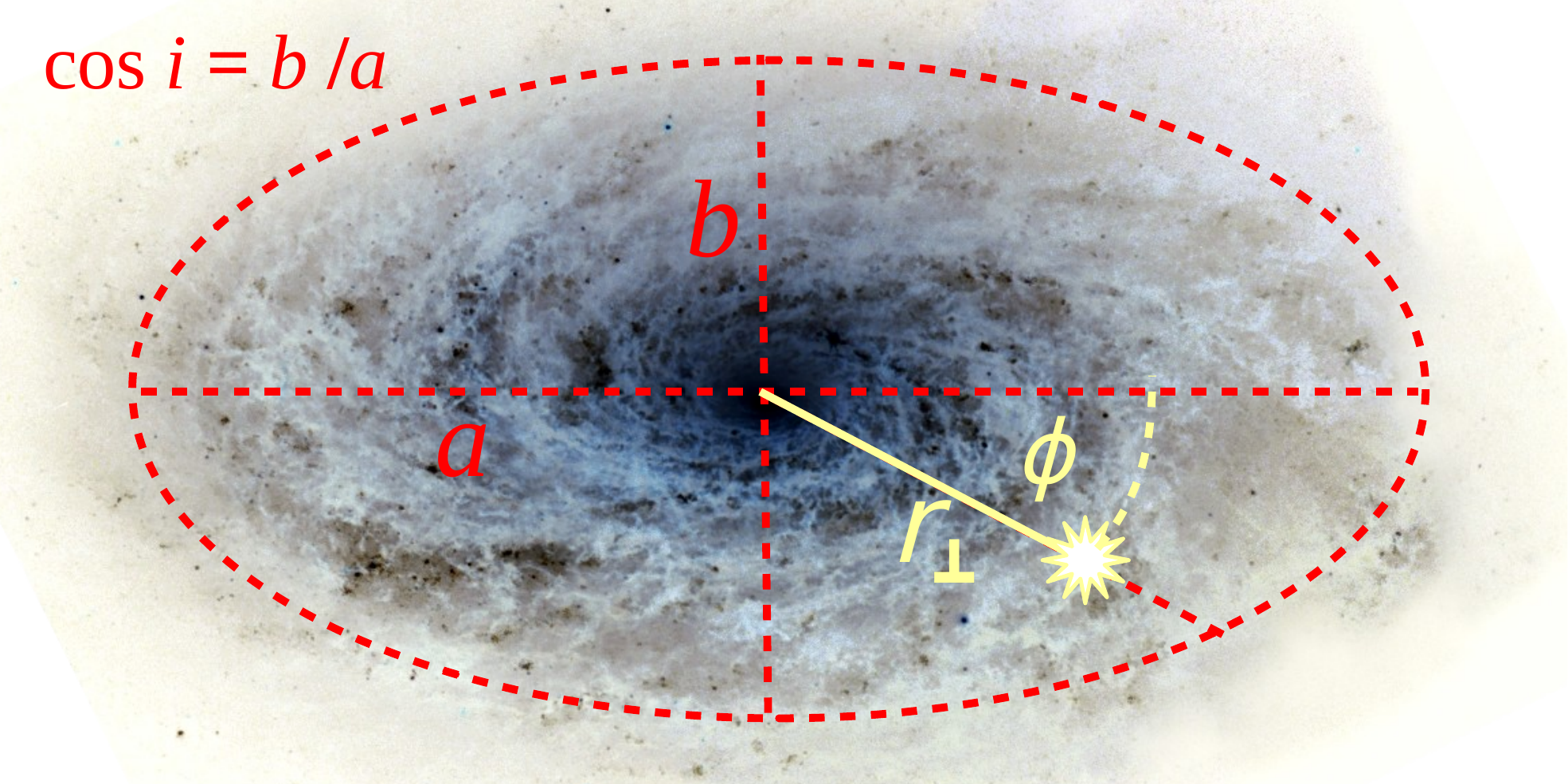}
    \caption{Schematic view of a galaxy showing the inclined disk (red dashed ellipse), major and minor axes $a$ and $b$ (red dashed lines) and the position angle $\phi$ and tangential offset $r_{\!\perp}$ (yellow plain line) of the SN (white star symbol).  Background picture credit: HST (\url{https://hubblesite.org/image/836/gallery}), color inverted.} 
    \label{fig:illustration}
\end{figure}

To construct the estimator, let us call $i$ the angle between the axis perpendicular to the disk and the line-of-sight and $\phi$ the position angle of the supernova with respect to the major axis of the inclined disc (see Fig.~\ref{fig:illustration}). 
In terms of these variables, the three-dimensional offset is given by
\begin{align}\label{eq:r}
    r = r_{\!\perp} \;\! \sqrt{\cos^2(\phi) + \sin^2(\phi) /\cos^2(i)}\,,
\end{align}
where all quantities on the right hand side are measurable. We expect that eq.~\eqref{eq:r} correctly estimates the radial offset for 80-90\% of all SNe Ia. As a precaution, since equation \eqref{eq:r} diverges when $i$ goes to $90^{\circ}$ and $\phi \neq 0^{\circ}$, we discard events where the host galaxy has an inclination within 10\% from edge-on, implying the cut $i\leq 81^{\circ}$.

\begin{figure}
    \centering
    \includegraphics[width=\columnwidth]{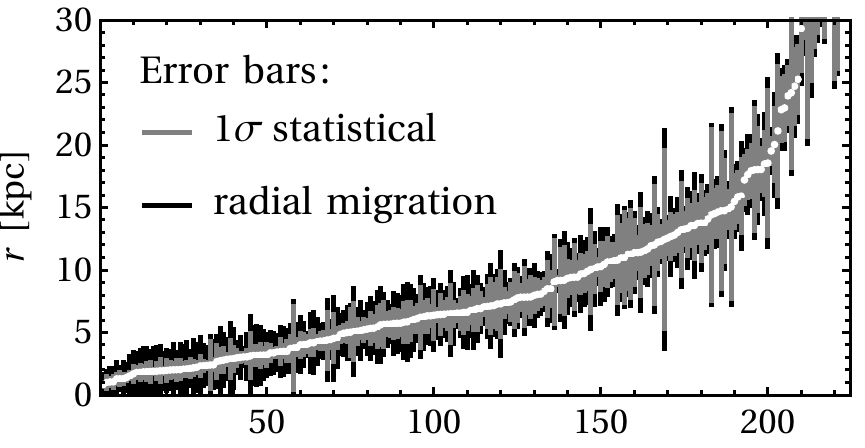}
    \caption{Average three-dimensional offset (white circles) of the progenitors of our SNe Ia from the sample selection of \texttt{TOSC}, together with statistical errors (gray bars) and systematic errors (black bars) due to radial migration of stars.} 
    \label{fig:sysOffsIa}
\end{figure}

Since what matters is not only the DM environment at the moment of explosion, but rather the integral effect over the progenitor star's lifetime, it is important to consider stellar movements on galactic scales. 
It is known that stars migrate radially over their lifetime, typically 3.6~kpc~$\sqrt{t/8\,{\rm Gyr}}$ \citep{2018ApJ...865...96F}, where $t$ is the age of the star. Since we are interested in the mean radial position over the star's lifetime, we take as systematic error half of this value, $\delta r_{\rm sys} = 1.8$~kpc~$\sqrt{t/8\,{\rm Gyr}}$. At small radii, the typical migration distance can exceed the central value. Since $r$ cannot become negative, for these cases, the expected inward migration is larger than the expected outward migration, and a non-zero net migration is expected. Therefore, whenever $r-\delta r-\delta r_{\rm sys}$ is negative, we increase the central value by half that quantity and decrease the systematic error equally by half that quantity. As can be seen in Fig.~\ref{fig:sysOffsIa}, systematic errors due to radial migration dominate at small radii, while statistical errors dominate at large radii.

In the absence of a technique to measure the progenitor age, $t$, from light curves and spectra, we use as proxy the mean local star age estimate given by the Hubble type modulated mean local star age at half-light radius and radial age gradient prescription of \cite{2015A&A...581A.103G}, using equations (2) to (5) from \cite{1998MNRAS.293..157R} to estimate the galaxy half-light radius. 
We have compared the estimations obtained by this method with age estimations based on local star light from \cite{2019ApJ...874...32R}, see Fig.~\ref{fig:6} panel (b). The correspondence is rather crude. In the absence of local star light measurements for the whole sample, we decide to continue to use the estimations based on the age gradient. Since our use is limited to the estimation of systematic errors, their quality is of secondary importance, and we emphasize that caution should be taken when these age estimations are further used.

\subsection{DM halo fitting}\label{sec:dm-environment}
Once the host galaxy is identified, we reconstruct the DM density profile, adopting the functional form $\rho_s (1+r/r_s)^{-1}[1+(r/r_s)^2]^{-1}$ of \citet{Navarro:1995iw}, where we use the halo concentration (see appendix~\ref{sec:density}) from \cite{Dutton:2014xda}, the halo mass from the halo abundance matching from \cite{Moster:2012fv} and the stellar mass from the mass-to-light ratio from \citet{Bell:2003cj} using preferentially $K$-band and if not $B$-band magnitudes of the host galaxies (see Table~\ref{tab:data-host} and appendix \ref{sec:halo-masses} for technical details). We estimate $\delta \rho_{\rm DM}  = 0.2\,$dex of systematic uncertainty due to the use of stellar mass-to-light relation and halo abundance matching (see appendix \ref{sec:density} for further details).
Assuming locally isotropic velocity dispersion, we adopt the $v_{\rm DM}(r)$, from eq.~(14) of \cite{Lokas:2000mu}, see also appendices \ref{sec:density} and \ref{sec:velocity} for technical details.

We take special care when choosing the distance measurements (see columns 6 to 8 in Table~\ref{tab:data-host}). Distance errors have a negligible effect on the estimation of $\rho_{\rm DM}$. 
Specifically, we have tested that for the hole range of parameters (angular separation $0.5'' < \theta < 278''$, host magnitude $1< K <14$, and distance $0.7\,{\rm Mpc} < d < 337\,$Mpc) an error of 20\% in the estimation of the distance, induces less than 1\% of error in the estimation of $\log_{10}(\rho_{\rm DM})$, with typical errors less than $0.1$\%. 

Let us briefly discuss the effects of baryonic feedback on the DM halo. Both in very small galaxies (from SN feedback) and in more massive galaxies (from active galactic nuclei, AGN), the core regions are expected to be shallower in DM density than modelled by the NFW profile.
First, for dwarf galaxies, we apply the prescriptions of \cite{DiCintio:2014xia}, and find that 5 DM halos of the presented sample are subjected to modifications, but only 2 SN Ia events (2012cg and 2012ht) of these 5 occur in the core region. The DM densities $\rho_{\rm DM}$ of these two events are diminished by $0.3$ and $0.2\,$dex respectively. 
Second, it is expected that in more massive galaxies active galactic nuclei (AGN) expand the DM halo in the core region. the sample we utilize contains 17 AGNs. The amplitude of the effect is still unknown, thus we present a crude estimation.  Assuming, for example, that AGN feedback efficiently flattens a core region of $5\,$kpc extension of halos with masses inferior to $10^{13}\,$M$_{\odot}$. Then, the DM densities $\rho_{\rm DM}$ of 4 SNe (2002de, 2002bf, 1998aq and 2003cg) in the presented sample are subjected to a modification. 
In view of the smallness of these modifications, it is reasonable to state that the DM density estimations presented in this section are robust under baryonic feedback effects.

\subsection{SN Ia sample selection}\label{sec:sn-sample}
Already discussed along the previous sections, we now briefly summarize our selection criteria.
We scrutinize the complete sample of \texttt{TOSC} (\url{https://sne.space/}), as of january 2020) with the following restrictions: 
(1) spectroscopically confirmed type Ia event; 
(2) peak $B$-band photometry available; 
(3) SN redshift $z_{\rm SN}$ available and at most $z_{\rm SN} < 0.08$; 
(4) host galaxy redshift $z_{\rm gal}$ available and $|z_{\rm SN}-z_{\rm gal}| < 0.005$; 
(5) host $K$-band or $B$-band magnitude available for determining the halo mass; 
(6) host inclination available on HyperLEDA\footnote{\url{http://leda.univ-lyon1.fr}} and at most $i \leq 81^{\circ}$; 
(7) SN events are excluded if the host galaxy has clear evidence of interaction (according to SIMBAD\footnote{\url{http://simbad.u-strasbg.fr/simbad/}});
(8) finally, we exclude SNe Ia with extremely low luminosity $>-15$ (excludes 2 events).

In total 222 events pass these cuts and their data is presented in Appendix~\ref{sec:data-information}, where we dedicate separate tables for host data and global properties (Table~\ref{tab:data-host}) and SN Ia data and local properties (Table~\ref{tab:data-sn}). The main result of this work are columns 9 and 10, the local DM density and velocity dispersion, that the progenitor stars were exposed to over their lifetime. The median is med$(\rho_{\rm DM,SN}) = (1.095\pm 0.002)\times 10^{-2}M_{\odot}$~pc$^{-3} = 0.473\pm 0.002$~GeV~cm$^{-3}$. This value is only slightly higher than the local DM density in the solar neighborhood, which is $\rho_{\rm DM,\odot} \simeq 0.4$~GeV~cm$^{-3}$ when inferred from global studies \citep[see, e.g.,][and references therein]{2021RPPh...84j4901D}.
Even though events are dispersed over almost three orders of magnitude, 50\% of progenitors were exposed to mean values comprised in a rather narrow range between $0.20-0.91$~GeV~cm$^{-3}$.

\begin{figure*}
    \centering
    \includegraphics[width=\textwidth]{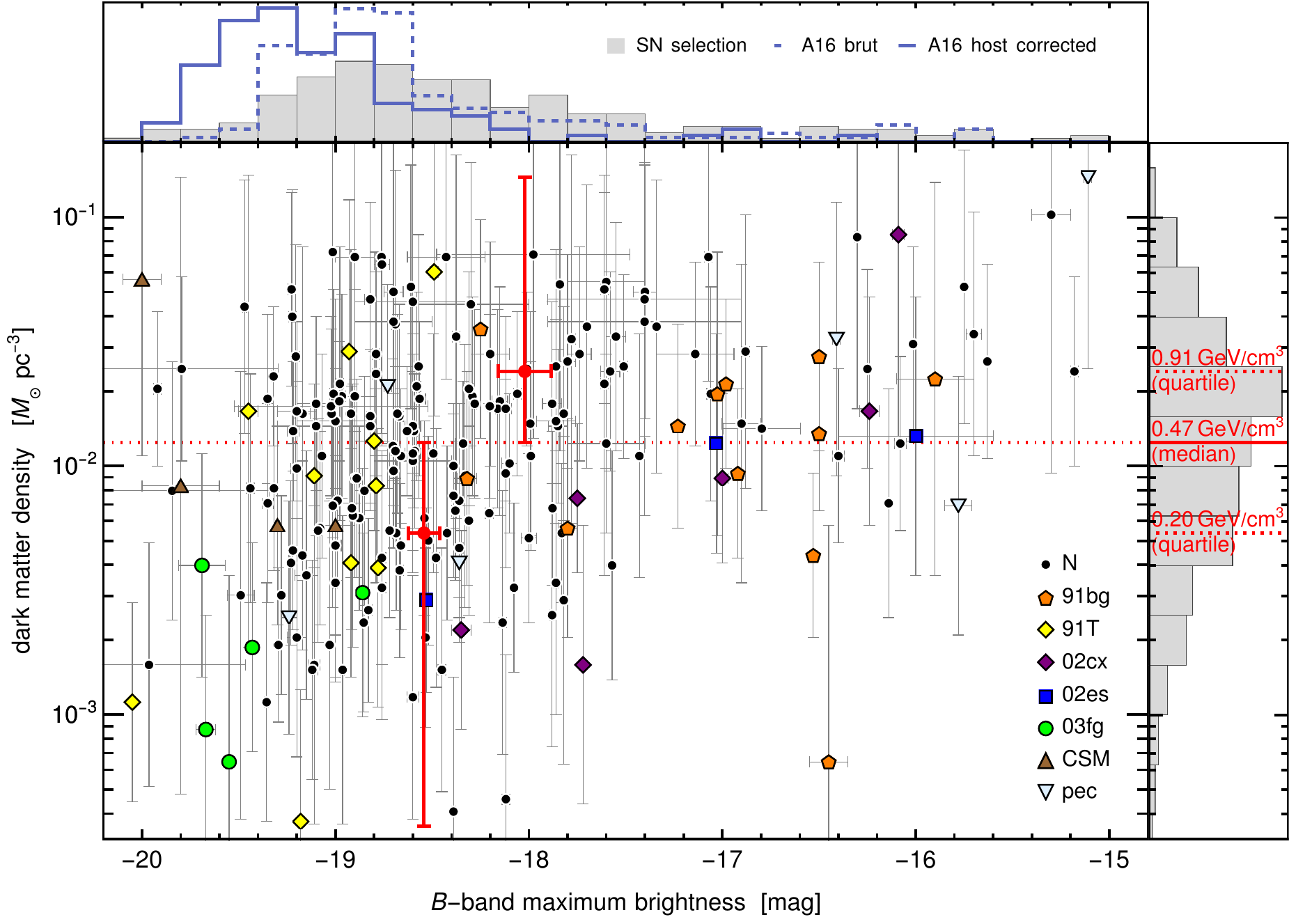}
    \caption{Main plot: Archival data of 222 SN Ia events that exploded with absolute $B$-band magnitude and evolved in a dark matter density $\rho_{\rm DM}$. Tick marks at the top assume bare detonations (see Sec.~\ref{sec:3} for details). Normal Ia events are represented by black disks and peculiar Ia events according to legend. Gray error bars represent $1\sigma$ statistical and systematic (radial migration + halo abundance matching) errors. The horizontal red dotted line divides into bins of high and low DM density. The thick red circles are the mean magnitudes
    in bins of DM density, drawn at the median value of the DM density in each bin. The error bars on these points represent the weighted $1\sigma$ error on the mean magnitude (horizontal) and the bin width in DM density (vertical). 
    Top panel: $M(B)_{\rm max}$-distribution compared with the sample of A16=\protect\cite{2016MNRAS.460.3529A}.
    Right panel: $\rho_{\rm DM}$-distribution with median (red thick) and quartiles (red dashed).
    }
    \label{fig:density-Bmax}
\end{figure*}

The distribution of events in the DM density-maximum brightness plane is shown in Fig.~\ref{fig:density-Bmax}. 
It can be noted that sub-luminous events ($M_B>-17$) occur preferentially in DM dense environments ($\rho_{\rm DM}>10^{-2}M_{\odot}$~pc$^{-3}$). It can also be noted that super-Chandrasekhar SNe Ia (03fg) occur systematically in DM shallow environments, however, statistical significance is low (only 5 events).

\subsection{Magnitude step}\label{sec:DM-density-step}

First, let us verify the existence of the stellar mass step in our data. Choosing two equal sized bins with high and low global stellar mass (with median $5.75\times 10^{10}\,$M$_{\odot}$ as threshold), we find the luminosity step $\Delta\langle M_B\rangle = 0.15\pm 0.11\,$mag ($1\sigma$ error bars), 
compatible with values from the literature \citep[0.116$\pm$0.030,][]{2021arXiv210902456B}.
Next, we divide our data into two equal bins with high and low local DM density environment, we find the luminosity step $\Delta \langle M_B\rangle = 0.52\pm 0.11\,$mag ($1\sigma$ error bars).
In comparison, with other magnitude steps from the literature, we find that the DM density step is much more pronounced \cite[see][for a recent comparison of magnitude steps]{2021arXiv210902456B}.

We note that if we use $10^{10}\,$M$_{\odot}$ ($10^{11}$M$_{\odot}$) as the threshold value, we find a much higher global host mass-step $0.39\pm 0.08$ ($0.26\pm 0.17\,$mag).
Since there seems to be a dependence on the choice of the threshold value, we test other values than the median DM density, which is $0.473\pm 0.002\,$GeV$\,$cm$^{-3}$ ($1\sigma$ statistical error) in our sample. For $0.1\,$GeV$\,$cm$^{-3}$ ($1.0\,$GeV$\,$cm$^{-3}$) as threshold, we find the local DM density-step $0.62\pm 0.09\,$mag ($0.58\pm 0.17\,$mag). Seemingly, the step is persistent in our data, irrespective of the threshold choice.
It can also be checked that the correlation between local DM density and global host mass is rather low (Pearson correlation coefficient is $0.16$, p-value for being uncorrelated $0.015$). 

When binned into groups of young and old local stellar age, we find the magnitude step $0.43\pm 0.11$ (1~$\sigma$ errors). This rather large step is also expected, as the local stellar age and local DM density are strongly correlated
(Pearson correlation coefficient is $0.65$, p-value for being uncorrelated $10^{-28}$). However, we caution since our determination of local stellar age is rather crude, and intrinsically correlated by the method of determination. A more reliable (and independent) method for the determination the local star age is from direct measurements of local stellar temperature. Therefore, this age step should be used with caution. 
Even though we argue that our DM density determinations are rather solid, we do not claim this for the DM density step, as possible selection effects could affect it. The large number of requirements (see Sec.~\ref{sec:sn-sample}) permitted only a small number of events finally being selected into the sample, and could have sorted out events according to certain patterns.


\section{Magnitude step from PBH mass segregation}\label{sec:3}
We repeat that this section serves as an amplitude-of-effect study, as we ignore important physical processes that influence the evolution of WD mass and composition and their effects on SN Ia brightnesses.
\subsection{Method}
The SN Ia brightness distribution is directly predicted from the WD and PBH mass functions, respectively \citep{2021PhRvL.127a1101S}.
The rate prediction requires additional knowledge about WD and PBH number densities, respectively. The PBH number densities are given by the DM density, assuming, for simplicity, that the fraction of DM in the form of PBHs is 1. 
However, to determine the WD number densities, stellar structure models for host galaxies would be necessary, which we do not have. To overcome this shortcoming, we make a very crude approximation, projecting our selected SNe Ia into the MW with corresponding DM density. This treatment is loosely motivated by the fact that the MW is generally considered as an `average galaxy'.

Since the two methods (rate and brightness) to determine the PBH mass function are independent, agreement between both is an imperative consistency check.  Here, we propose to perform this test independently at different DM densities, allowing for variable PBH masses in each bin. For simplicity, however, we assume a single PBH mass scale for each bin. This is also a rather crude approximation, in view of the theoretically better motivated extended mass functions.

As mentioned earlier, the main components of the MW containing WDs are the thin disc, the thick disc and the stellar halo, while the bulge contributes only negligibly (see Table~\ref{tab:wd-population}). We use the stellar structure model of \cite{2003A&A...409..523R} for the WD distributions in each component (see Appendix~\ref{app:stellar-structure} for details).
For the dark component we use parameters $\rho_s=4.88\times 10^6\,$M$_{\odot}$kpc$^{-3}$ and $r_s=21.5\,$kpc of \cite[table 2]{2002ApJ...573..597K}.
According to \cite{2019MNRAS.484.5453C}, the DM profile with these parameters fairly reproduces the observed rotation curves in the MW, but there might still be a residual uncertainty of the total mass within a factor of 2.

\subsection{Data analysis}\label{sec:data-binning}
We slice the MW into radial shells with $8\,$kpc thickness, defining 3 bins. We discard the outskirts because it is difficult to define a mean DM density in a bin with arbitrary limits.
Using the aforementioned density profile, we determine for each bin its DM density threshold values, as well as mean DM density, and mean velocity dispersion (see Table~\ref{tab:rho-DM}).

\begin{table}
    \centering
    \begin{tabular}{l|rrr}
    \hline
    Bin  & \#1 & \#2  & \#3  \\
    \hline 
    $r$ [kpc] $\in$  & $[0,8]$ & $[8,16]$ & $[16,24]$ \\
    $\rho_{\rm DM}$ [GeV$\,$cm$^{-3}$] $\in\!\!\!$  & $\!\![0.265,\infty]$ & $\!\![0.082,0.265]$ & $\!\![0.037,0.082]$ \\
    $\langle \rho_{\rm DM}\rangle$ [GeV$\,$cm$^{-3}$] & $0.49$ & $0.13$ & $0.05$ \\
    $\langle v_{\rm DM}\rangle$ [km$\,$s$^{-1}$] & $143.9$ & $193.7$ & $165.5$ \\ 
    \hline
    \end{tabular}
    \caption{DM parameters for each bin: $\langle \rho_{\rm DM}\rangle$ and $\langle v_{\rm DM}\rangle$, are the volume integrated mean densities and velocity dispersions, respectively.}
    \label{tab:rho-DM}
\end{table}

Next, we attribute each SN Ia event to one of the bins according to the local DM density found in the original host galaxy. This classification leads to 150, 52, and 14 events in bins \#1, \#2, and \#3, respectively, while 5 events lie in the outskirts and are discarded. 
Then, we compute the number of WDs in each bin, integrating the three contributions to the WD number distribution of \cite{2003A&A...409..523R}  over the radial extent of each shell (see Table~\ref{tab:wd-bin}).
\begin{table}
    \centering
    \begin{tabular}{lccccr}
    \hline
    Bin  &  thin disc  &  thick disc  &  stellar halo  &  total & fraction \\
    \hline 
    \#1  & $1.41\times 10^9$ & $3.19\times 10^9$ & $1.06\times 10^9$ & $5.66\times 10^9$ & 70.5\% \\
    \#2  & $5.33\times 10^8$ & $6.60\times 10^8$ & $6.01\times 10^8$ & $1.79\times 10^9$ & 22.3\% \\
    \#3 & $5.66\times 10^7$ & $4.81\times 10^7$ & $4.76\times 10^8$ & $5.77\times 10^8$ & 7.2\% \\
    \hline
    \end{tabular}
    \caption{Number of WDs in each population and bin. (0-8,8-16,16-24)kpc. Note that the fractions are relative, not including the (small) population of WDs beyond bin \#3.}
    \label{tab:wd-bin}
\end{table}
Finally, we determine, for each bin, the PBH mass scale from the two aforementioned independent SN Ia observations.

For the determination from the median brightness data, as a first step, we take into account the effects of host extinction.
The median brightness at maximum light in the $B$-band of our full sample is $-18.57$, which is comparable to the brute median, $-18.7\pm 0.1$, of the low-redshift sample of \cite{2016MNRAS.460.3529A}. Applying a host extinction correction technique based on SN spectra,  \cite{2016MNRAS.460.3529A} find that the actual median is much lower, $-19.3\pm 0.1$ (see Fig.~\ref{fig:density-Bmax}, top panel). Therefore, to correctly determine the progenitor masses, we subtract globally $0.6\,$mag from our data. We note that this is a very crude correction, because extinction is much lower or even unimportant in passive galaxies, while mostly important in star-forming galaxies.
Once luminosities are corrected, we determine the amount $^{56}$Ni from the relation $\log_{10}(M_{\rm Ni}) = -0.4 (M_{B} + 19.841 \mp 0.020)$ \citep[eq.~2]{Scalzo:2014wxa}, and the WD masses from the fitting relation $M_{\rm wd} = (0.796\pm 0.008) + 0.383 M_{\rm Ni}$ \citep{2020MNRAS.499.4725K} which assumes pure detonations, in agreement with \cite{2021PhRvL.127a1101S}. 
Once we have the SN Ia explosion masses, we determine the PBH mass scale according to \cite{2021PhRvL.127a1101S} assuming the WD mass function of \cite{2020ApJ...898...84K}. As conservative assumptions, we adopt the detonability parameter $\alpha=1$ \citep[see][]{2021PhRvL.127a1101S} and detonation speed $v_{\rm CJ}$ at nuclear-statistical equilibrium (NSE, \citealt{1999ApJ...512..827G}). As optimistic assumptions, we adopt $\alpha=10$ and $v_{\rm CJ}$ at quasi-NSE \citep{2014ApJ...785...61S}. The results are shown in Table~\ref{tab:median-brightness} and Fig.~\ref{fig:5}, where systematic errors correspond to the conservative versus optimistic assumptions.

\begin{table}
    \centering
    \begin{tabular}{lrr}
    \hline
    Bin  &  median $M_B\pm$(stat)$\pm$(syst)  &  $\log_{\rm 10} (M_{\rm PBH}/$g$)\pm$(stat)$\pm$(syst) \\
    \hline 
    \#1 & $-18.768\pm 0.005\pm 0.2$ & $24.20\pm 0.62\pm 0.20 $ \\
    \#2 & $-19.103\pm 0.003\pm 0.2$ & $23.05\pm 0.45\pm 0.35$ \\
    \#3 & $-19.515\pm 0.015\pm 0.2$ & $21.20\pm 0.45\pm 0.30$ \\
    \hline
    \end{tabular}
    \caption{PBH peak mass determined from SN Ia median brightness. Systematic errors are correspond to conservative (+) and optimistic (-) assumptions.}
    \label{tab:median-brightness}
\end{table}

The second method to determine the mass scale $M_{{\rm bh},j}$ of bin~$j$ is based on the constraint that the computed rate in each bin with given mean DM parameters $\langle \rho_{{\rm dm},j}\rangle $ and $\langle v_{{\rm dm},j}\rangle$ (see Table~\ref{tab:rho-DM}) must be equal to $\Gamma_{\rm SN}\;\!N_{{\rm wd},j}/N_{\rm wd,tot}$ where $\Gamma_{\rm SN}=0.45\pm 0.15$ is the observed SN Ia rate per century per MW-like galaxy \citep{1991ARA&A..29..363V,2011MNRAS.412.1473L}, $N_{{\rm wd},j}$ is the number of WDs in bin $j$ (see Table~\ref{tab:wd-bin}) and $N_{\rm wd,tot}\simeq 10^{10}$ is the total number of WDs in a MW-like galaxy \citep{2009JPhCS.172a2004N}. 
The results are shown in Table~\ref{tab:rate} and Fig.~\ref{fig:5}.
\begin{table}
    \centering
    \begin{tabular}{lrrr}
    \hline
    Bin  &  $N_{\rm SN}$ & fraction &  $\log_{\rm 10} (M_{\rm PBH}/$g$)\pm$(stat)$\pm$(syst) \\
    \hline 
    \#1 & 150 & 67.6\% & $22.80\pm 0.40\pm 0.20$ \\
    \#2 & 52 & 23.4\% & $21.88\pm 0.30\pm 0.22$ \\
    \#3 & 14 & 6.3\% & $21.63\pm 0.30\pm 0.03$ \\
    \hline
    \end{tabular}
    \caption{PBH peak mass determined from the observed SN Ia rate. Systematic errors are correspond to conservative (-) and optimistic (+) assumptions.
    Note that the fractions in column 3 do not sum up to 100\% because 6 events lie beyond bin \#3 and were excluded.
    }
    \label{tab:rate}
\end{table}

\subsection{Results and outlook}
\begin{figure}
    \centering
    \includegraphics[width=\columnwidth]{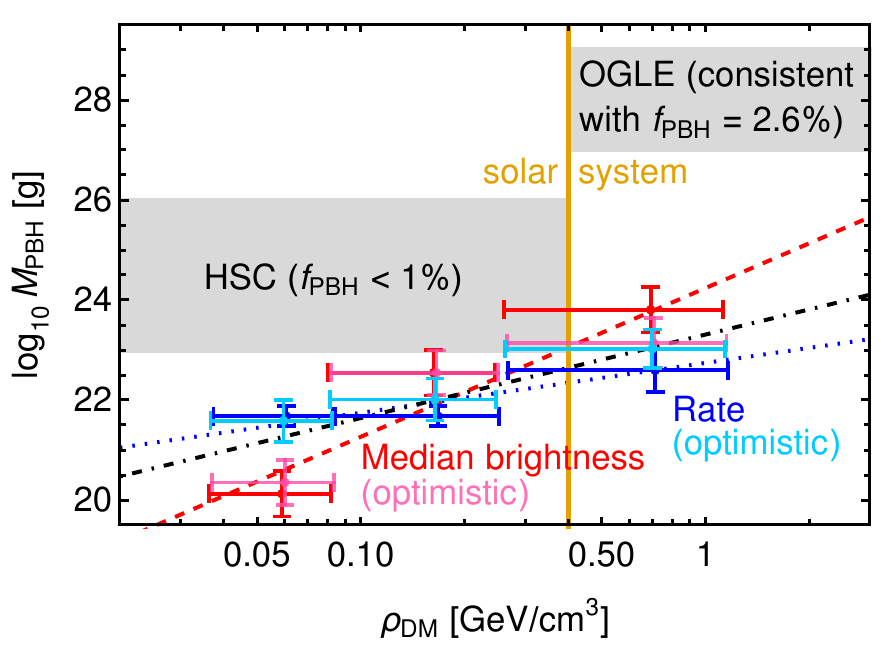}
    \caption{PBH mass scale as a function of DM density. Red (blue) data points correspond to constraints from the SN Ia brightness (rate) from Table~\ref{tab:median-brightness} (Table~\ref{tab:rate}). Dark (light) colors indicate conservative (optimistic) assumptions.  Horizontal error bars correspond to bin sizes. Data points are slightly offset horizontally for better reading. The red dashed, blue dotted, and black dotdashed lines are best-fits to the luminosity, the rate, and the combination, respectively. Also drawn is the DM density in the vicinity of the solar system (yellow vertical line), and microlensing constraints on the fraction of DM in the form of PBHs, $f_{\rm PBH}$,  (gray shaded areas).}
    \label{fig:5}
\end{figure}

As can be seen in Fig.~\ref{fig:5}, there are tensions between both methods in all three bins for the conservative analysis. For the optimistic case, the tensions are relieved in bins \#1 and \#2, but persist in bin \#3. Let us underline here that bin \#3 is only comprised of 14 events, which is much less than the other two bins, and it can therefore be regarded as the least reliable. 

To capture the evolution mathematically, let us suppose a simple parametric relation of the form $M_{\rm PBH} = M_{\rm PBH,0}\;\!(\rho_{\rm DM}/0.4\,{\rm GeV}\,{\rm cm}^{-3})^{\gamma}$,
where $M_{\rm PBH,0}$ is the local solar system PBH mass scale and $\gamma$ is a slope parameter that characterizes the steepness of the PBH mass scale evolution. We find for the local PBH mass scale $\log_{10}(M_{\rm PBH,0}/$g$)=(23.1\pm 0.3,\,22.3\pm 0.2,\,22.6\pm 0.2)$
and for the slope $\gamma = (3.0\pm 0.6,\, 1.0\pm 0.3,\,1.7\pm 0.4)$, where the first, second and third result correspond to fitting of the brightness, the rate, and the combination, respectively. As can be seen, the brightness alone requires strong segregation, while the rate requires much less.

Even though unlikely and ignoring important physics, such a steep mass segregation is still permitted by microlensing observations.
HSC survey has excluded (only 1 candidate event)
the existence of more than 1\% of DM in the form of PBHs in the mass range $[10^{23},10^{26}]$ in the region between the solar system and the Andromeda galaxy \citep{2019NatAs...3..524N,2020PhRvD.101f3005S}.  
The 5-year \texttt{OGLE} survey has detected two events ($10^{27}\,$g and $10^{29}\,$g) that are difficult to be attributed to free-floating planets \citep{2019A&A...624A.120V} and would be consistent with 2.6\% of DM in the form of PBHs in the mass range [$10^{27}\,$g, $10^{29}\,$g] in the region between the solar system and the Galactic bulge \citep{2019PhRvD..99h3503N}.


\section{Discussion, summary, and outlook}\label{sec:4}
First the discussion. About Sec.~\ref{sec:2}, the main drawback is the non-consideration of selection effects, when deriving the magnitude step. If local DM density was to be used to standardize SNe Ia, then this issue should be settled first.
Sec.~\ref{sec:3} mainly suffers from the lack of host galaxy stellar structure models. This can easily be overcome if in the future more stellar luminosity profiles were available. Only two were found in the literature when cross-matching our host galaxy sample with the SPARC catalog. 
In addition, for the determination of $M_{\rm PBH}$ from the median brightness, host extinction has been taken into account globally to correct our median values. Ideally, one would correct SN Ia brightnesses individually, based on their spectra, and then calculate the median. Finally, it should be recalled that all the modeling about WD populations (including the mass function, spatial distribution, and total number) is based on extrapolations from volume limited surveys in the vicinity of the solar system, usually not exceeding a range of $100\,$pc, which is only a tiny fraction of the total population.

Summary. In Sec.~\ref{sec:2}, we have presented a method to estimate the DM density and velocity dispersion that the progenitors of SNe Ia where exposed to over their lifetime. Analyzing 222 low redshift ($z<0.08$) SNe Ia from the \texttt{Open Supernova Catalog}, we find a luminosity step of $0.52\pm 0.11\,$mag when divided into high and low DM density bins. Even though there are uncertainties related to the choice of the binning threshold, the step seems robust and more pronounced than due to global stellar host mass binning for the same data.
This finding may suggest a different modeling of the magnitude step in supernova cosmological analyses.

In Sec.~\ref{sec:3}, assuming DM in the form of asteroid mass PBHs and allowing for a spatially inhomogeneous mass function, we have found that strong mass segregation is required to account for the observed SN Ia magnitude step. 
Given that stellar properties are knowingly environment dependent and can independently account for the magnitude step \citep{SMRP21}, such strong mass segregation is unlikely, in addition to not being expected on galactic scales in standard formation scenarios. Nevertheless, current microlensing constraints cannot exclude it (Fig.~\ref{fig:5}), and a subdominant contribution to the magnitude step is still a possibility.

Looking to the future, a reanalysis of this study with the expected $10^4$ low redshift SNe Ia from extended ZTF survey \citep{2019PASP..131g8001G,2020ApJ...904...35P} will significantly improve its constraining power. Additional benefit can be gained from  more accurate host galaxy modeling (especially the stellar distribution).  Finally, more insight in the mass function of PBHs, should they exist, can be gained with HSC-like microlensing survey towards the Galactic bulge, with the additional benefit of testing its spatial homogeneity.

\section*{Acknowledgements}

We are grateful for useful discussions with Richard Parker, Kurt Anderson, Wayne Bagget, Andrea V. Maccio, Arianna Di Cintio, Martin Barstow, Detlev Koester, Howard Bond, Simon Joyce, Benjamin Rose, and Jack Lissauer.

HS is thankful for FAPES/CAPES DCR grant n$^{\circ}$ 009/2014.
SP is partly supported by the U.S.A. Department of Energy, grant number de-sc0010107.
DR and VM thank CNPq and FAPES for partial financial support.
This project has received funding from the European Union’s Horizon 2020 research and innovation programme under the Marie Skłodowska-Curie grant agreement No 888258.

We acknowledge the usage of The Open Supernova Catalogue (\url{https://sne.space/}) HyperLeda database (\url{http://leda.univ-lyon1.fr}), NASA/IPAC Extragalactic Database (\url{https://ned.ipac.caltech.edu/}), SDSS SkyServer (\url{http://skyserver.sdss.org/dr13/en/home.aspx}) and Aladin Sky Atlas (\url{https://aladin.u-strasbg.fr/}).

\section*{Data availability}

The data underlying this article will be shared on reasonable request to the corresponding author.



\bibliographystyle{mnras}
\bibliography{biblio} 




\appendix

\section{Estimation of halo masses}\label{sec:halo-masses}

Given the apparent $K$-band magnitude $m_K$ of a galaxy, the total luminosity of stars is 
\begin{align}\label{eq:M-stars}
    L_{*}(K) = \Big(\frac{d}{1\,{\rm Mpc}} \Big)^{2}  10^{\:\! 0.4\:\!\big(M_{K\odot} - m_{K} +25 \big)} \quad [{\rm L}_{\odot}]\,,
\end{align}
where $d$ is the luminosity distance and $M_{K\odot} = 3.286$ \citep{2000asqu.book.....C} is the absolute $K$-band magnitude of the sun. The total mass in stars is 
\begin{align}\label{eq:mass-stars}
    M_{*} = \mathit{\Upsilon}_{*}(K) \;\! L_{*}(K) \quad [{\rm M}_{\odot}]\,,
\end{align}
where $\mathit{\Upsilon}_{*}(K) = 0.6\pm 0.1$ according to \citep{2014AJ....148...77M} or $\mathit{\Upsilon}_{*}(K) = 0.31\pm 0.07$ according to \cite{2013A&A...557A.131M} is the color independent $K$-band mass-to-light ratio.

Alternatively, given the apparent $B$-band and $V$-band magnitudes, the total luminosity of stars $L_{*}(B)$ in the $B$-band is determined from equations analog to \eqref{eq:M-stars} and \eqref{eq:mass-stars} with  $M_{B\odot}=5.44$ \citep{2000asqu.book.....C} and the 
 color dependent $B$-band mass-to-light ratio \citep[table~A7]{Bell:2003cj}
\begin{align}
    \log_{10} \mathit{\Upsilon}_{*}(B) = -0.942 + 1.737 \;\! (B\!-\!V) \pm 0.1 \,.
\end{align}

\begin{figure}
    \centering
    \includegraphics[width=0.49\columnwidth]{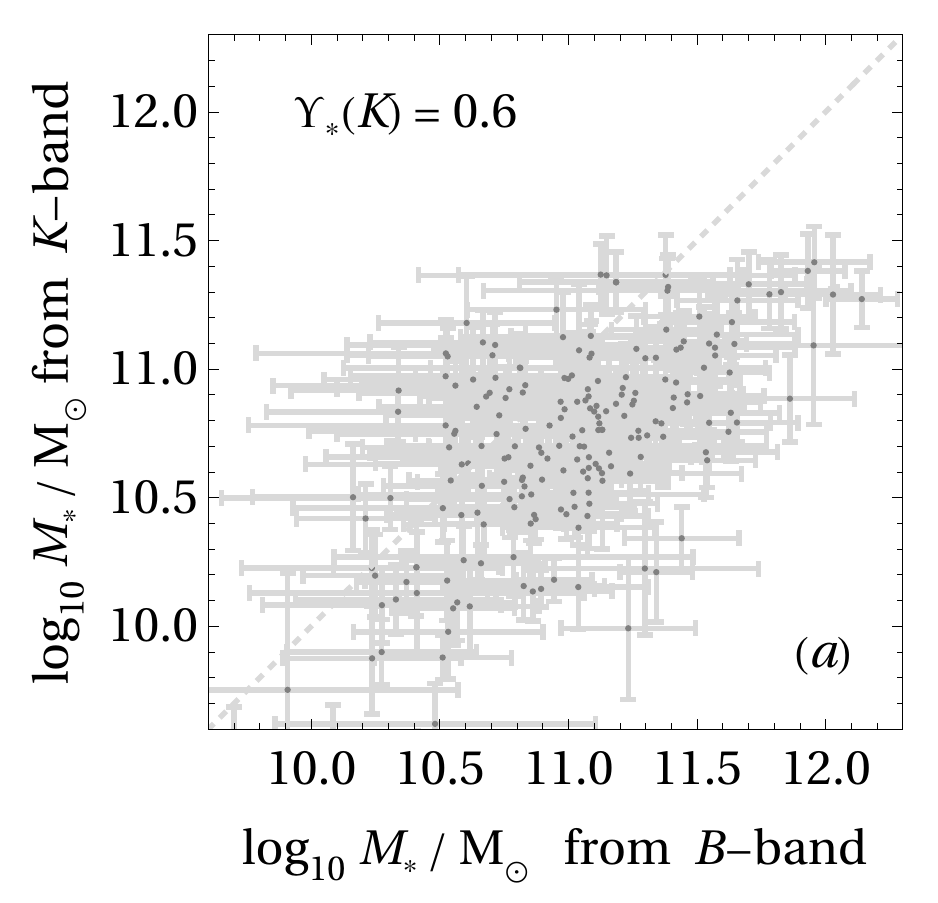}
    \includegraphics[width=0.49\columnwidth]{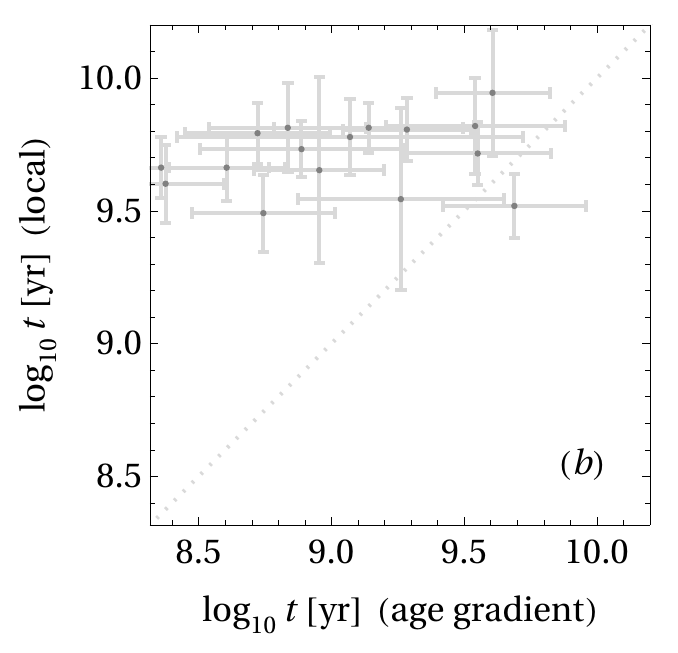}
    \caption{(a) Stellar masses of those galaxies where both $K$-band and $B$-band estimations are available. (b) Mean local star age estimated from age gradient procedure of \protect\cite{2015A&A...581A.103G} compared to the estimation based on local star age from \protect\cite{2019ApJ...874...32R}.
    }    
    \label{fig:6}
\end{figure}

In this work we use magnitudes and colors from HyperLEDA. For some rare cases, we use the the transformations of \citet[table~1]{Jester:2005id} to convert magnitudes and colors from the ugriz to the UBV photometric systems.  For galaxies where both $K$-band and $B$-band magnitudes are available, we use the $K$-band procedure as default. Since we find a better match between the $K$-band and $B$-band procedures (see Fig.~\ref{fig:6} a) with the mass-to-light ratio from  (cite), we use $\mathit{\Upsilon}_{*}(K) = 0.6\pm 0.05$ throughout. If, instead we would use the mass-to-light ratio from \cite{2013A&A...557A.131M}, the final DM density estimations vary about $0.1\,$dex.

Finally, we estimate the halo masses using the halo abundance matching relation $M_{*}$-$M_{200}$ from  \cite{Moster:2012fv}. If instead we would use the one of \cite{Kravtsov:2014sra}, the final DM density estimations vary about $0.1\,$dex. Therefore, we estimate the systematic error on the estimation of the DM density by the combined use of the mass-to-light ratio and the halo abundance matching to be $0.2\,$dex.

\section{Dark Matter density profile}\label{sec:density}
The universal DM density profile is of the form \citep{Navarro:1995iw}
\begin{align}\label{eq:rho-chi}
\rho_{\rm DM} (r) = \frac{\rho_{\rm s}\;\! r_{\!\rm s}^{\,3}}{r\;\! (r_{\rm s}+r)^2}\,,
\end{align}
where $r_{\rm s}$ and $\rho_{\rm s}$ are scaling constants  determined from the halo mass $M_{200}$ in the following way: First, the scale radius 
\begin{align}\label{eq:r-s}
r_{\rm s} \equiv  r_{200}/c_{200}
\end{align}
is determined from the virial radius
\begin{align}\label{eq:r-200}
r_{200} = \Big(\frac{3\;\! M_{200}}{4\;\! \pi\;\! 200\;\! \rho_{\rm cr,0}}\Big)^{\! 1/3}\,,
\end{align}
where $\rho_{\rm cr,0}=1.34\times 10^{11}\,$M$_{\odot}\,$Mpc$^{-3}$ is the critical density today, and the concentration parameter is \citep{Dutton:2014xda}
\begin{align}\label{eq:c-200}
c_{200} = 10^{0.905} \;\! \Big( \frac{M_{200}}{10^{12}h^{-1}{\rm M}_{\odot}}\Big)^{-0.101} \,,
\end{align}
where $h=0.7$ is the reduced Hubble constant. Then, the scaling density is given by
\begin{align}\label{eq:rho-s}
\rho_{\rm s} = \frac{200\;\! \rho_{\rm cr,0}}{3}\;\! c_{200}^{\;3} \;\! g(c_{200}) \,,
\end{align}
with the auxiliary function
\begin{align}\label{eq:g(x)}
g(x) = \frac{1}{\ln(1+x) -x/(1+x)}\,.
\end{align}

\section{Dark Matter velocity dispersion}\label{sec:velocity}
Assuming DM particles distributed in a spherically symmetric density profile and in steady-state hydrodynamical equilibrium, the radial Jeans equation is
\begin{align}\label{eq:jeans}
    \frac{1}{\rho_{\rm DM}}\frac{\partial}{\partial r}(\rho_{\rm DM}\;\! v_{{\rm dm}, r}^{\,2}) + 2 \frac{(v_{{\rm dm},r}^{\, 2} - v_{{\rm dm},t}^{\,2})}{r} = - \frac{\partial \Phi}{\partial r} \,,
\end{align}
where $v_{{\rm dm},r}$ and $v_{{\rm dm},t}$ are the radial and transverse velocity dispersions and $\Phi$ is the gravitational potential. For the NFW profile given by equation~\eqref{eq:rho-chi}, the gravitational potential is
\citep{1996MNRAS.281..716C}
\begin{align}\label{eq:Phi}
\Phi(r) = - 4\;\! \pi\;\! G \;\! \rho_{\rm s}\;\! r_{\!\rm s}^{2}\;\!  \frac{\ln(1+r/r_{\rm s})}{r/r_{\rm s}}\,,
\end{align}
where $\rho_{\rm s}$ and $r_{\rm s}$ are given by equations \eqref{eq:rho-s} an \eqref{eq:r-s}. Further assuming isotropic velocity dispersion ($v_{{\rm dm},t} = v_{{\rm dm},r}$), and introducing equation \eqref{eq:Phi} in equation \eqref{eq:jeans}, the total square isotropic velocity dispersion is  \citep{Lokas:2000mu}
\begin{align}\label{eq:v}
v_{\rm DM}^{\;2}(r) = &\; \frac{3\;\! G\;\! M_{200}\;\!  g(c_{200})\;\! r}{2\;\! r_{\! s}^{\,2}} \Big(1+\frac{r}{r_{\;\!\!\rm s}}\Big)^2 \Bigg[\pi^2 - \frac{r_{\;\!\!\rm s}}{r} - \frac{6}{1+r/r_{\;\!\!\rm s}} \nonumber \\
&\; - \frac{1}{(1+r/r_{\!\rm s})^2} + \ln \Big(1+r/r_{\;\!\!\rm s} \Big)\Bigg(1+\frac{r_{\;\!\!\rm s}^{\,2}}{r^2}- \frac{4\;\! r_{\!\rm s}}{r} - \frac{2}{1+r/r_{\!\rm s}} \Bigg) \nonumber \\
&\; -\ln\Big(\frac{r}{r_{\;\!\!\rm s}}\Big)  + 3\ln^2(1+r/r_{\;\!\!\rm s}) + 6\, {\rm Li}_2\Big(\!-\frac{r}{r_{\;\!\!\rm s}}\Big)
   \Bigg]\,,
\end{align}
where $r_{200}$,  $c_{200}$ and $g(x)$ are given by equations \eqref{eq:r-200}, \eqref{eq:c-200} and \eqref{eq:g(x)} and ${\rm Li}_2(x)= \int_{x}^{0}[\ln(1-t)/t]\;\! {\rm d}t$ is the dilogarithm function. Strictly speaking, \cite{Lokas:2000mu} derived the radial part of the isotropic velocity dispersion, $v_{{\rm dm},r}$. In this work, we are interested in the total (three-dimensional) velocity dispersion, which is $v_{\rm DM} = \sqrt{3}\;\!v_{{\rm dm},r}$. Therefore, equation \eqref{eq:v} differs from equation (14) of \cite{Lokas:2000mu} by a factor 3.

\section{Stellar structure}\label{app:stellar-structure}

The number density laws for the WD population used by \cite{2003A&A...409..523R}. Thin disk
\begin{align}
    n(x,y,z)=n_0\Big[e^{-(0.5^2+a^2/h_+^2)^{1/2}} - e^{-(0.5^2+a^2/h_-^2)^{1/2}}\Big] 
\end{align}
where $a=(R^2+z^2/\epsilon^2)^{1/2}$ and $R=x^2+y^2$ is the galactocentric distance, $z$ is the  height above the galactic plane, and $\epsilon$ is the axis ratio. For the WD population of the thin disc, 
$\rho_0 = 3.96\times 10^{-3}$M$_{\odot}$pc$^{-3}$, $h_+=2530\,$pc, $h_-=1320\,$pc, and we assume $\epsilon=0.08$, and we assume $r_{\odot}=8000\,$pc, and we find $n_0=0.6069$.

For the thick disc
\begin{align}
    n(x,y,z) =&\; n_0 e^{-(r-r_{\odot})/h_r}\times \begin{cases}
        \Big(1-\frac{1/h_z}{x_l(2+x_l/h_z)}\;\!z^2\Big) & (|z|\leq x_l) \\
        \frac{e^{x_l/h_z}}{1+x_l/2 h_z}\;\!e^{-|z|/h_z} & (|z|>x_l)
    \end{cases}
\end{align}
For the WD population of the thick disc, $\rho_0 = 3.04\times 10^{-4}$, $h_r=2500\,$pc, $h_z=800\,$pc, and $x_l = 400\,$pc, and $n_0=0.0170$

For the spheroid (stellar halo)
\begin{align}
    n(x,y,z) =  n_0 \Big(\frac{{\rm max}(a,a_c)}{r_{\odot}}\Big)^{-2.44}
\end{align}
For the stellar halo $\rho_0 = 9.32\times 10^{-6}$, and $\epsilon = 0.76$, and we find $n_0=0.00111$.
Using the total numbers of WDs in each population from Table~\ref{tab:wd-population}, we find the normalizations $n_0$.


\section{Supernova Ia data}\label{sec:data-information}
In the following two Tables are presented the data. In Table~\ref{tab:data-host} we specify SN Ia host information. In Table~\ref{tab:data-sn} we specify particular SN Ia information.

\clearpage
\onecolumn

\begin{footnotesize}
\begin{longtable}{rllrrrllrrr}
\caption{Global SN host galaxy information}\\
\hline
ID & SN name & host name & Hubble type & $z$ & $\mu\pm$stat & method & ref & $i$ & $M_{*}\pm $stat & $M_{200}\pm $tot  \\
   &         &           &             &     & [mag] &  &  & [$^{\circ}$] & [dex(M$_{\odot}$)] & [dex(M$_{\odot}$)]\\
\hline\\[-3mm]
\endfirsthead
\multicolumn{4}{c}%
{\tablename\ \thetable\ -- \textit{Continued from previous page}} \\
\hline
ID & SN name & host name & Hubble type & $z$ & $\mu\pm$stat & method & ref & $i$ & $M_{*}\pm $stat & $M_{200}\pm $tot  \\
   &         &           &             &     & [mag] &  &  & [$^{\circ}$] & [dex(M$_{\odot}$)] & [dex(M$_{\odot}$)]\\
\hline
\endhead
\hline \multicolumn{10}{r}{\textit{Continued on next page}} \\
\endfoot
\hline
\endlastfoot
$1$ & $\text{1885A}$ & $\text{M31}$ & $3.0\pm 0.4$ & $0.0000$ & $24.40\pm 0.12$ & $\text{CTLS}$ & $\text{T16}$ & $72$ & $10.5\pm 0.1$ & $11.9\pm 0.4$ \\
$2$ & $\text{1939A}$ & $\text{NGC 4636}$ & $-4.8\pm 0.5$ & $0.0031$ & $30.90\pm 0.24$ & $\text{SF}$ & $\text{T16}$ & $64$ & $10.9\pm 0.2$ & $12.7\pm 0.7$ \\
$3$ & $\text{1939B}$ & $\text{M59}$ & $-4.8\pm 0.4$ & $0.0030$ & $30.90\pm 0.24$ & $\text{SF}$ & $\text{T16}$ & $72$ & $10.8\pm 0.2$ & $12.4\pm 0.7$ \\
$4$ & $\text{1980N}$ & $\text{NGC 1316}$ & $-1.8\pm 0.7$ & $0.0059$ & $31.20\pm 0.15$ & $\text{SNF}$ & $\text{T16}$ & $65$ & $11.3\pm 0.1$ & $13.9\pm 0.6$ \\
$5$ & $\text{1981B}$ & $\text{NGC 4536}$ & $4.3\pm 0.7$ & $0.0060$ & $30.90\pm 0.05$ & $\text{C}$ & $\text{R16}$ & $77$ & $10.4\pm 0.1$ & $11.9\pm 0.4$ \\
$6$ & $\text{1981D}$ & $\text{NGC 1316}$ & $-1.8\pm 0.7$ & $0.0059$ & $31.20\pm 0.15$ & $\text{SNF}$ & $\text{T16}$ & $65$ & $11.3\pm 0.1$ & $13.9\pm 0.6$ \\
$7$ & $\text{1986A}$ & $\text{NGC 3367}$ & $5.1\pm 0.6$ & $0.0100$ & $33.30\pm 0.35$ & $\text{M}$ & $\text{M14}$ & $48$ & $10.9\pm 0.2$ & $12.8\pm 0.8$ \\
$8$ & $\text{1986G}$ & $\text{NGC 5128}$ & $-2.1\pm 0.6$ & $0.0012$ & $27.80\pm 0.12$ & $\text{CTLS}$ & $\text{T16}$ & $70$ & $10.6\pm 0.1$ & $12.2\pm 0.6$ \\
$9$ & $\text{1989A}$ & $\text{NGC 3687}$ & $3.8\pm 0.7$ & $0.0084$ & $32.90\pm 0.41$ & $\text{M}$ & $\text{M14}$ & $27$ & $10.4\pm 0.2$ & $11.8\pm 0.6$ \\
$10$ & $\text{1989B}$ & $\text{NGC 3627}$ & $3.1\pm 0.4$ & $0.0024$ & $29.80\pm 0.13$ & $\text{CTHI}$ & $\text{T16}$ & $60$ & $10.6\pm 0.1$ & $12.2\pm 0.5$ \\
$11$ & $\text{1990N}$ & $\text{NGC 4639}$ & $3.6\pm 0.7$ & $0.0034$ & $31.50\pm 0.07$ & $\text{C}$ & $\text{R16}$ & $49$ & $10.2\pm 0.1$ & $11.6\pm 0.4$ \\
$12$ & $\text{1991bg}$ & $\text{NGC 4374}$ & $-4.4\pm 1.2$ & $0.0035$ & $31.10\pm 0.15$ & $\text{SNF}$ & $\text{T16}$ & $32$ & $11.0\pm 0.1$ & $13.1\pm 0.6$ \\
$13$ & $\text{1991T}$ & $\text{NGC 4527}$ & $4.0\pm 0.2$ & $0.0058$ & $30.60\pm 0.17$ & $\text{CHI}$ & $\text{T16}$ & $79$ & $10.6\pm 0.1$ & $12.1\pm 0.5$ \\
$14$ & $\text{1992al}$ & $\text{PGC 65335}$ & $5.1\pm 0.5$ & $0.0140$ & $33.70\pm 0.17$ & $\text{N}$ & $\text{T16}$ & $79$ & $10.2\pm 0.2$ & $11.6\pm 0.4$ \\
$15$ & $\text{1992bo}$ & $\text{PGC 4972}$ & $-1.5\pm 1.4$ & $0.0185$ & $34.50\pm 0.17$ & $\text{N}$ & $\text{T16}$ & $60$ & $10.5\pm 0.1$ & $12.0\pm 0.5$ \\
$16$ & $\text{1993H}$ & $\text{PGC 49304}$ & $1.9\pm 0.5$ & $0.0242$ & $34.80\pm 0.17$ & $\text{N}$ & $\text{T16}$ & $18$ & $10.7\pm 0.2$ & $12.4\pm 0.7$ \\
$17$ & $\text{1994ae}$ & $\text{NGC 3370}$ & $5.1\pm 1.1$ & $0.0043$ & $32.10\pm 0.05$ & $\text{C}$ & $\text{R16}$ & $68$ & $10.1\pm 0.1$ & $11.6\pm 0.4$ \\
$18$ & $\text{1994S}$ & $\text{NGC 4495}$ & $2.1\pm 0.6$ & $0.0152$ & $34.00\pm 0.17$ & $\text{N}$ & $\text{T16}$ & $69$ & $10.6\pm 0.1$ & $12.1\pm 0.5$ \\
$19$ & $\text{1995ak}$ & $\text{IC 1844}$ & $4.0\pm 2.0$ & $0.0227$ & $34.50\pm 0.15$ & $\text{NHI}$ & $\text{T16}$ & $68$ & $10.3\pm 0.1$ & $11.8\pm 0.4$ \\
$20$ & $\text{1995al}$ & $\text{NGC 3021}$ & $4.0\pm 0.2$ & $0.0068$ & $32.50\pm 0.09$ & $\text{C}$ & $\text{R16}$ & $60$ & $10.4\pm 0.1$ & $11.8\pm 0.4$ \\
$21$ & $\text{1995bd}$ & $\text{UGC 3151}$ & $4.0\pm 2.0$ & $0.0151$ & $33.70\pm 0.17$ & $\text{N}$ & $\text{T16}$ & $76$ & $10.6\pm 0.1$ & $12.1\pm 0.5$ \\
$22$ & $\text{1995D}$ & $\text{NGC 2962}$ & $-1.0\pm 0.7$ & $0.0066$ & $32.50\pm 0.20$ & $\text{N}$ & $\text{T16}$ & $50$ & $10.6\pm 0.1$ & $12.1\pm 0.5$ \\
$23$ & $\text{1995E}$ & $\text{NGC 2441}$ & $3.3\pm 0.9$ & $0.0116$ & $33.30\pm 0.17$ & $\text{N}$ & $\text{T16}$ & $3$ & $10.4\pm 0.1$ & $11.8\pm 0.5$ \\
$24$ & $\text{1996bl}$ & $\text{PGC 1392929}$ & $4.0\pm 1.5$ & $0.0349$ & $35.70\pm 0.17$ & $\text{N}$ & $\text{T16}$ & $25$ & $10.5\pm 0.1$ & $12.0\pm 0.5$ \\
$25$ & $\text{1996bo}$ & $\text{NGC 673}$ & $5.3\pm 0.6$ & $0.0173$ & $33.90\pm 0.15$ & $\text{NHI}$ & $\text{T16}$ & $48$ & $10.8\pm 0.1$ & $12.5\pm 0.6$ \\
$26$ & $\text{1997bp}$ & $\text{NGC 4680}$ & $0.3\pm 1.5$ & $0.0083$ & $32.40\pm 0.16$ & $\text{NH}$ & $\text{T16}$ & $55$ & $10.1\pm 0.1$ & $11.6\pm 0.4$ \\
$27$ & $\text{1997bq}$ & $\text{NGC 3147}$ & $3.9\pm 0.5$ & $0.0099$ & $33.00\pm 0.17$ & $\text{N}$ & $\text{T16}$ & $22$ & $11.3\pm 0.1$ & $13.8\pm 0.6$ \\
$28$ & $\text{1997do}$ & $\text{UGC 3845}$ & $3.7\pm 0.6$ & $0.0101$ & $33.10\pm 0.15$ & $\text{NHI}$ & $\text{T16}$ & $45$ & $10.0\pm 0.2$ & $11.5\pm 0.4$ \\
$29$ & $\text{1997E}$ & $\text{NGC 2258}$ & $-2.0\pm 0.4$ & $0.0135$ & $33.80\pm 0.17$ & $\text{N}$ & $\text{T16}$ & $45$ & $11.3\pm 0.1$ & $13.8\pm 0.6$ \\
$30$ & $\text{1998ab}$ & $\text{NGC 4704}$ & $3.8\pm 0.6$ & $0.0272$ & $34.90\pm 0.17$ & $\text{N}$ & $\text{T16}$ & $4$ & $10.7\pm 0.1$ & $12.4\pm 0.6$ \\
$31$ & $\text{1998aq}$ & $\text{NGC 3982}$ & $3.2\pm 0.6$ & $0.0037$ & $31.70\pm 0.07$ & $\text{C}$ & $\text{R16}$ & $20$ & $10.2\pm 0.1$ & $11.7\pm 0.4$ \\
$32$ & $\text{1998bp}$ & $\text{NGC 6495}$ & $-4.9\pm 0.4$ & $0.0104$ & $33.00\pm 0.17$ & $\text{N}$ & $\text{T16}$ & $65$ & $10.5\pm 0.1$ & $11.9\pm 0.5$ \\
$33$ & $\text{1998bu}$ & $\text{NGC 3368}$ & $2.1\pm 0.7$ & $0.0030$ & $30.10\pm 0.15$ & $\text{CSHI}$ & $\text{T16}$ & $52$ & $10.6\pm 0.1$ & $12.1\pm 0.5$ \\
$34$ & $\text{1998de}$ & $\text{NGC 252}$ & $-1.2\pm 0.7$ & $0.0166$ & $34.10\pm 0.17$ & $\text{N}$ & $\text{T16}$ & $35$ & $11.1\pm 0.1$ & $13.3\pm 0.6$ \\
$35$ & $\text{1998dh}$ & $\text{NGC 7541}$ & $4.7\pm 0.9$ & $0.0089$ & $32.50\pm 0.15$ & $\text{NHI}$ & $\text{T16}$ & $72$ & $10.8\pm 0.1$ & $12.4\pm 0.5$ \\
$36$ & $\text{1998ef}$ & $\text{UGC 646}$ & $2.5\pm 1.3$ & $0.0177$ & $33.90\pm 0.15$ & $\text{NHI}$ & $\text{T16}$ & $72$ & $10.5\pm 0.1$ & $11.9\pm 0.5$ \\
$37$ & $\text{1998es}$ & $\text{NGC 632}$ & $-1.3\pm 1.7$ & $0.0106$ & $32.80\pm 0.17$ & $\text{N}$ & $\text{T16}$ & $32$ & $10.2\pm 0.1$ & $11.6\pm 0.4$ \\
$38$ & $\text{1999aa}$ & $\text{NGC 2595}$ & $5.0\pm 0.4$ & $0.0144$ & $34.00\pm 0.17$ & $\text{N}$ & $\text{T16}$ & $22$ & $10.8\pm 0.2$ & $12.6\pm 0.7$ \\
$39$ & $\text{1999ac}$ & $\text{NGC 6063}$ & $5.8\pm 0.5$ & $0.0105$ & $33.10\pm 0.15$ & $\text{NHI}$ & $\text{T16}$ & $40$ & $10.1\pm 0.1$ & $11.5\pm 0.4$ \\
$40$ & $\text{1999bh}$ & $\text{NGC 3435}$ & $3.0\pm 0.4$ & $0.0172$ & $34.90\pm 0.40$ & $\text{H}$ & $\text{T16}$ & $54$ & $10.8\pm 0.2$ & $12.4\pm 0.8$ \\
$41$ & $\text{1999by}$ & $\text{NGC 2841}$ & $2.9\pm 0.5$ & $0.0021$ & $30.80\pm 0.14$ & $\text{CNH}$ & $\text{T16}$ & $68$ & $11.0\pm 0.1$ & $13.0\pm 0.6$ \\
$42$ & $\text{1999cc}$ & $\text{NGC 6038}$ & $5.1\pm 0.7$ & $0.0313$ & $35.50\pm 0.17$ & $\text{N}$ & $\text{T16}$ & $42$ & $11.2\pm 0.1$ & $13.5\pm 0.7$ \\
$43$ & $\text{1999cl}$ & $\text{NGC 4501}$ & $3.3\pm 0.6$ & $0.0076$ & $31.50\pm 0.30$ & $\text{HI}$ & $\text{T16}$ & $62$ & $11.2\pm 0.2$ & $13.5\pm 0.8$ \\
$44$ & $\text{1999cp}$ & $\text{NGC 5468}$ & $6.0\pm 0.3$ & $0.0098$ & $33.20\pm 0.17$ & $\text{N}$ & $\text{T16}$ & $15$ & $10.2\pm 0.2$ & $11.7\pm 0.5$ \\
$45$ & $\text{1999dk}$ & $\text{UGC 1087}$ & $5.3\pm 0.6$ & $0.0150$ & $33.90\pm 0.16$ & $\text{NH}$ & $\text{T16}$ & $14$ & $10.2\pm 0.2$ & $11.6\pm 0.4$ \\
$46$ & $\text{1999dq}$ & $\text{NGC 976}$ & $4.1\pm 0.6$ & $0.0143$ & $33.30\pm 0.16$ & $\text{NH}$ & $\text{T16}$ & $21$ & $10.8\pm 0.1$ & $12.4\pm 0.6$ \\
$47$ & $\text{1999ee}$ & $\text{IC 5179}$ & $4.0\pm 0.2$ & $0.0114$ & $33.10\pm 0.15$ & $\text{NHI}$ & $\text{T16}$ & $69$ & $10.9\pm 0.1$ & $12.8\pm 0.6$ \\
$48$ & $\text{1999ej}$ & $\text{NGC 495}$ & $0.2\pm 0.9$ & $0.0137$ & $34.10\pm 0.17$ & $\text{N}$ & $\text{T16}$ & $42$ & $10.7\pm 0.1$ & $12.4\pm 0.6$ \\
$49$ & $\text{1999ek}$ & $\text{UGC 3329}$ & $4.1\pm 0.8$ & $0.0175$ & $34.10\pm 0.15$ & $\text{NHI}$ & $\text{T16}$ & $72$ & $10.8\pm 0.1$ & $12.6\pm 0.6$ \\
$50$ & $\text{1999gh}$ & $\text{NGC 2986}$ & $-4.7\pm 0.6$ & $0.0077$ & $32.40\pm 0.17$ & $\text{NFP}$ & $\text{T16}$ & $52$ & $11.0\pm 0.1$ & $13.0\pm 0.6$ \\
$51$ & $\text{2000cn}$ & $\text{UGC 11064}$ & $5.9\pm 0.4$ & $0.0235$ & $34.90\pm 0.17$ & $\text{N}$ & $\text{T16}$ & $7$ & $11.0\pm 0.2$ & $13.0\pm 0.7$ \\
$52$ & $\text{2000cw}$ & $\text{PGC 72411}$ & $4.2\pm 2.5$ & $0.0301$ & $35.10\pm 0.30$ & $\text{HI}$ & $\text{T16}$ & $63$ & $10.7\pm 0.2$ & $12.3\pm 0.7$ \\
$53$ & $\text{2000cx}$ & $\text{NGC 524}$ & $-1.2\pm 0.6$ & $0.0080$ & $31.90\pm 0.28$ & $\text{S}$ & $\text{T16}$ & $4$ & $11.0\pm 0.2$ & $13.0\pm 0.7$ \\
$54$ & $\text{2000E}$ & $\text{NGC 6951}$ & $3.9\pm 0.6$ & $0.0052$ & $31.40\pm 0.20$ & $\text{N}$ & $\text{T16}$ & $44$ & $10.7\pm 0.2$ & $12.3\pm 0.7$ \\
$55$ & $\text{2001ah}$ & $\text{UGC 6211}$ & $4.1\pm 0.5$ & $0.0578$ & $36.80\pm 0.17$ & $\text{N}$ & $\text{T16}$ & $19$ & $11.1\pm 0.2$ & $13.2\pm 0.8$ \\
$56$ & $\text{2001ba}$ & $\text{PGC 36028}$ & $3.8\pm 0.6$ & $0.0294$ & $35.50\pm 0.16$ & $\text{NH}$ & $\text{T16}$ & $20$ & $11.1\pm 0.1$ & $13.2\pm 0.6$ \\
$57$ & $\text{2001bt}$ & $\text{IC 4830}$ & $3.9\pm 0.5$ & $0.0146$ & $33.50\pm 0.16$ & $\text{NFP}$ & $\text{T16}$ & $31$ & $10.9\pm 0.1$ & $12.7\pm 0.6$ \\
$58$ & $\text{2001cj}$ & $\text{UGC 8399}$ & $3.0\pm 0.4$ & $0.0242$ & $35.40\pm 0.03$ & $\text{M}$ & $\text{M14}$ & $12$ & $10.6\pm 0.1$ & $12.2\pm 0.4$ \\
$59$ & $\text{2001cz}$ & $\text{NGC 4679}$ & $4.9\pm 0.6$ & $0.0154$ & $33.80\pm 0.15$ & $\text{NHI}$ & $\text{T16}$ & $75$ & $10.8\pm 0.1$ & $12.6\pm 0.6$ \\
$60$ & $\text{2001da}$ & $\text{NGC 7780}$ & $2.0\pm 0.4$ & $0.0172$ & $34.10\pm 0.16$ & $\text{NH}$ & $\text{T16}$ & $71$ & $10.5\pm 0.1$ & $12.0\pm 0.5$ \\
$61$ & $\text{2001dl}$ & $\text{UGC 11725}$ & $8.1\pm 0.4$ & $0.0207$ & $35.30\pm 0.25$ & $\text{M}$ & $\text{M14}$ & $78$ & $10.6\pm 0.2$ & $12.1\pm 0.6$ \\
$62$ & $\text{2001eh}$ & $\text{UGC 1162}$ & $3.0\pm 0.4$ & $0.0370$ & $35.80\pm 0.17$ & $\text{N}$ & $\text{T16}$ & $9$ & $11.1\pm 0.2$ & $13.3\pm 0.7$ \\
$63$ & $\text{2001ep}$ & $\text{NGC 1699}$ & $3.1\pm 0.6$ & $0.0130$ & $33.50\pm 0.17$ & $\text{N}$ & $\text{T16}$ & $50$ & $10.2\pm 0.1$ & $11.7\pm 0.4$ \\
$64$ & $\text{2001gb}$ & $\text{IC 582}$ & $3.2\pm 2.1$ & $0.0257$ & $35.20\pm 0.17$ & $\text{N}$ & $\text{T16}$ & $13$ & $10.8\pm 0.1$ & $12.5\pm 0.6$ \\
$65$ & $\text{2002bf}$ & $\text{PGC 29953}$ & $3.0\pm 0.6$ & $0.0242$ & $34.90\pm 0.17$ & $\text{N}$ & $\text{T16}$ & $44$ & $10.7\pm 0.1$ & $12.2\pm 0.6$ \\
$66$ & $\text{2002bo}$ & $\text{NGC 3190}$ & $0.9\pm 0.6$ & $0.0054$ & $31.70\pm 0.17$ & $\text{N}$ & $\text{T16}$ & $73$ & $10.8\pm 0.1$ & $12.5\pm 0.6$ \\
$67$ & $\text{2002cd}$ & $\text{NGC 6916}$ & $4.0\pm 0.3$ & $0.0103$ & $33.20\pm 0.18$ & $\text{NH}$ & $\text{T16}$ & $52$ & $10.7\pm 0.1$ & $12.2\pm 0.5$ \\
$68$ & $\text{2002cf}$ & $\text{NGC 4786}$ & $-4.3\pm 0.5$ & $0.0155$ & $33.70\pm 0.50$ & $\text{P}$ & $\text{T16}$ & $55$ & $11.1\pm 0.3$ & $13.2\pm 0.9$ \\
$69$ & $\text{2002cr}$ & $\text{NGC 5468}$ & $6.0\pm 0.3$ & $0.0098$ & $33.20\pm 0.17$ & $\text{N}$ & $\text{T16}$ & $4$ & $10.2\pm 0.2$ & $11.7\pm 0.5$ \\
$70$ & $\text{2002cs}$ & $\text{NGC 6702}$ & $-4.9\pm 0.4$ & $0.0158$ & $33.60\pm 0.28$ & $\text{S}$ & $\text{T16}$ & $42$ & $10.8\pm 0.2$ & $12.6\pm 0.7$ \\
$71$ & $\text{2002cx}$ & $\text{PGC 45981}$ & $5.1\pm 2.6$ & $0.0240$ & $36.10\pm 0.03$ & $\text{M}$ & $\text{M14}$ & $79$ & $10.2\pm 0.1$ & $11.6\pm 0.4$ \\
$72$ & $\text{2002de}$ & $\text{NGC 6104}$ & $3.0\pm 2.4$ & $0.0281$ & $35.30\pm 0.17$ & $\text{N}$ & $\text{T16}$ & $39$ & $10.9\pm 0.1$ & $12.9\pm 0.6$ \\
$73$ & $\text{2002do}$ & $\text{PGC 63832}$ & $-5.0\pm 0.5$ & $0.0159$ & $34.10\pm 0.17$ & $\text{N}$ & $\text{T16}$ & $43$ & $11.1\pm 0.1$ & $13.3\pm 0.6$ \\
$74$ & $\text{2002dp}$ & $\text{NGC 7678}$ & $4.9\pm 0.5$ & $0.0116$ & $33.20\pm 0.17$ & $\text{N}$ & $\text{T16}$ & $27$ & $10.7\pm 0.1$ & $12.4\pm 0.6$ \\
$75$ & $\text{2002eb}$ & $\text{PGC 68560}$ & $2.0\pm 2.0$ & $0.0275$ & $35.50\pm 0.04$ & $\text{M}$ & $\text{M14}$ & $38$ & $10.7\pm 0.1$ & $12.3\pm 0.5$ \\
$76$ & $\text{2002ef}$ & $\text{NGC 7761}$ & $-2.0\pm 0.5$ & $0.0240$ & $33.70\pm 0.50$ & $\text{P}$ & $\text{T16}$ & $33$ & $10.7\pm 0.3$ & $12.3\pm 0.8$ \\
$77$ & $\text{2002el}$ & $\text{NGC 6986}$ & $-2.8\pm 0.8$ & $0.0287$ & $35.40\pm 0.04$ & $\text{M}$ & $\text{M14}$ & $62$ & $11.2\pm 0.1$ & $13.5\pm 0.5$ \\
$78$ & $\text{2002er}$ & $\text{UGC 10743}$ & $1.0\pm 0.5$ & $0.0086$ & $32.40\pm 0.17$ & $\text{NHI}$ & $\text{T16}$ & $73$ & $9.9\pm 0.1$ & $11.4\pm 0.4$ \\
$79$ & $\text{2002es}$ & $\text{UGC 2708}$ & $-2.0\pm 0.4$ & $0.0284$ & $34.50\pm 0.17$ & $\text{N}$ & $\text{T16}$ & $12$ & $11.2\pm 0.3$ & $13.4\pm 1.1$ \\
$80$ & $\text{2002fb}$ & $\text{NGC 759}$ & $-4.8\pm 0.4$ & $0.0156$ & $34.10\pm 0.17$ & $\text{NF}$ & $\text{T16}$ & $26$ & $11.1\pm 0.1$ & $13.2\pm 0.6$ \\
$81$ & $\text{2002fk}$ & $\text{NGC 1309}$ & $3.9\pm 0.6$ & $0.0071$ & $32.50\pm 0.06$ & $\text{C}$ & $\text{R16}$ & $5$ & $10.5\pm 0.1$ & $11.9\pm 0.4$ \\
$82$ & $\text{2002G}$ & $\text{PGC 45498}$ & $-3.5\pm 2.1$ & $0.0337$ & $35.80\pm 0.17$ & $\text{N}$ & $\text{T16}$ & $37$ & $11.0\pm 0.1$ & $12.9\pm 0.6$ \\
$83$ & $\text{2002ha}$ & $\text{NGC 6962}$ & $1.7\pm 0.6$ & $0.0140$ & $33.80\pm 0.15$ & $\text{NHI}$ & $\text{T16}$ & $42$ & $11.1\pm 0.1$ & $13.3\pm 0.6$ \\
$84$ & $\text{2002hw}$ & $\text{UGC 52}$ & $5.3\pm 0.6$ & $0.0175$ & $34.30\pm 0.17$ & $\text{N}$ & $\text{T16}$ & $20$ & $10.6\pm 0.1$ & $12.2\pm 0.6$ \\
$85$ & $\text{2002ic}$ & $\text{A013002+2153}$ & $9.5\pm 1.0$ & $0.0660$ & $37.20\pm 0.20$ & $\text{M}$ & $\text{M14}$ & $60$ & $9.3\pm 0.8$ & $11.1\pm 0.7$ \\
$86$ & $\text{2002jy}$ & $\text{NGC 477}$ & $5.0\pm 0.4$ & $0.0203$ & $34.70\pm 0.16$ & $\text{NH}$ & $\text{T16}$ & $65$ & $10.8\pm 0.2$ & $12.5\pm 0.8$ \\
$87$ & $\text{2003cg}$ & $\text{NGC 3169}$ & $1.2\pm 0.5$ & $0.0041$ & $31.80\pm 0.20$ & $\text{N}$ & $\text{T16}$ & $40$ & $10.9\pm 0.1$ & $12.8\pm 0.6$ \\
$88$ & $\text{2003cq}$ & $\text{NGC 3978}$ & $3.8\pm 0.6$ & $0.0333$ & $35.70\pm 0.17$ & $\text{N}$ & $\text{T16}$ & $3$ & $11.4\pm 0.1$ & $14.0\pm 0.7$ \\
$89$ & $\text{2003du}$ & $\text{UGC 9391}$ & $8.1\pm 0.5$ & $0.0064$ & $32.90\pm 0.06$ & $\text{C}$ & $\text{R16}$ & $41$ & $9.5\pm 0.1$ & $11.2\pm 0.4$ \\
$90$ & $\text{2003fa}$ & $\text{PGC 60771}$ & $2.7\pm 3.0$ & $0.0060$ & $35.90\pm 0.17$ & $\text{N}$ & $\text{T16}$ & $60$ & $10.9\pm 0.2$ & $12.7\pm 0.7$ \\
$91$ & $\text{2003gn}$ & $\text{PGC 69160}$ & $4.0\pm 0.3$ & $0.0345$ & $36.20\pm 0.15$ & $\text{M}$ & $\text{M14}$ & $72$ & $10.9\pm 0.1$ & $12.7\pm 0.6$ \\
$92$ & $\text{2003gq}$ & $\text{NGC 7407}$ & $4.7\pm 0.9$ & $0.0215$ & $35.20\pm 0.40$ & $\text{H}$ & $\text{T16}$ & $63$ & $11.0\pm 0.2$ & $13.0\pm 0.9$ \\
$93$ & $\text{2003he}$ & $\text{PGC 73123}$ & $4.0\pm 0.6$ & $0.0255$ & $35.10\pm 0.40$ & $\text{H}$ & $\text{T16}$ & $73$ & $10.6\pm 0.3$ & $12.0\pm 0.7$ \\
$94$ & $\text{2003W}$ & $\text{UGC 5234}$ & $5.4\pm 1.1$ & $0.0201$ & $34.50\pm 0.17$ & $\text{N}$ & $\text{T16}$ & $47$ & $10.8\pm 0.1$ & $12.4\pm 0.6$ \\
$95$ & $\text{2003Y}$ & $\text{IC 522}$ & $-1.9\pm 0.5$ & $0.0169$ & $34.20\pm 0.04$ & $\text{M}$ & $\text{M14}$ & $38$ & $10.8\pm 0.1$ & $12.4\pm 0.5$ \\
$96$ & $\text{2004at}$ & $\text{PGC 33043}$ & $3.6\pm 2.2$ & $0.0231$ & $35.20\pm 0.03$ & $\text{M}$ & $\text{M14}$ & $69$ & $10.9\pm 0.3$ & $12.7\pm 1.0$ \\
$97$ & $\text{2004bg}$ & $\text{UGC 6363}$ & $6.0\pm 0.5$ & $0.0210$ & $34.70\pm 0.17$ & $\text{N}$ & $\text{T16}$ & $81$ & $10.5\pm 0.1$ & $12.0\pm 0.5$ \\
$98$ & $\text{2004bk}$ & $\text{NGC 5246}$ & $3.1\pm 0.4$ & $0.0231$ & $35.20\pm 0.04$ & $\text{M}$ & $\text{M14}$ & $6$ & $10.7\pm 0.1$ & $12.4\pm 0.5$ \\
$99$ & $\text{2004bw}$ & $\text{PGC 53750}$ & $5.9\pm 0.6$ & $0.0212$ & $35.10\pm 0.10$ & $\text{M}$ & $\text{M14}$ & $29$ & $10.7\pm 0.2$ & $12.4\pm 0.8$ \\
$100$ & $\text{2004dt}$ & $\text{NGC 799}$ & $1.1\pm 0.6$ & $0.0193$ & $34.50\pm 0.04$ & $\text{M}$ & $\text{M14}$ & $29$ & $10.8\pm 0.1$ & $12.6\pm 0.5$ \\
$101$ & $\text{2004ef}$ & $\text{UGC 12158}$ & $3.1\pm 0.4$ & $0.0304$ & $35.40\pm 0.17$ & $\text{N}$ & $\text{T16}$ & $2$ & $10.8\pm 0.2$ & $12.6\pm 0.7$ \\
$102$ & $\text{2004eo}$ & $\text{NGC 6928}$ & $2.0\pm 0.2$ & $0.0152$ & $33.80\pm 0.17$ & $\text{N}$ & $\text{T16}$ & $77$ & $11.1\pm 0.1$ & $13.4\pm 0.6$ \\
$103$ & $\text{2004ey}$ & $\text{UGC 11816}$ & $4.3\pm 0.5$ & $0.0152$ & $34.00\pm 0.17$ & $\text{N}$ & $\text{T16}$ & $10$ & $10.1\pm 0.1$ & $11.6\pm 0.4$ \\
$104$ & $\text{2004fz}$ & $\text{NGC 783}$ & $5.3\pm 0.6$ & $0.0173$ & $33.90\pm 0.04$ & $\text{M}$ & $\text{M14}$ & $42$ & $10.8\pm 0.1$ & $12.5\pm 0.5$ \\
$105$ & $\text{2004gs}$ & $\text{PGC 24286}$ & $-3.6\pm 1.5$ & $0.0271$ & $35.40\pm 0.17$ & $\text{N}$ & $\text{T16}$ & $35$ & $10.8\pm 0.1$ & $12.6\pm 0.6$ \\
$106$ & $\text{2004L}$ & $\text{PGC 30719}$ & $4.9\pm 2.3$ & $0.0323$ & $35.70\pm 0.17$ & $\text{N}$ & $\text{T16}$ & $5$ & $10.6\pm 0.2$ & $12.1\pm 0.6$ \\
$107$ & $\text{2005A}$ & $\text{NGC 958}$ & $4.9\pm 0.5$ & $0.0187$ & $34.30\pm 0.15$ & $\text{NHIFP}$ & $\text{T16}$ & $71$ & $11.3\pm 0.1$ & $13.7\pm 0.6$ \\
$108$ & $\text{2005ag}$ & $\text{PGC 200171}$ & $3.6\pm 3.0$ & $0.0794$ & $37.60\pm 0.17$ & $\text{N}$ & $\text{T16}$ & $38$ & $11.1\pm 0.1$ & $13.4\pm 0.6$ \\
$109$ & $\text{2005al}$ & $\text{NGC 5304}$ & $-3.2\pm 1.1$ & $0.0128$ & $34.00\pm 0.15$ & $\text{NFP}$ & $\text{T16}$ & $50$ & $10.9\pm 0.1$ & $12.7\pm 0.6$ \\
$110$ & $\text{2005am}$ & $\text{NGC 2811}$ & $1.1\pm 0.4$ & $0.0083$ & $32.20\pm 0.16$ & $\text{NFP}$ & $\text{T16}$ & $70$ & $10.8\pm 0.1$ & $12.5\pm 0.6$ \\
$111$ & $\text{2005bg}$ & $\text{PGC 39401}$ & $3.4\pm 2.8$ & $0.0236$ & $35.00\pm 0.17$ & $\text{N}$ & $\text{T16}$ & $9$ & $10.3\pm 0.1$ & $11.7\pm 0.4$ \\
$112$ & $\text{2005bo}$ & $\text{NGC 4708}$ & $2.1\pm 0.8$ & $0.0145$ & $33.80\pm 0.17$ & $\text{N}$ & $\text{T16}$ & $40$ & $10.5\pm 0.1$ & $12.0\pm 0.5$ \\
$113$ & $\text{2005cc}$ & $\text{NGC 5383}$ & $3.1\pm 0.5$ & $0.0074$ & $32.80\pm 0.44$ & $\text{M}$ & $\text{M14}$ & $21$ & $10.8\pm 0.3$ & $12.4\pm 0.9$ \\
$114$ & $\text{2005de}$ & $\text{UGC 11097}$ & $6.0\pm 1.4$ & $0.0152$ & $34.50\pm 0.03$ & $\text{M}$ & $\text{M14}$ & $74$ & $10.7\pm 0.1$ & $12.3\pm 0.4$ \\
$115$ & $\text{2005el}$ & $\text{NGC 1819}$ & $-2.0\pm 0.4$ & $0.0149$ & $33.90\pm 0.17$ & $\text{N}$ & $\text{T16}$ & $40$ & $11.0\pm 0.1$ & $12.9\pm 0.6$ \\
$116$ & $\text{2005eq}$ & $\text{PGC 11767}$ & $1.4\pm 4.0$ & $0.0287$ & $35.40\pm 0.17$ & $\text{NI}$ & $\text{T16}$ & $73$ & $10.9\pm 0.2$ & $12.8\pm 0.7$ \\
$117$ & $\text{2005gj}$ & $\text{SDSS J030111.99-003313.5}$ & $4.5\pm 0.5$ & $0.0616$ & $37.50\pm 0.88$ & $\text{M}$ & $\text{M14}$ & $79$ & $9.0\pm 0.9$ & $11.0\pm 0.7$ \\
$118$ & $\text{2005hc}$ & $\text{PGC 7299}$ & $4.4\pm 2.5$ & $0.0450$ & $36.40\pm 0.17$ & $\text{N}$ & $\text{T16}$ & $50$ & $10.8\pm 0.1$ & $12.7\pm 0.7$ \\
$119$ & $\text{2005hk}$ & $\text{UGC 272}$ & $6.5\pm 0.8$ & $0.0130$ & $33.90\pm 0.30$ & $\text{HI}$ & $\text{T16}$ & $67$ & $9.5\pm 0.2$ & $11.2\pm 0.4$ \\
$120$ & $\text{2005iq}$ & $\text{PGC 73098}$ & $2.0\pm 1.4$ & $0.0335$ & $35.80\pm 0.17$ & $\text{N}$ & $\text{T16}$ & $60$ & $10.7\pm 0.1$ & $12.3\pm 0.6$ \\
$121$ & $\text{2005ir}$ & $\text{PGC 3116670}$ & $-2.1\pm 4.6$ & $0.0753$ & $37.60\pm 0.17$ & $\text{N}$ & $\text{T16}$ & $40$ & $10.6\pm 0.3$ & $12.1\pm 0.8$ \\
$122$ & $\text{2005kc}$ & $\text{NGC 7311}$ & $2.0\pm 0.3$ & $0.0147$ & $33.90\pm 0.17$ & $\text{N}$ & $\text{T16}$ & $59$ & $11.1\pm 0.1$ & $13.2\pm 0.6$ \\
$123$ & $\text{2005ke}$ & $\text{NGC 1371}$ & $1.1\pm 0.3$ & $0.0047$ & $32.30\pm 0.30$ & $\text{HI}$ & $\text{T16}$ & $42$ & $11.0\pm 0.2$ & $12.9\pm 0.8$ \\
$124$ & $\text{2005ki}$ & $\text{NGC 3332}$ & $-2.7\pm 1.0$ & $0.0192$ & $34.60\pm 0.17$ & $\text{N}$ & $\text{T16}$ & $32$ & $11.2\pm 0.1$ & $13.4\pm 0.6$ \\
$125$ & $\text{2005M}$ & $\text{NGC 2930}$ & $4.4\pm 2.7$ & $0.0225$ & $35.00\pm 0.17$ & $\text{N}$ & $\text{T16}$ & $60$ & $10.3\pm 0.2$ & $11.7\pm 0.4$ \\
$126$ & $\text{2005mc}$ & $\text{UGC 4414}$ & $0.2\pm 0.8$ & $0.0256$ & $35.20\pm 0.17$ & $\text{N}$ & $\text{T16}$ & $18$ & $10.9\pm 0.1$ & $12.8\pm 0.6$ \\
$127$ & $\text{2005ms}$ & $\text{UGC 4614}$ & $3.0\pm 1.8$ & $0.0252$ & $35.20\pm 0.17$ & $\text{N}$ & $\text{T16}$ & $8$ & $10.7\pm 0.2$ & $12.2\pm 0.6$ \\
$128$ & $\text{2005na}$ & $\text{UGC 3634}$ & $1.0\pm 0.4$ & $0.0266$ & $35.10\pm 0.17$ & $\text{N}$ & $\text{T16}$ & $51$ & $11.1\pm 0.1$ & $13.3\pm 0.6$ \\
$129$ & $\text{2005W}$ & $\text{NGC 691}$ & $4.0\pm 0.2$ & $0.0089$ & $32.90\pm 0.30$ & $\text{HI}$ & $\text{T16}$ & $51$ & $10.7\pm 0.2$ & $12.4\pm 0.7$ \\
$130$ & $\text{2006ac}$ & $\text{NGC 4619}$ & $3.1\pm 0.5$ & $0.0231$ & $34.90\pm 0.17$ & $\text{N}$ & $\text{T16}$ & $5$ & $11.1\pm 0.1$ & $13.2\pm 0.6$ \\
$131$ & $\text{2006ax}$ & $\text{NGC 3663}$ & $3.9\pm 0.5$ & $0.0167$ & $34.30\pm 0.16$ & $\text{NFP}$ & $\text{T16}$ & $32$ & $10.8\pm 0.2$ & $12.6\pm 0.7$ \\
$132$ & $\text{2006az}$ & $\text{NGC 4172}$ & $2.0\pm 2.0$ & $0.0309$ & $35.50\pm 0.17$ & $\text{N}$ & $\text{T16}$ & $30$ & $11.3\pm 0.1$ & $13.9\pm 0.6$ \\
$133$ & $\text{2006bh}$ & $\text{NGC 7329}$ & $3.6\pm 1.0$ & $0.0106$ & $33.30\pm 0.15$ & $\text{NHI}$ & $\text{T16}$ & $42$ & $10.9\pm 0.1$ & $12.7\pm 0.6$ \\
$134$ & $\text{2006bq}$ & $\text{NGC 6685}$ & $-3.0\pm 0.5$ & $0.0219$ & $34.80\pm 0.17$ & $\text{N}$ & $\text{T16}$ & $53$ & $10.9\pm 0.1$ & $12.7\pm 0.6$ \\
$135$ & $\text{2006br}$ & $\text{NGC 5185}$ & $3.0\pm 0.4$ & $0.0251$ & $35.50\pm 0.18$ & $\text{NH}$ & $\text{T16}$ & $74$ & $11.2\pm 0.1$ & $13.5\pm 0.6$ \\
$136$ & $\text{2006bt}$ & $\text{PGC 56443}$ & $-0.2\pm 1.0$ & $0.0322$ & $35.60\pm 0.17$ & $\text{N}$ & $\text{T16}$ & $28$ & $11.1\pm 0.1$ & $13.2\pm 0.6$ \\
$137$ & $\text{2006cp}$ & $\text{UGC 7357}$ & $5.3\pm 0.6$ & $0.0223$ & $34.70\pm 0.17$ & $\text{N}$ & $\text{T16}$ & $15$ & $10.1\pm 0.2$ & $11.5\pm 0.4$ \\
$138$ & $\text{2006D}$ & $\text{PGC 43690}$ & $2.1\pm 0.7$ & $0.0089$ & $32.80\pm 0.17$ & $\text{N}$ & $\text{T16}$ & $40$ & $10.0\pm 0.1$ & $11.5\pm 0.4$ \\
$139$ & $\text{2006ef}$ & $\text{NGC 809}$ & $-1.7\pm 0.7$ & $0.0174$ & $34.40\pm 0.17$ & $\text{N}$ & $\text{T16}$ & $45$ & $10.6\pm 0.1$ & $12.1\pm 0.5$ \\
$140$ & $\text{2006em}$ & $\text{NGC 911}$ & $-4.8\pm 0.6$ & $0.0192$ & $34.50\pm 0.46$ & $\text{F}$ & $\text{T16}$ & $75$ & $10.9\pm 0.2$ & $12.9\pm 0.9$ \\
$141$ & $\text{2006en}$ & $\text{PGC 70600}$ & $5.0\pm 2.0$ & $0.0319$ & $35.60\pm 0.17$ & $\text{N}$ & $\text{T16}$ & $11$ & $11.0\pm 0.1$ & $13.1\pm 0.6$ \\
$142$ & $\text{2006et}$ & $\text{NGC 232}$ & $1.1\pm 0.5$ & $0.0217$ & $34.60\pm 0.17$ & $\text{N}$ & $\text{T16}$ & $47$ & $11.0\pm 0.1$ & $12.9\pm 0.6$ \\
$143$ & $\text{2006gr}$ & $\text{UGC 12071}$ & $3.1\pm 0.5$ & $0.0346$ & $35.80\pm 0.17$ & $\text{N}$ & $\text{T16}$ & $41$ & $11.6\pm 0.2$ & $14.5\pm 0.9$ \\
$144$ & $\text{2006gz}$ & $\text{IC 1277}$ & $5.8\pm 0.6$ & $0.0280$ & $35.50\pm 0.13$ & $\text{M}$ & $\text{M14}$ & $48$ & $10.5\pm 0.2$ & $12.0\pm 0.6$ \\
$145$ & $\text{2006hb}$ & $\text{PGC 16595}$ & $-3.2\pm 1.2$ & $0.0153$ & $33.70\pm 0.17$ & $\text{NFP}$ & $\text{T16}$ & $55$ & $10.5\pm 0.1$ & $11.9\pm 0.5$ \\
$146$ & $\text{2006mr}$ & $\text{NGC 1316}$ & $-1.8\pm 0.7$ & $0.0055$ & $31.20\pm 0.15$ & $\text{SNF}$ & $\text{T16}$ & $64$ & $11.3\pm 0.1$ & $13.9\pm 0.6$ \\
$147$ & $\text{2006ob}$ & $\text{UGC 1333}$ & $2.9\pm 0.7$ & $0.0589$ & $36.90\pm 0.17$ & $\text{N}$ & $\text{T16}$ & $44$ & $11.3\pm 0.1$ & $13.8\pm 0.7$ \\
$148$ & $\text{2006ot}$ & $\text{PGC 8610}$ & $1.4\pm 1.3$ & $0.0526$ & $36.60\pm 0.16$ & $\text{NFP}$ & $\text{T16}$ & $65$ & $11.4\pm 0.2$ & $14.0\pm 0.7$ \\
$149$ & $\text{2006S}$ & $\text{UGC 7934}$ & $1.5\pm 3.5$ & $0.0321$ & $35.80\pm 0.17$ & $\text{N}$ & $\text{T16}$ & $69$ & $10.6\pm 0.1$ & $12.1\pm 0.6$ \\
$150$ & $\text{2006X}$ & $\text{M100}$ & $4.0\pm 0.3$ & $0.0053$ & $30.70\pm 0.12$ & $\text{CNH}$ & $\text{T16}$ & $24$ & $10.7\pm 0.2$ & $12.3\pm 0.7$ \\
$151$ & $\text{2007A}$ & $\text{NGC 105}$ & $1.7\pm 0.7$ & $0.0169$ & $34.10\pm 0.17$ & $\text{N}$ & $\text{T16}$ & $40$ & $10.6\pm 0.1$ & $12.1\pm 0.5$ \\
$152$ & $\text{2007af}$ & $\text{NGC 5584}$ & $5.9\pm 0.3$ & $0.0055$ & $31.80\pm 0.05$ & $\text{C}$ & $\text{R16}$ & $40$ & $9.0\pm 0.2$ & $11.0\pm 0.4$ \\
$153$ & $\text{2007ax}$ & $\text{NGC 2577}$ & $-2.9\pm 0.5$ & $0.0072$ & $32.50\pm 0.51$ & $\text{M}$ & $\text{M14}$ & $74$ & $10.4\pm 0.3$ & $11.9\pm 0.6$ \\
$154$ & $\text{2007ba}$ & $\text{UGC 9798}$ & $-0.1\pm 0.4$ & $0.0387$ & $36.40\pm 0.20$ & $\text{N}$ & $\text{T16}$ & $70$ & $11.4\pm 0.2$ & $14.0\pm 0.7$ \\
$155$ & $\text{2007bc}$ & $\text{UGC 6332}$ & $1.0\pm 0.4$ & $0.0211$ & $34.70\pm 0.17$ & $\text{N}$ & $\text{T16}$ & $23$ & $10.9\pm 0.2$ & $12.7\pm 0.7$ \\
$156$ & $\text{2007bd}$ & $\text{UGC 4455}$ & $1.0\pm 0.4$ & $0.0315$ & $35.50\pm 0.16$ & $\text{NFP}$ & $\text{T16}$ & $42$ & $10.9\pm 0.2$ & $12.8\pm 0.7$ \\
$157$ & $\text{2007bj}$ & $\text{NGC 6172}$ & $-4.2\pm 0.8$ & $0.0167$ & $34.60\pm 0.13$ & $\text{M}$ & $\text{M14}$ & $13$ & $11.0\pm 0.1$ & $13.0\pm 0.6$ \\
$158$ & $\text{2007bm}$ & $\text{NGC 3672}$ & $5.0\pm 0.4$ & $0.0066$ & $32.20\pm 0.17$ & $\text{NHI}$ & $\text{T16}$ & $67$ & $10.6\pm 0.1$ & $12.2\pm 0.5$ \\
$159$ & $\text{2007ci}$ & $\text{NGC 3873}$ & $-4.9\pm 0.4$ & $0.0181$ & $34.40\pm 0.17$ & $\text{NF}$ & $\text{T16}$ & $28$ & $10.9\pm 0.1$ & $12.8\pm 0.6$ \\
$160$ & $\text{2007F}$ & $\text{UGC 8162}$ & $5.9\pm 0.7$ & $0.0236$ & $35.10\pm 0.17$ & $\text{N}$ & $\text{T16}$ & $1$ & $10.2\pm 0.2$ & $11.6\pm 0.4$ \\
$161$ & $\text{2007hj}$ & $\text{NGC 7461}$ & $-1.9\pm 0.5$ & $0.0137$ & $33.80\pm 0.05$ & $\text{M}$ & $\text{M14}$ & $34$ & $10.6\pm 0.1$ & $12.1\pm 0.4$ \\
$162$ & $\text{2007kk}$ & $\text{UGC 2828}$ & $3.2\pm 0.4$ & $0.0410$ & $36.10\pm 0.20$ & $\text{N}$ & $\text{T16}$ & $32$ & $11.3\pm 0.1$ & $13.8\pm 0.6$ \\
$163$ & $\text{2007le}$ & $\text{NGC 7721}$ & $4.9\pm 0.5$ & $0.0061$ & $31.50\pm 0.30$ & $\text{HI}$ & $\text{T16}$ & $81$ & $10.2\pm 0.2$ & $11.6\pm 0.5$ \\
$164$ & $\text{2007N}$ & $\text{PGC 43319}$ & $0.9\pm 1.8$ & $0.0129$ & $34.10\pm 0.20$ & $\text{N}$ & $\text{T16}$ & $72$ & $10.4\pm 0.2$ & $11.8\pm 0.5$ \\
$165$ & $\text{2007nq}$ & $\text{UGC 595}$ & $-4.4\pm 1.2$ & $0.0443$ & $36.20\pm 0.19$ & $\text{F}$ & $\text{T16}$ & $24$ & $11.1\pm 0.3$ & $13.3\pm 1.1$ \\
$166$ & $\text{2007O}$ & $\text{UGC 9612}$ & $4.9\pm 0.5$ & $0.0362$ & $35.80\pm 0.17$ & $\text{N}$ & $\text{T16}$ & $8$ & $10.9\pm 0.2$ & $12.8\pm 0.7$ \\
$167$ & $\text{2007on}$ & $\text{NGC 1404}$ & $-4.8\pm 0.5$ & $0.0063$ & $31.40\pm 0.22$ & $\text{SFP}$ & $\text{T16}$ & $44$ & $10.9\pm 0.1$ & $12.8\pm 0.7$ \\
$168$ & $\text{2007S}$ & $\text{UGC 5378}$ & $3.1\pm 0.5$ & $0.0145$ & $33.90\pm 0.17$ & $\text{N}$ & $\text{T16}$ & $49$ & $10.1\pm 0.1$ & $11.6\pm 0.4$ \\
$169$ & $\text{2007st}$ & $\text{NGC 692}$ & $4.1\pm 0.7$ & $0.0212$ & $34.70\pm 0.18$ & $\text{M}$ & $\text{M14}$ & $45$ & $11.2\pm 0.1$ & $13.6\pm 0.6$ \\
$170$ & $\text{2008ae}$ & $\text{IC 577}$ & $4.8\pm 1.8$ & $0.0301$ & $35.60\pm 0.12$ & $\text{M}$ & $\text{M14}$ & $17$ & $10.7\pm 0.1$ & $12.3\pm 0.6$ \\
$171$ & $\text{2008ar}$ & $\text{IC 3284}$ & $1.8\pm 2.0$ & $0.0261$ & $35.30\pm 0.20$ & $\text{N}$ & $\text{T16}$ & $35$ & $10.4\pm 0.2$ & $11.9\pm 0.5$ \\
$172$ & $\text{2008bc}$ & $\text{PGC 90108}$ & $3.0\pm 2.0$ & $0.0157$ & $33.90\pm 0.20$ & $\text{N}$ & $\text{T16}$ & $43$ & $10.1\pm 0.1$ & $11.6\pm 0.4$ \\
$173$ & $\text{2008bf}$ & $\text{NGC 4055}$ & $-5.0\pm 0.5$ & $0.0240$ & $35.30\pm 0.03$ & $\text{M}$ & $\text{M14}$ & $45$ & $11.3\pm 0.2$ & $13.7\pm 0.7$ \\
$174$ & $\text{2008bi}$ & $\text{NGC 2618}$ & $1.5\pm 0.6$ & $0.0137$ & $33.80\pm 0.28$ & $\text{M}$ & $\text{M14}$ & $41$ & $10.9\pm 0.2$ & $12.8\pm 0.8$ \\
$175$ & $\text{2008cc}$ & $\text{PGC 66000}$ & $-4.0\pm 0.7$ & $0.0104$ & $32.00\pm 0.50$ & $\text{P}$ & $\text{T16}$ & $46$ & $10.2\pm 0.3$ & $11.7\pm 0.5$ \\
$176$ & $\text{2008fp}$ & $\text{PGC 20551}$ & $-1.8\pm 0.9$ & $0.0060$ & $31.60\pm 0.75$ & $\text{M}$ & $\text{M14}$ & $62$ & $10.2\pm 0.4$ & $11.7\pm 0.6$ \\
$177$ & $\text{2008fu}$ & $\text{PGC 11470}$ & $5.0\pm 3.0$ & $0.0524$ & $36.80\pm 0.07$ & $\text{M}$ & $\text{M14}$ & $54$ & $10.9\pm 0.2$ & $12.8\pm 0.7$ \\
$178$ & $\text{2008fw}$ & $\text{NGC 3261}$ & $3.6\pm 0.9$ & $0.0084$ & $32.70\pm 0.46$ & $\text{M}$ & $\text{M14}$ & $41$ & $10.9\pm 0.3$ & $12.7\pm 0.9$ \\
$179$ & $\text{2008gl}$ & $\text{UGC 881}$ & $-4.8\pm 0.6$ & $0.0340$ & $35.60\pm 0.20$ & $\text{N}$ & $\text{T16}$ & $64$ & $11.0\pm 0.2$ & $13.1\pm 0.7$ \\
$180$ & $\text{2008gp}$ & $\text{PGC 12669}$ & $1.5\pm 1.9$ & $0.0328$ & $35.60\pm 0.20$ & $\text{N}$ & $\text{T16}$ & $38$ & $10.9\pm 0.2$ & $12.7\pm 0.7$ \\
$181$ & $\text{2008ha}$ & $\text{UGC 12682}$ & $9.8\pm 0.5$ & $0.0046$ & $31.70\pm 0.74$ & $\text{M}$ & $\text{M14}$ & $36$ & $9.4\pm 0.4$ & $11.2\pm 0.5$ \\
$182$ & $\text{2008hj}$ & $\text{PGC 282}$ & $1.3\pm 1.3$ & $0.0379$ & $36.00\pm 0.20$ & $\text{N}$ & $\text{T16}$ & $38$ & $10.5\pm 0.2$ & $11.9\pm 0.5$ \\
$183$ & $\text{2008J}$ & $\text{PGC 9800}$ & $3.9\pm 0.6$ & $0.0159$ & $34.10\pm 0.30$ & $\text{HI}$ & $\text{T16}$ & $70$ & $10.7\pm 0.2$ & $12.2\pm 0.7$ \\
$184$ & $\text{2008L}$ & $\text{NGC 1259}$ & $-3.0\pm 2.0$ & $0.0193$ & $34.20\pm 0.17$ & $\text{N}$ & $\text{T16}$ & $57$ & $10.6\pm 0.1$ & $12.1\pm 0.5$ \\
$185$ & $\text{2008R}$ & $\text{NGC 1200}$ & $-2.9\pm 0.6$ & $0.0129$ & $33.50\pm 0.19$ & $\text{P}$ & $\text{T16}$ & $64$ & $11.1\pm 0.1$ & $13.2\pm 0.6$ \\
$186$ & $\text{2008s1}$ & $\text{UGC 8472}$ & $-1.9\pm 0.6$ & $0.0221$ & $35.20\pm 0.14$ & $\text{M}$ & $\text{M14}$ & $69$ & $10.9\pm 0.1$ & $12.7\pm 0.7$ \\
$187$ & $\text{2009al}$ & $\text{NGC 3425}$ & $-2.0\pm 0.4$ & $0.0231$ & $34.90\pm 0.20$ & $\text{N}$ & $\text{T16}$ & $34$ & $10.8\pm 0.1$ & $12.6\pm 0.6$ \\
$188$ & $\text{2009an}$ & $\text{NGC 4332}$ & $1.1\pm 0.5$ & $0.0092$ & $33.20\pm 0.36$ & $\text{M}$ & $\text{M14}$ & $41$ & $10.7\pm 0.2$ & $12.3\pm 0.7$ \\
$189$ & $\text{2009cz}$ & $\text{NGC 2789}$ & $-0.2\pm 0.8$ & $0.0211$ & $34.90\pm 0.17$ & $\text{M}$ & $\text{M14}$ & $31$ & $11.1\pm 0.1$ & $13.2\pm 0.6$ \\
$190$ & $\text{2009D}$ & $\text{PGC 14075}$ & $3.2\pm 1.3$ & $0.0250$ & $35.00\pm 0.18$ & $\text{H}$ & $\text{T16}$ & $63$ & $10.6\pm 0.1$ & $12.0\pm 0.5$ \\
$191$ & $\text{2009dc}$ & $\text{UGC 10064}$ & $-1.8\pm 1.1$ & $0.0216$ & $34.90\pm 0.16$ & $\text{M}$ & $\text{M14}$ & $79$ & $10.8\pm 0.1$ & $12.5\pm 0.6$ \\
$192$ & $\text{2009ds}$ & $\text{NGC 3905}$ & $4.7\pm 0.5$ & $0.0192$ & $34.70\pm 0.20$ & $\text{N}$ & $\text{T16}$ & $48$ & $11.0\pm 0.2$ & $12.9\pm 0.7$ \\
$193$ & $\text{2009F}$ & $\text{NGC 1725}$ & $-2.6\pm 0.6$ & $0.0129$ & $33.60\pm 0.30$ & $\text{M}$ & $\text{M14}$ & $25$ & $10.8\pm 0.2$ & $12.5\pm 0.7$ \\
$194$ & $\text{2009fv}$ & $\text{NGC 6173}$ & $-4.8\pm 0.5$ & $0.0293$ & $34.90\pm 0.44$ & $\text{F}$ & $\text{T16}$ & $77$ & $11.3\pm 0.2$ & $13.8\pm 0.9$ \\
$195$ & $\text{2009I}$ & $\text{NGC 1080}$ & $4.7\pm 0.5$ & $0.0262$ & $35.30\pm 0.14$ & $\text{M}$ & $\text{M14}$ & $33$ & $10.8\pm 0.1$ & $12.6\pm 0.6$ \\
$196$ & $\text{2009ig}$ & $\text{NGC 1015}$ & $1.0\pm 0.4$ & $0.0086$ & $32.50\pm 0.08$ & $\text{C}$ & $\text{R16}$ & $2$ & $10.4\pm 0.2$ & $11.8\pm 0.5$ \\
$197$ & $\text{2009le}$ & $\text{PGC 8223}$ & $4.2\pm 0.7$ & $0.0178$ & $34.20\pm 0.18$ & $\text{H}$ & $\text{T16}$ & $57$ & $10.9\pm 0.1$ & $12.8\pm 0.6$ \\
$198$ & $\text{2009Y}$ & $\text{NGC 5728}$ & $1.2\pm 0.7$ & $0.0093$ & $32.90\pm 0.18$ & $\text{I}$ & $\text{T16}$ & $53$ & $11.0\pm 0.1$ & $12.9\pm 0.6$ \\
$199$ & $\text{2010B}$ & $\text{NGC 5370}$ & $-1.8\pm 0.7$ & $0.0102$ & $33.40\pm 0.33$ & $\text{M}$ & $\text{M14}$ & $16$ & $10.4\pm 0.2$ & $11.9\pm 0.5$ \\
$200$ & $\text{2010kg}$ & $\text{NGC 1633}$ & $2.1\pm 0.7$ & $0.0166$ & $34.20\pm 0.22$ & $\text{M}$ & $\text{M14}$ & $36$ & $10.7\pm 0.1$ & $12.2\pm 0.6$ \\
$201$ & $\text{2011by}$ & $\text{NGC 3972}$ & $4.0\pm 0.3$ & $0.0028$ & $31.60\pm 0.07$ & $\text{C}$ & $\text{R16}$ & $76$ & $9.9\pm 0.1$ & $11.4\pm 0.3$ \\
$202$ & $\text{2011de}$ & $\text{UGC 10018}$ & $3.6\pm 0.6$ & $0.0292$ & $35.60\pm 0.12$ & $\text{M}$ & $\text{M14}$ & $15$ & $10.9\pm 0.1$ & $12.8\pm 0.7$ \\
$203$ & $\text{2011fe}$ & $\text{NGC 5457}$ & $5.9\pm 0.3$ & $0.0008$ & $29.10\pm 0.05$ & $\text{C}$ & $\text{R16}$ & $16$ & $10.5\pm 0.1$ & $11.9\pm 0.5$ \\
$204$ & $\text{2011hb}$ & $\text{NGC 7674}$ & $3.8\pm 0.6$ & $0.0289$ & $35.50\pm 0.12$ & $\text{M}$ & $\text{M14}$ & $18$ & $11.4\pm 0.1$ & $14.0\pm 0.6$ \\
$205$ & $\text{2011im}$ & $\text{NGC 7364}$ & $0.2\pm 1.1$ & $0.0162$ & $34.20\pm 0.40$ & $\text{H}$ & $\text{T16}$ & $52$ & $11.0\pm 0.2$ & $12.9\pm 0.8$ \\
$206$ & $\text{2012cg}$ & $\text{NGC 4424}$ & $1.3\pm 1.6$ & $0.0015$ & $31.10\pm 0.29$ & $\text{C}$ & $\text{R16}$ & $64$ & $9.9\pm 0.2$ & $11.4\pm 0.4$ \\
$207$ & $\text{2012dn}$ & $\text{PGC 64605}$ & $5.9\pm 0.5$ & $0.0102$ & $33.00\pm 0.40$ & $\text{H}$ & $\text{T16}$ & $45$ & $10.0\pm 0.3$ & $11.5\pm 0.5$ \\
$208$ & $\text{2012fr}$ & $\text{NGC 1365}$ & $3.2\pm 0.7$ & $0.0054$ & $31.30\pm 0.06$ & $\text{C}$ & $\text{R16}$ & $59$ & $11.0\pm 0.1$ & $13.2\pm 0.6$ \\
$209$ & $\text{2012hr}$ & $\text{PGC 18880}$ & $3.9\pm 0.5$ & $0.0080$ & $33.00\pm 0.40$ & $\text{H}$ & $\text{T16}$ & $43$ & $10.7\pm 0.2$ & $12.3\pm 0.7$ \\
$210$ & $\text{2012ht}$ & $\text{NGC 3447}$ & $8.8\pm 0.6$ & $0.0036$ & $31.90\pm 0.04$ & $\text{C}$ & $\text{R16}$ & $55$ & $8.8\pm 0.1$ & $10.9\pm 0.3$ \\
$211$ & $\text{2012Z}$ & $\text{NGC 1309}$ & $3.9\pm 0.6$ & $0.0071$ & $32.50\pm 0.14$ & $\text{CN}$ & $\text{T16}$ & $9$ & $10.4\pm 0.1$ & $11.9\pm 0.4$ \\
$212$ & $\text{2013aa}$ & $\text{NGC 5643}$ & $5.0\pm 0.3$ & $0.0040$ & $31.00\pm 1.01$ & $\text{M}$ & $\text{M14}$ & $27$ & $10.6\pm 0.5$ & $12.2\pm 1.1$ \\
$213$ & $\text{2013cv}$ & $\text{PGC 5068206}$ & $9.5\pm 1.5$ & $0.0350$ & $36.00\pm 1.00$ & $\text{M}$ & $\text{M14}$ & $20$ & $9.1\pm 0.9$ & $11.0\pm 0.7$ \\
$214$ & $\text{2013dy}$ & $\text{NGC 7250}$ & $7.2\pm 2.0$ & $0.0039$ & $31.50\pm 0.08$ & $\text{C}$ & $\text{R16}$ & $79$ & $9.6\pm 0.1$ & $11.3\pm 0.4$ \\
$215$ & $\text{2013gs}$ & $\text{UGC 5066}$ & $4.1\pm 0.5$ & $0.0169$ & $34.40\pm 0.21$ & $\text{M}$ & $\text{M14}$ & $69$ & $9.9\pm 0.2$ & $11.4\pm 0.4$ \\
$216$ & $\text{2013gy}$ & $\text{NGC 1418}$ & $3.1\pm 0.7$ & $0.0140$ & $33.30\pm 0.40$ & $\text{H}$ & $\text{T16}$ & $57$ & $10.1\pm 0.2$ & $11.6\pm 0.5$ \\
$217$ & $\text{2013Q}$ & $\text{NGC 7753}$ & $3.8\pm 0.6$ & $0.0172$ & $34.40\pm 0.21$ & $\text{M}$ & $\text{M14}$ & $32$ & $11.4\pm 0.1$ & $14.1\pm 0.7$ \\
$218$ & $\text{2014dg}$ & $\text{UGC 2855}$ & $5.0\pm 0.5$ & $0.0040$ & $30.80\pm 0.40$ & $\text{H}$ & $\text{T16}$ & $66$ & $10.5\pm 0.2$ & $12.0\pm 0.6$ \\
$219$ & $\text{2015F}$ & $\text{NGC 2442}$ & $3.7\pm 0.6$ & $0.0054$ & $31.50\pm 0.05$ & $\text{C}$ & $\text{R16}$ & $30$ & $10.9\pm 0.1$ & $12.9\pm 0.5$ \\
$220$ & $\text{2016bln}$ & $\text{NGC 5221}$ & $2.9\pm 0.5$ & $0.0233$ & $35.10\pm 0.15$ & $\text{M}$ & $\text{M14}$ & $62$ & $11.1\pm 0.1$ & $13.3\pm 0.6$ \\
$221$ & $\text{2016dwu}$ & $\text{PGC 11768}$ & $7.0\pm 0.4$ & $0.0149$ & $33.90\pm 0.26$ & $\text{M}$ & $\text{M14}$ & $34$ & $9.8\pm 0.5$ & $11.4\pm 0.6$ \\
$222$ & $\text{2018oh}$ & $\text{UGC 4780}$ & $8.0\pm 0.5$ & $0.0110$ & $33.40\pm 0.33$ & $\text{M}$ & $\text{M14}$ & $39$ & $9.5\pm 0.4$ & $11.2\pm 0.5$ \\
$223$ & $\text{LSQ12gdj}$ & $\text{PGC 72841}$ & $4.6\pm 2.2$ & $0.0300$ & $35.10\pm 0.20$ & $\text{N}$ & $\text{T16}$ & $38$ & $9.6\pm 0.2$ & $11.3\pm 0.4$ \\
$224$ & $\text{PTF11kx}$ & $\text{PGC 3131180}$ & $4.3\pm 2.6$ & $0.0466$ & $36.60\pm 0.08$ & $\text{M}$ & $\text{M14}$ & $65$ & $10.2\pm 0.1$ & $11.6\pm 0.4$ \\
\\[-3mm]
\hline\\[-3mm]
\caption{List of host galaxies used in the analysis. $z$ = redshift (the error $\delta z$ is globally neglected); $\mu$ = distance modulus; method for distance modulus determination: C = cepheid, T = tip of RGB from HST, L = tip of RGB from literature, M = miscellaneous, S = surface brightness fluctuation, N = SN Ia, H = Tully-Fisher relation with optical photometry, I = Tully-Fisher relation with 3.6 Spitzer, F = fundamental plane from CF2, P = fundamental plane from 6dFDS; reference for the distance modulus: T16 = \citet{Tully:2016ppz} (cosmicflows-3), R16 = \citet{Riess:2016jrr}, M14 = \citet{Makarov:2014txa} (HyperLEDA: \url{http://leda.univ-lyon1.fr/}); $i$ = inclination of galactic disk ($0^{\circ}$ = face-on, $90^{\circ}$ = edge-on) from HyperLEDA (error estimation globally $\delta i = 2^{\circ}$); $M_{*}$ = stellar mass;  $M_{200}$ = halo mass; stat = statistical error; tot = total error (statistical + halo abundance matching)}
\label{tab:data-host}
\end{longtable}
\end{footnotesize}

\begin{footnotesize}
\begin{longtable}{rllrrrrrrrr}
\caption{Local SN information}\\
\hline
ID & SN name & Ia-type & $M_B\pm$stat & $\theta$ & $r_{\:\!\!\perp}\pm$stat  & $\phi$ & $r\pm$stat$\pm$sys & $\rho_{\rm DM}\pm$stat$\pm$sys & $v_{\rm DM}\pm $stat & $t\pm$stat \\
  &         &      & [mag]              & 
  ["]  & [kpc]  & [$^{\circ}$] &  [kpc]  & $\!\!\!\!\!$[dex(M$_{\odot}\,$Mpc$^{-3}$)]$\!\!\!$ & $\!\!\!$[$100\,$km$\,$s$^{-1}$]$\!\!\!\!\!$  & [dex(Gyr)] \\
\hline\\[-3mm]
\endfirsthead
\multicolumn{4}{c}%
{\tablename\ \thetable\ -- \textit{Continued from previous page}} \\
\hline
ID & SN name & Ia-type & $M_B\pm$stat & $\theta$ & $r_{\:\!\!\perp}\pm$stat  & $\phi$ & $r\pm$stat$\pm$sys & $\rho_{\rm DM}\pm$stat$\pm$sys & $v_{\rm DM}\pm $stat & $t\pm$stat \\
  &         &      & [mag]              & 
  ["]  & [kpc]  & [$^{\circ}$] &  [kpc]  & $\!\!\!\!\!$[dex(M$_{\odot}\,$Mpc$^{-3}$)]$\!\!\!$ & $\!\!\!$[$100\,$km$\,$s$^{-1}$]$\!\!\!\!\!$  & [dex(Gyr)] \\
\hline
\endhead
\hline \multicolumn{10}{r}{\textit{Continued on next page}} \\
\endfoot
\hline
\endlastfoot
$1$ & $\text{1885A}$ & $\text{N}$ & $-15.30\pm 0.10$ & $14.5$ & $0.05\pm 0.0$ & $35$ & $0.8\pm 0.0\pm 0.8$ & $17.0\pm 0.2\pm 0.8$ & $0.6\pm 0.1$ & $9.8\pm 0.2$ \\
$2$ & $\text{1939A}$ & $\text{N}$ & $-17.60\pm 0.20$ & $22.1$ & $1.61\pm 0.3$ & $24$ & $2.2\pm 0.5\pm 1.7$ & $16.7\pm 0.3\pm 0.5$ & $2.0\pm 0.8$ & $9.9\pm 0.1$ \\
$3$ & $\text{1939B}$ & $\text{N}$ & $-19.00\pm 0.20$ & $57.3$ & $4.24\pm 0.5$ & $18$ & $5.8\pm 1.5\pm 1.4$ & $16.2\pm 0.4\pm 0.1$ & $2.3\pm 0.9$ & $9.7\pm 0.1$ \\
$4$ & $\text{1980N}$ & $\text{N}$ & $-18.00\pm 0.04$ & $210.0$ & $17.56\pm 1.3$ & $43$ & $31.1\pm 6.0\pm 1.2$ & $15.7\pm 0.3\pm 0.0$ & $6.8\pm 2.6$ & $9.6\pm 0.2$ \\
$5$ & $\text{1981B}$ & $\text{N}$ & $-18.50\pm 0.02$ & $55.8$ & $4.05\pm 0.2$ & $88$ & $18.0\pm 4.9\pm 0.5$ & $15.2\pm 0.5\pm 0.0$ & $1.7\pm 0.6$ & $8.8\pm 0.3$ \\
$6$ & $\text{1981D}$ & $\text{N}$ & $-17.90\pm 0.05$ & $97.9$ & $8.19\pm 0.6$ & $38$ & $13.6\pm 2.6\pm 1.4$ & $16.2\pm 0.3\pm 0.1$ & $5.7\pm 2.0$ & $9.7\pm 0.1$ \\
$7$ & $\text{1986A}$ & $\text{N}$ & $-19.10\pm 0.20$ & $22.8$ & $4.90\pm 1.0$ & $82$ & $7.3\pm 1.9\pm 0.6$ & $16.2\pm 0.4\pm 0.0$ & $2.8\pm 1.3$ & $9.0\pm 0.3$ \\
$8$ & $\text{1986G}$ & $\text{91bg}$ & $-14.70\pm 0.03$ & $104.0$ & $1.84\pm 0.1$ & $5$ & $1.9\pm 0.2\pm 1.7$ & $16.7\pm 0.2\pm 0.6$ & $1.5\pm 0.5$ & $9.8\pm 0.1$ \\
$9$ & $\text{1989A}$ & $\text{N}$ & $-18.90\pm 0.05$ & $32.8$ & $6.06\pm 1.3$ & $45$ & $6.4\pm 1.5\pm 0.8$ & $15.9\pm 0.4\pm 0.1$ & $1.6\pm 0.6$ & $9.2\pm 0.4$ \\
$10$ & $\text{1989B}$ & $\text{N}$ & $-18.20\pm 0.10$ & $51.8$ & $2.25\pm 0.2$ & $32$ & $3.1\pm 0.5\pm 1.3$ & $16.5\pm 0.2\pm 0.2$ & $1.7\pm 0.5$ & $9.6\pm 0.2$ \\
$11$ & $\text{1990N}$ & $\text{N}$ & $-18.70\pm 0.02$ & $61.3$ & $5.98\pm 0.3$ & $39$ & $7.4\pm 0.7\pm 0.7$ & $15.7\pm 0.2\pm 0.1$ & $1.4\pm 0.4$ & $9.1\pm 0.4$ \\
$12$ & $\text{1991bg}$ & $\text{91bg}$ & $-16.50\pm 0.02$ & $57.6$ & $4.67\pm 0.4$ & $48$ & $5.1\pm 0.6\pm 1.7$ & $16.4\pm 0.2\pm 0.2$ & $3.2\pm 1.0$ & $9.8\pm 0.1$ \\
$13$ & $\text{1991T}$ & $\text{91T}$ & $-18.80\pm 0.02$ & $51.4$ & $3.29\pm 0.3$ & $43$ & $12.0\pm 4.7\pm 0.6$ & $15.6\pm 0.5\pm 0.0$ & $2.0\pm 0.7$ & $9.0\pm 0.2$ \\
$14$ & $\text{1992al}$ & $\text{N}$ & $-19.10\pm 0.03$ & $20.8$ & $5.48\pm 0.7$ & $29$ & $14.7\pm 6.2\pm 0.5$ & $15.2\pm 0.6\pm 0.0$ & $1.5\pm 0.5$ & $8.8\pm 0.3$ \\
$15$ & $\text{1992bo}$ & $\text{N}$ & $-18.40\pm 0.01$ & $74.3$ & $27.22\pm 2.5$ & $34$ & $37.9\pm 6.6\pm 0.3$ & $14.6\pm 0.5\pm 0.0$ & $1.9\pm 0.8$ & $8.3\pm 0.7$ \\
$16$ & $\text{1993H}$ & $\text{N}$ & $-17.90\pm 0.10$ & $13.2$ & $5.68\pm 0.9$ & $27$ & $5.7\pm 0.9\pm 1.2$ & $16.2\pm 0.3\pm 0.1$ & $2.2\pm 0.9$ & $9.5\pm 0.2$ \\
$17$ & $\text{1994ae}$ & $\text{N}$ & $-19.00\pm 0.03$ & $31.2$ & $3.89\pm 0.2$ & $45$ & $7.8\pm 1.5\pm 0.6$ & $15.7\pm 0.3\pm 0.1$ & $1.4\pm 0.3$ & $8.9\pm 0.4$ \\
$18$ & $\text{1994S}$ & $\text{N}$ & $-19.20\pm 0.06$ & $16.0$ & $4.74\pm 0.7$ & $88$ & $13.2\pm 3.7\pm 0.9$ & $15.6\pm 0.4\pm 0.0$ & $2.1\pm 0.8$ & $9.3\pm 0.3$ \\
$19$ & $\text{1995ak}$ & $\text{N}$ & $-18.20\pm 0.04$ & $8.0$ & $2.87\pm 0.6$ & $16$ & $3.5\pm 1.0\pm 1.0$ & $16.3\pm 0.3\pm 0.2$ & $1.4\pm 0.4$ & $9.4\pm 0.6$ \\
$20$ & $\text{1995al}$ & $\text{N}$ & $-18.80\pm 0.02$ & $16.0$ & $2.42\pm 0.3$ & $50$ & $4.0\pm 0.8\pm 0.9$ & $16.2\pm 0.2\pm 0.1$ & $1.5\pm 0.4$ & $9.3\pm 0.2$ \\
$21$ & $\text{1995bd}$ & $\text{N}$ & $-16.40\pm 0.03$ & $23.2$ & $5.90\pm 0.7$ & $4$ & $6.1\pm 1.0\pm 0.7$ & $16.0\pm 0.3\pm 0.1$ & $1.9\pm 0.6$ & $9.1\pm 0.6$ \\
$22$ & $\text{1995D}$ & $\text{N}$ & $-19.30\pm 0.01$ & $91.6$ & $13.62\pm 1.4$ & $18$ & $14.5\pm 1.9\pm 0.7$ & $15.5\pm 0.4\pm 0.0$ & $2.1\pm 0.8$ & $9.1\pm 0.3$ \\
$23$ & $\text{1995E}$ & $\text{N}$ & $-16.10\pm 0.03$ & $22.9$ & $4.84\pm 0.6$ & $25$ & $4.8\pm 0.6\pm 0.9$ & $16.1\pm 0.2\pm 0.1$ & $1.6\pm 0.5$ & $9.3\pm 0.4$ \\
$24$ & $\text{1996bl}$ & $\text{N}$ & $-18.90\pm 0.05$ & $5.9$ & $3.70\pm 0.9$ & $13$ & $3.7\pm 0.9\pm 0.9$ & $16.3\pm 0.3\pm 0.1$ & $1.6\pm 0.5$ & $9.3\pm 0.5$ \\
$25$ & $\text{1996bo}$ & $\text{N}$ & $-17.60\pm 0.02$ & $5.2$ & $1.45\pm 0.4$ & $75$ & $2.1\pm 0.7\pm 0.7$ & $16.7\pm 0.3\pm 0.2$ & $1.9\pm 0.7$ & $9.1\pm 0.4$ \\
$26$ & $\text{1997bp}$ & $\text{N}$ & $-18.70\pm 0.02$ & $24.0$ & $3.44\pm 0.4$ & $7$ & $3.5\pm 0.4\pm 1.4$ & $16.2\pm 0.2\pm 0.2$ & $1.3\pm 0.3$ & $9.7\pm 0.2$ \\
$27$ & $\text{1997bq}$ & $\text{N}$ & $-18.30\pm 0.04$ & $78.5$ & $14.65\pm 1.3$ & $23$ & $14.8\pm 1.4\pm 0.7$ & $16.1\pm 0.2\pm 0.0$ & $5.6\pm 1.9$ & $9.1\pm 0.3$ \\
$28$ & $\text{1997do}$ & $\text{N}$ & $-18.70\pm 0.05$ & $3.0$ & $0.60\pm 0.2$ & $80$ & $1.3\pm 0.4\pm 0.9$ & $16.7\pm 0.3\pm 0.4$ & $0.9\pm 0.3$ & $9.6\pm 0.3$ \\
$29$ & $\text{1997E}$ & $\text{N}$ & $-18.10\pm 0.04$ & $64.9$ & $17.37\pm 1.6$ & $10$ & $17.6\pm 1.8\pm 1.2$ & $16.0\pm 0.2\pm 0.0$ & $5.9\pm 2.1$ & $9.6\pm 0.1$ \\
$30$ & $\text{1998ab}$ & $\text{91T}$ & $-18.80\pm 0.10$ & $15.1$ & $6.46\pm 0.9$ & $24$ & $6.5\pm 0.9\pm 0.8$ & $16.1\pm 0.3\pm 0.1$ & $2.2\pm 0.8$ & $9.2\pm 0.3$ \\
$31$ & $\text{1998aq}$ & $\text{N}$ & $-19.20\pm 0.02$ & $20.7$ & $2.22\pm 0.2$ & $88$ & $2.4\pm 0.2\pm 1.2$ & $16.4\pm 0.2\pm 0.3$ & $1.3\pm 0.3$ & $9.6\pm 0.3$ \\
$32$ & $\text{1998bp}$ & $\text{N}$ & $-17.40\pm 0.01$ & $12.9$ & $2.47\pm 0.4$ & $72$ & $5.6\pm 1.5\pm 1.4$ & $16.0\pm 0.3\pm 0.1$ & $1.7\pm 0.5$ & $9.7\pm 0.1$ \\
$33$ & $\text{1998bu}$ & $\text{N}$ & $-17.70\pm 0.06$ & $55.6$ & $2.86\pm 0.2$ & $15$ & $3.0\pm 0.3\pm 1.3$ & $16.4\pm 0.2\pm 0.2$ & $1.7\pm 0.5$ & $9.6\pm 0.2$ \\
$34$ & $\text{1998de}$ & $\text{91bg}$ & $-16.50\pm 0.02$ & $72.2$ & $22.37\pm 2.1$ & $42$ & $24.7\pm 2.9\pm 1.0$ & $15.6\pm 0.3\pm 0.0$ & $4.6\pm 1.8$ & $9.4\pm 0.2$ \\
$35$ & $\text{1998dh}$ & $\text{N}$ & $-18.30\pm 0.01$ & $54.5$ & $8.25\pm 0.7$ & $18$ & $11.4\pm 2.5\pm 0.6$ & $15.8\pm 0.4\pm 0.0$ & $2.5\pm 0.9$ & $8.9\pm 0.4$ \\
$36$ & $\text{1998ef}$ & $\text{N}$ & $-18.70\pm 0.02$ & $7.3$ & $2.04\pm 0.4$ & $3$ & $2.1\pm 0.5\pm 1.3$ & $16.6\pm 0.2\pm 0.4$ & $1.4\pm 0.4$ & $9.6\pm 0.3$ \\
$37$ & $\text{1998es}$ & $\text{91T}$ & $-18.90\pm 0.03$ & $11.9$ & $2.08\pm 0.3$ & $37$ & $2.2\pm 0.4\pm 1.4$ & $16.5\pm 0.2\pm 0.4$ & $1.2\pm 0.3$ & $9.7\pm 0.2$ \\
$38$ & $\text{1999aa}$ & $\text{91T}$ & $-19.10\pm 0.01$ & $30.2$ & $9.14\pm 1.0$ & $39$ & $9.4\pm 1.2\pm 0.6$ & $16.0\pm 0.3\pm 0.0$ & $2.7\pm 1.2$ & $8.9\pm 0.3$ \\
$39$ & $\text{1999ac}$ & $\text{pec}$ & $-18.40\pm 0.03$ & $39.7$ & $7.68\pm 0.7$ & $29$ & $8.3\pm 1.0\pm 0.5$ & $15.6\pm 0.3\pm 0.0$ & $1.4\pm 0.4$ & $8.7\pm 0.3$ \\
$40$ & $\text{1999bh}$ & $\text{02es}$ & $-16.00\pm 0.40$ & $10.3$ & $4.70\pm 1.3$ & $44$ & $6.5\pm 2.3\pm 1.0$ & $16.1\pm 0.5\pm 0.1$ & $2.3\pm 1.2$ & $9.4\pm 0.3$ \\
$41$ & $\text{1999by}$ & $\text{91bg}$ & $-16.90\pm 0.01$ & $134.4$ & $9.47\pm 0.7$ & $17$ & $11.7\pm 1.9\pm 0.9$ & $16.0\pm 0.3\pm 0.0$ & $3.5\pm 1.3$ & $9.3\pm 0.3$ \\
$42$ & $\text{1999cc}$ & $\text{N}$ & $-18.60\pm 0.02$ & $17.4$ & $10.19\pm 1.4$ & $36$ & $11.5\pm 2.0\pm 0.6$ & $16.1\pm 0.3\pm 0.0$ & $4.7\pm 1.8$ & $8.9\pm 0.3$ \\
$43$ & $\text{1999cl}$ & $\text{N}$ & $-15.20\pm 0.02$ & $54.4$ & $5.18\pm 0.8$ & $28$ & $6.9\pm 1.7\pm 0.8$ & $16.4\pm 0.3\pm 0.1$ & $4.0\pm 1.7$ & $9.2\pm 0.3$ \\
$44$ & $\text{1999cp}$ & $\text{N}$ & $-18.90\pm 0.01$ & $59.4$ & $12.28\pm 1.2$ & $10$ & $12.3\pm 1.2\pm 0.4$ & $15.4\pm 0.3\pm 0.0$ & $1.5\pm 0.5$ & $8.7\pm 0.3$ \\
$45$ & $\text{1999dk}$ & $\text{N}$ & $-18.70\pm 0.01$ & $27.4$ & $7.60\pm 0.8$ & $79$ & $7.8\pm 1.0\pm 0.5$ & $15.7\pm 0.3\pm 0.0$ & $1.4\pm 0.4$ & $8.8\pm 0.3$ \\
$46$ & $\text{1999dq}$ & $\text{91T}$ & $-18.50\pm 0.02$ & $7.7$ & $1.68\pm 0.3$ & $80$ & $1.8\pm 0.4\pm 1.1$ & $16.8\pm 0.2\pm 0.3$ & $1.7\pm 0.6$ & $9.5\pm 0.4$ \\
$47$ & $\text{1999ee}$ & $\text{N}$ & $-18.20\pm 0.01$ & $11.9$ & $2.39\pm 0.4$ & $72$ & $6.4\pm 1.9\pm 0.9$ & $16.2\pm 0.3\pm 0.1$ & $2.7\pm 1.0$ & $9.3\pm 0.2$ \\
$48$ & $\text{1999ej}$ & $\text{N}$ & $-18.00\pm 0.01$ & $21.7$ & $6.78\pm 0.8$ & $28$ & $7.4\pm 1.1\pm 1.3$ & $16.0\pm 0.3\pm 0.1$ & $2.3\pm 0.8$ & $9.6\pm 0.2$ \\
$49$ & $\text{1999ek}$ & $\text{N}$ & $-16.10\pm 0.01$ & $15.7$ & $4.92\pm 0.7$ & $42$ & $11.3\pm 3.5\pm 0.7$ & $15.8\pm 0.4\pm 0.0$ & $2.7\pm 1.1$ & $9.0\pm 0.4$ \\
$50$ & $\text{1999gh}$ & $\text{N}$ & $-17.60\pm 0.20$ & $54.1$ & $7.94\pm 0.8$ & $33$ & $9.7\pm 1.4\pm 1.4$ & $16.1\pm 0.3\pm 0.1$ & $3.4\pm 1.3$ & $9.7\pm 0.1$ \\
$51$ & $\text{2000cn}$ & $\text{N}$ & $-18.00\pm 0.02$ & $11.8$ & $5.21\pm 0.8$ & $16$ & $5.2\pm 0.9\pm 0.5$ & $16.4\pm 0.3\pm 0.0$ & $2.9\pm 1.1$ & $8.8\pm 0.3$ \\
$52$ & $\text{2000cw}$ & $\text{N}$ & $-17.90\pm 0.01$ & $22.4$ & $10.84\pm 2.0$ & $28$ & $14.7\pm 4.0\pm 0.6$ & $15.5\pm 0.5\pm 0.0$ & $2.3\pm 1.1$ & $9.0\pm 0.6$ \\
$53$ & $\text{2000cx}$ & $\text{91T}$ & $-18.80\pm 0.01$ & $110.8$ & $12.69\pm 1.8$ & $10$ & $12.7\pm 1.8\pm 1.3$ & $15.9\pm 0.3\pm 0.1$ & $3.5\pm 1.5$ & $9.6\pm 0.2$ \\
$54$ & $\text{2000E}$ & $\text{N}$ & $-17.60\pm 0.05$ & $30.5$ & $2.75\pm 0.3$ & $16$ & $2.8\pm 0.4\pm 1.2$ & $16.5\pm 0.2\pm 0.2$ & $1.8\pm 0.7$ & $9.5\pm 0.3$ \\
$55$ & $\text{2001ah}$ & $\text{pec}$ & $-19.20\pm 0.04$ & $32.9$ & $33.26\pm 3.6$ & $48$ & $34.3\pm 4.2\pm 0.5$ & $15.4\pm 0.5\pm 0.0$ & $4.6\pm 2.6$ & $8.8\pm 0.3$ \\
$56$ & $\text{2001ba}$ & $\text{N}$ & $-19.00\pm 0.01$ & $26.8$ & $15.39\pm 1.7$ & $42$ & $15.8\pm 1.9\pm 0.7$ & $15.9\pm 0.3\pm 0.0$ & $4.1\pm 1.5$ & $9.1\pm 0.3$ \\
$57$ & $\text{2001bt}$ & $\text{N}$ & $-18.10\pm 0.02$ & $21.4$ & $5.03\pm 0.6$ & $59$ & $5.7\pm 0.8\pm 0.9$ & $16.3\pm 0.2\pm 0.1$ & $2.7\pm 0.9$ & $9.3\pm 0.3$ \\
$58$ & $\text{2001cj}$ & $\text{N}$ & $-19.30\pm 0.01$ & $34.5$ & $19.10\pm 0.8$ & $81$ & $19.5\pm 1.0\pm 0.7$ & $15.3\pm 0.3\pm 0.0$ & $2.2\pm 0.7$ & $9.0\pm 0.3$ \\
$59$ & $\text{2001cz}$ & $\text{N}$ & $-18.40\pm 0.01$ & $33.0$ & $8.88\pm 0.9$ & $12$ & $11.2\pm 2.7\pm 0.5$ & $15.9\pm 0.4\pm 0.0$ & $2.8\pm 1.1$ & $8.8\pm 0.3$ \\
$60$ & $\text{2001da}$ & $\text{N}$ & $-18.50\pm 0.02$ & $11.3$ & $3.51\pm 0.6$ & $89$ & $10.8\pm 3.4\pm 0.9$ & $15.6\pm 0.5\pm 0.1$ & $1.9\pm 0.7$ & $9.3\pm 0.2$ \\
$61$ & $\text{2001dl}$ & $\text{N}$ & $-17.80\pm 0.01$ & $11.1$ & $5.94\pm 1.2$ & $28$ & $14.4\pm 6.8\pm 0.4$ & $15.5\pm 0.6\pm 0.0$ & $2.0\pm 0.9$ & $8.5\pm 0.2$ \\
$62$ & $\text{2001eh}$ & $\text{N}$ & $-19.20\pm 0.01$ & $38.0$ & $25.23\pm 2.6$ & $15$ & $25.3\pm 2.7\pm 0.7$ & $15.6\pm 0.4\pm 0.0$ & $4.8\pm 2.3$ & $9.1\pm 0.3$ \\
$63$ & $\text{2001ep}$ & $\text{N}$ & $-18.70\pm 0.01$ & $18.6$ & $4.41\pm 0.6$ & $5$ & $4.4\pm 0.6\pm 0.8$ & $16.1\pm 0.2\pm 0.1$ & $1.4\pm 0.4$ & $9.2\pm 0.3$ \\
$64$ & $\text{2001gb}$ & $\text{N}$ & $-18.60\pm 0.20$ & $15.1$ & $7.64\pm 1.1$ & $83$ & $7.8\pm 1.2\pm 0.9$ & $16.0\pm 0.3\pm 0.1$ & $2.4\pm 0.9$ & $9.3\pm 0.5$ \\
$65$ & $\text{2002bf}$ & $\text{N}$ & $-18.60\pm 0.30$ & $4.0$ & $1.73\pm 0.6$ & $34$ & $2.0\pm 0.7\pm 1.3$ & $16.7\pm 0.3\pm 0.4$ & $1.6\pm 0.6$ & $9.7\pm 0.2$ \\
$66$ & $\text{2002bo}$ & $\text{N}$ & $-17.70\pm 0.01$ & $16.0$ & $1.71\pm 0.2$ & $25$ & $2.9\pm 0.9\pm 1.4$ & $16.6\pm 0.3\pm 0.3$ & $2.0\pm 0.7$ & $9.7\pm 0.2$ \\
$67$ & $\text{2002cd}$ & $\text{N}$ & $-16.30\pm 0.01$ & $12.9$ & $2.70\pm 0.4$ & $40$ & $3.5\pm 0.8\pm 0.9$ & $16.4\pm 0.3\pm 0.1$ & $1.8\pm 0.6$ & $9.3\pm 0.3$ \\
$68$ & $\text{2002cf}$ & $\text{91bg}$ & $-17.00\pm 0.02$ & $20.0$ & $5.17\pm 1.5$ & $44$ & $7.3\pm 2.6\pm 1.5$ & $16.3\pm 0.5\pm 0.1$ & $3.6\pm 2.0$ & $9.8\pm 0.1$ \\
$69$ & $\text{2002cr}$ & $\text{N}$ & $-18.50\pm 0.01$ & $64.2$ & $13.27\pm 1.2$ & $20$ & $13.3\pm 1.3\pm 0.4$ & $15.3\pm 0.3\pm 0.0$ & $1.5\pm 0.5$ & $8.6\pm 0.3$ \\
$70$ & $\text{2002cs}$ & $\text{N}$ & $-18.00\pm 0.01$ & $25.3$ & $6.18\pm 1.0$ & $25$ & $6.6\pm 1.3\pm 1.5$ & $16.2\pm 0.3\pm 0.1$ & $2.6\pm 1.1$ & $9.8\pm 0.1$ \\
$71$ & $\text{2002cx}$ & $\text{02cx}$ & $-18.40\pm 0.05$ & $15.2$ & $11.50\pm 0.9$ & $4$ & $12.2\pm 2.1\pm 0.5$ & $15.3\pm 0.4\pm 0.0$ & $1.4\pm 0.4$ & $8.8\pm 0.6$ \\
$72$ & $\text{2002de}$ & $\text{N}$ & $-18.80\pm 0.01$ & $2.3$ & $1.22\pm 0.6$ & $78$ & $1.9\pm 0.9\pm 1.0$ & $16.8\pm 0.4\pm 0.3$ & $2.0\pm 0.9$ & $9.7\pm 0.6$ \\
$73$ & $\text{2002do}$ & $\text{N}$ & $-17.40\pm 0.06$ & $8.8$ & $2.76\pm 0.5$ & $37$ & $3.2\pm 0.7\pm 1.6$ & $16.7\pm 0.3\pm 0.3$ & $2.9\pm 1.0$ & $9.8\pm 0.1$ \\
$74$ & $\text{2002dp}$ & $\text{N}$ & $-18.70\pm 0.03$ & $37.1$ & $7.82\pm 0.8$ & $30$ & $8.1\pm 0.9\pm 0.6$ & $16.0\pm 0.3\pm 0.0$ & $2.3\pm 0.8$ & $8.9\pm 0.3$ \\
$75$ & $\text{2002eb}$ & $\text{N}$ & $-19.20\pm 0.01$ & $19.5$ & $11.36\pm 0.8$ & $60$ & $13.7\pm 1.5\pm 0.8$ & $15.6\pm 0.3\pm 0.0$ & $2.3\pm 0.7$ & $9.2\pm 0.4$ \\
$76$ & $\text{2002ef}$ & $\text{N}$ & $-18.40\pm 0.02$ & $10.9$ & $2.73\pm 0.9$ & $9$ & $2.7\pm 0.9\pm 1.7$ & $16.5\pm 0.4\pm 0.4$ & $1.8\pm 0.9$ & $9.8\pm 0.1$ \\
$77$ & $\text{2002el}$ & $\text{N}$ & $-18.40\pm 0.02$ & $27.3$ & $14.80\pm 0.8$ & $29$ & $20.0\pm 2.9\pm 1.2$ & $15.8\pm 0.2\pm 0.0$ & $5.0\pm 1.5$ & $9.5\pm 0.2$ \\
$78$ & $\text{2002er}$ & $\text{N}$ & $-17.80\pm 0.02$ & $11.2$ & $1.63\pm 0.3$ & $9$ & $1.8\pm 0.5\pm 1.4$ & $16.5\pm 0.2\pm 0.5$ & $1.1\pm 0.3$ & $9.7\pm 0.2$ \\
$79$ & $\text{2002es}$ & $\text{02es}$ & $-17.00\pm 0.02$ & $32.0$ & $11.70\pm 1.3$ & $70$ & $11.9\pm 1.4\pm 1.0$ & $16.1\pm 0.4\pm 0.0$ & $4.4\pm 2.7$ & $9.4\pm 0.2$ \\
$80$ & $\text{2002fb}$ & $\text{91bg}$ & $-17.00\pm 0.01$ & $20.5$ & $6.29\pm 0.8$ & $33$ & $6.5\pm 0.9\pm 1.4$ & $16.3\pm 0.2\pm 0.1$ & $3.4\pm 1.1$ & $9.7\pm 0.1$ \\
$81$ & $\text{2002fk}$ & $\text{N}$ & $-19.20\pm 0.02$ & $12.7$ & $1.94\pm 0.2$ & $12$ & $1.9\pm 0.2\pm 1.2$ & $16.6\pm 0.2\pm 0.4$ & $1.4\pm 0.3$ & $9.5\pm 0.3$ \\
$82$ & $\text{2002G}$ & $\text{N}$ & $-17.90\pm 0.05$ & $8.8$ & $5.70\pm 1.1$ & $60$ & $6.8\pm 1.6\pm 1.5$ & $16.2\pm 0.3\pm 0.1$ & $3.1\pm 1.2$ & $9.7\pm 0.2$ \\
$83$ & $\text{2002ha}$ & $\text{N}$ & $-18.80\pm 0.01$ & $30.2$ & $8.20\pm 0.8$ & $45$ & $9.7\pm 1.4\pm 1.1$ & $16.2\pm 0.2\pm 0.1$ & $3.9\pm 1.3$ & $9.5\pm 0.3$ \\
$84$ & $\text{2002hw}$ & $\text{N}$ & $-17.10\pm 0.20$ & $8.8$ & $2.93\pm 0.6$ & $29$ & $3.0\pm 0.6\pm 0.6$ & $16.5\pm 0.3\pm 0.1$ & $1.7\pm 0.6$ & $9.0\pm 0.4$ \\
$85$ & $\text{2002ic}$ & $\text{CSM}$ & $-19.00\pm 0.10$ & $4.0$ & $4.81\pm 1.6$ & $10$ & $5.0\pm 1.8\pm 0.3$ & $15.8\pm 0.6\pm 0.0$ & $1.0\pm 0.5$ & $8.3\pm 0.2$ \\
$86$ & $\text{2002jy}$ & $\text{N}$ & $-19.20\pm 0.05$ & $59.5$ & $23.92\pm 2.2$ & $1$ & $23.9\pm 2.2\pm 0.5$ & $15.3\pm 0.5\pm 0.0$ & $2.8\pm 1.5$ & $8.8\pm 0.3$ \\
$87$ & $\text{2003cg}$ & $\text{N}$ & $-15.80\pm 0.01$ & $19.8$ & $2.21\pm 0.3$ & $31$ & $2.4\pm 0.4\pm 1.4$ & $16.7\pm 0.2\pm 0.3$ & $2.2\pm 0.7$ & $9.7\pm 0.2$ \\
$88$ & $\text{2003cq}$ & $\text{N}$ & $-18.70\pm 0.02$ & $28.8$ & $17.94\pm 2.0$ & $1$ & $17.9\pm 2.0\pm 0.7$ & $16.0\pm 0.3\pm 0.0$ & $6.5\pm 2.4$ & $9.1\pm 0.3$ \\
$89$ & $\text{2003du}$ & $\text{N}$ & $-19.20\pm 0.03$ & $16.4$ & $3.01\pm 0.3$ & $10$ & $3.0\pm 0.3\pm 0.4$ & $16.1\pm 0.2\pm 0.1$ & $1.0\pm 0.2$ & $8.6\pm 0.2$ \\
$90$ & $\text{2003fa}$ & $\text{N}$ & $-19.40\pm 0.01$ & $51.0$ & $37.49\pm 3.7$ & $6$ & $38.1\pm 4.3\pm 0.6$ & $15.1\pm 0.5\pm 0.0$ & $3.2\pm 1.6$ & $8.9\pm 0.4$ \\
$91$ & $\text{2003gn}$ & $\text{N}$ & $-18.40\pm 0.01$ & $18.7$ & $14.56\pm 1.8$ & $2$ & $14.6\pm 2.0\pm 0.7$ & $15.7\pm 0.3\pm 0.0$ & $3.0\pm 1.2$ & $9.0\pm 0.3$ \\
$92$ & $\text{2003gq}$ & $\text{02cx}$ & $-16.20\pm 0.05$ & $14.9$ & $7.40\pm 1.9$ & $7$ & $7.6\pm 2.1\pm 0.7$ & $16.2\pm 0.4\pm 0.0$ & $3.3\pm 1.8$ & $9.1\pm 0.4$ \\
$93$ & $\text{2003he}$ & $\text{N}$ & $-18.90\pm 0.01$ & $6.0$ & $2.93\pm 1.0$ & $20$ & $4.4\pm 2.2\pm 1.0$ & $16.2\pm 0.5\pm 0.1$ & $1.7\pm 0.9$ & $9.4\pm 0.3$ \\
$94$ & $\text{2003W}$ & $\text{N}$ & $-18.60\pm 0.01$ & $4.2$ & $1.56\pm 0.5$ & $50$ & $2.0\pm 0.7\pm 0.6$ & $16.7\pm 0.3\pm 0.2$ & $1.8\pm 0.7$ & $9.0\pm 0.6$ \\
$95$ & $\text{2003Y}$ & $\text{91bg}$ & $-16.50\pm 0.01$ & $20.0$ & $6.41\pm 0.4$ & $4$ & $6.4\pm 0.5\pm 1.3$ & $16.1\pm 0.2\pm 0.1$ & $2.3\pm 0.6$ & $9.6\pm 0.1$ \\
$96$ & $\text{2004at}$ & $\text{N}$ & $-19.40\pm 0.02$ & $19.7$ & $9.95\pm 0.6$ & $9$ & $10.7\pm 1.3\pm 0.7$ & $15.9\pm 0.4\pm 0.0$ & $2.9\pm 1.7$ & $9.0\pm 0.5$ \\
$97$ & $\text{2004bg}$ & $\text{N}$ & $-19.40\pm 0.03$ & $17.9$ & $7.37\pm 1.0$ & $3$ & $7.8\pm 1.8\pm 0.5$ & $15.8\pm 0.4\pm 0.0$ & $1.8\pm 0.6$ & $8.8\pm 0.3$ \\
$98$ & $\text{2004bk}$ & $\text{N}$ & $-19.10\pm 0.10$ & $9.9$ & $4.92\pm 0.6$ & $42$ & $4.9\pm 0.6\pm 1.1$ & $16.3\pm 0.2\pm 0.1$ & $2.1\pm 0.7$ & $9.5\pm 0.3$ \\
$99$ & $\text{2004bw}$ & $\text{N}$ & $-18.50\pm 0.01$ & $23.7$ & $11.28\pm 1.0$ & $75$ & $12.8\pm 1.5\pm 0.5$ & $15.7\pm 0.4\pm 0.0$ & $2.5\pm 1.3$ & $8.7\pm 0.3$ \\
$100$ & $\text{2004dt}$ & $\text{N}$ & $-19.40\pm 0.01$ & $12.5$ & $4.63\pm 0.5$ & $80$ & $5.3\pm 0.7\pm 1.3$ & $16.3\pm 0.2\pm 0.1$ & $2.4\pm 0.7$ & $9.6\pm 0.2$ \\
$101$ & $\text{2004ef}$ & $\text{N}$ & $-18.30\pm 0.01$ & $9.9$ & $5.35\pm 1.0$ & $50$ & $5.3\pm 1.0\pm 1.0$ & $16.3\pm 0.3\pm 0.1$ & $2.5\pm 1.0$ & $9.4\pm 0.3$ \\
$102$ & $\text{2004eo}$ & $\text{N}$ & $-18.10\pm 0.02$ & $59.8$ & $16.50\pm 1.6$ & $29$ & $38.4\pm 13.1\pm 0.6$ & $15.4\pm 0.5\pm 0.0$ & $5.0\pm 2.2$ & $9.0\pm 0.2$ \\
$103$ & $\text{2004ey}$ & $\text{N}$ & $-18.70\pm 0.01$ & $14.6$ & $4.32\pm 0.6$ & $3$ & $4.3\pm 0.6\pm 0.7$ & $16.1\pm 0.2\pm 0.1$ & $1.3\pm 0.3$ & $9.1\pm 0.3$ \\
$104$ & $\text{2004fz}$ & $\text{N}$ & $-19.30\pm 0.01$ & $12.8$ & $3.67\pm 0.4$ & $43$ & $4.3\pm 0.6\pm 0.7$ & $16.4\pm 0.2\pm 0.1$ & $2.2\pm 0.6$ & $9.0\pm 0.4$ \\
$105$ & $\text{2004gs}$ & $\text{N}$ & $-18.10\pm 0.01$ & $16.3$ & $9.16\pm 1.3$ & $16$ & $9.3\pm 1.4\pm 1.3$ & $16.0\pm 0.3\pm 0.1$ & $2.7\pm 1.0$ & $9.6\pm 0.1$ \\
$106$ & $\text{2004L}$ & $\text{N}$ & $-18.30\pm 0.30$ & $3.2$ & $1.97\pm 0.8$ & $1$ & $2.0\pm 0.8\pm 0.8$ & $16.7\pm 0.4\pm 0.2$ & $1.5\pm 0.6$ & $9.2\pm 0.7$ \\
$107$ & $\text{2005A}$ & $\text{N}$ & $-16.00\pm 0.01$ & $6.3$ & $2.09\pm 0.5$ & $75$ & $6.2\pm 2.4\pm 0.8$ & $16.5\pm 0.3\pm 0.1$ & $4.4\pm 1.6$ & $9.1\pm 0.4$ \\
$108$ & $\text{2005ag}$ & $\text{N}$ & $-19.20\pm 0.03$ & $9.2$ & $12.33\pm 2.3$ & $39$ & $13.7\pm 3.0\pm 0.7$ & $16.0\pm 0.3\pm 0.0$ & $4.4\pm 1.8$ & $9.1\pm 0.6$ \\
$109$ & $\text{2005al}$ & $\text{N}$ & $-18.30\pm 0.01$ & $18.6$ & $5.55\pm 0.7$ & $21$ & $6.0\pm 0.9\pm 1.4$ & $16.2\pm 0.3\pm 0.1$ & $2.7\pm 0.9$ & $9.7\pm 0.1$ \\
$110$ & $\text{2005am}$ & $\text{N}$ & $-18.70\pm 0.01$ & $35.7$ & $4.76\pm 0.5$ & $14$ & $5.7\pm 1.1\pm 1.3$ & $16.2\pm 0.3\pm 0.1$ & $2.3\pm 0.8$ & $9.6\pm 0.2$ \\
$111$ & $\text{2005bg}$ & $\text{N}$ & $-19.00\pm 0.01$ & $0.7$ & $0.33\pm 0.5$ & $85$ & $1.0\pm 0.3\pm 0.7$ & $16.9\pm 0.7\pm 0.4$ & $0.8\pm 0.4$ & $9.7\pm 0.7$ \\
$112$ & $\text{2005bo}$ & $\text{N}$ & $-17.10\pm 0.02$ & $13.9$ & $3.76\pm 0.6$ & $11$ & $3.8\pm 0.6\pm 1.3$ & $16.3\pm 0.2\pm 0.2$ & $1.7\pm 0.5$ & $9.6\pm 0.3$ \\
$113$ & $\text{2005cc}$ & $\text{02cx}$ & $-16.10\pm 0.03$ & $4.7$ & $0.80\pm 0.3$ & $26$ & $1.3\pm 0.3\pm 0.9$ & $16.9\pm 0.4\pm 0.4$ & $1.4\pm 0.7$ & $9.7\pm 0.2$ \\
$114$ & $\text{2005de}$ & $\text{N}$ & $-18.40\pm 0.02$ & $32.6$ & $12.01\pm 0.6$ & $5$ & $12.6\pm 1.2\pm 0.5$ & $15.7\pm 0.3\pm 0.0$ & $2.3\pm 0.7$ & $8.7\pm 0.4$ \\
$115$ & $\text{2005el}$ & $\text{N}$ & $-18.40\pm 0.01$ & $44.7$ & $12.74\pm 1.3$ & $14$ & $13.0\pm 1.4\pm 1.2$ & $15.9\pm 0.3\pm 0.1$ & $3.4\pm 1.3$ & $9.5\pm 0.1$ \\
$116$ & $\text{2005eq}$ & $\text{N}$ & $-18.80\pm 0.02$ & $29.8$ & $16.12\pm 1.8$ & $15$ & $21.1\pm 5.2\pm 0.9$ & $15.5\pm 0.5\pm 0.0$ & $3.2\pm 1.6$ & $9.3\pm 0.6$ \\
$117$ & $\text{2005gj}$ & $\text{CSM}$ & $-19.80\pm 0.20$ & $0.5$ & $0.70\pm 1.6$ & $88$ & $3.8\pm 3.3\pm 0.4$ & $15.9\pm 1.2\pm 0.1$ & $0.9\pm 0.6$ & $8.8\pm 0.3$ \\
$118$ & $\text{2005hc}$ & $\text{N}$ & $-19.10\pm 0.01$ & $8.2$ & $7.01\pm 1.4$ & $35$ & $8.5\pm 2.1\pm 0.7$ & $16.0\pm 0.4\pm 0.0$ & $2.7\pm 1.1$ & $9.1\pm 0.7$ \\
$119$ & $\text{2005hk}$ & $\text{02cx}$ & $-17.70\pm 0.02$ & $18.8$ & $5.27\pm 1.0$ & $51$ & $11.0\pm 3.6\pm 0.3$ & $15.2\pm 0.5\pm 0.0$ & $1.1\pm 0.3$ & $8.4\pm 0.4$ \\
$120$ & $\text{2005iq}$ & $\text{N}$ & $-18.80\pm 0.02$ & $18.9$ & $12.55\pm 1.6$ & $12$ & $13.3\pm 2.2\pm 0.9$ & $15.6\pm 0.4\pm 0.0$ & $2.3\pm 1.0$ & $9.3\pm 0.4$ \\
$121$ & $\text{2005ir}$ & $\text{N}$ & $-19.20\pm 0.04$ & $3.1$ & $4.37\pm 1.8$ & $19$ & $4.5\pm 1.9\pm 1.4$ & $16.2\pm 0.5\pm 0.2$ & $1.8\pm 1.0$ & $9.7\pm 0.3$ \\
$122$ & $\text{2005kc}$ & $\text{N}$ & $-17.60\pm 0.05$ & $11.9$ & $3.33\pm 0.5$ & $62$ & $5.9\pm 1.5\pm 1.3$ & $16.4\pm 0.3\pm 0.1$ & $3.3\pm 1.2$ & $9.6\pm 0.2$ \\
$123$ & $\text{2005ke}$ & $\text{91bg}$ & $-17.20\pm 0.01$ & $56.4$ & $7.96\pm 1.2$ & $3$ & $8.0\pm 1.3\pm 1.1$ & $16.2\pm 0.3\pm 0.1$ & $3.1\pm 1.5$ & $9.5\pm 0.2$ \\
$124$ & $\text{2005ki}$ & $\text{N}$ & $-18.70\pm 0.02$ & $71.0$ & $27.04\pm 2.5$ & $42$ & $29.3\pm 3.4\pm 1.0$ & $15.6\pm 0.3\pm 0.0$ & $5.1\pm 2.1$ & $9.4\pm 0.2$ \\
$125$ & $\text{2005M}$ & $\text{91T}$ & $-18.90\pm 0.01$ & $9.8$ & $4.59\pm 0.8$ & $85$ & $9.2\pm 2.5\pm 0.6$ & $15.6\pm 0.4\pm 0.0$ & $1.5\pm 0.5$ & $9.0\pm 0.6$ \\
$126$ & $\text{2005mc}$ & $\text{N}$ & $-17.80\pm 0.01$ & $4.5$ & $2.26\pm 0.7$ & $46$ & $2.3\pm 0.7\pm 1.6$ & $16.7\pm 0.3\pm 0.4$ & $2.2\pm 0.8$ & $9.8\pm 0.2$ \\
$127$ & $\text{2005ms}$ & $\text{N}$ & $-19.10\pm 0.03$ & $43.9$ & $22.61\pm 2.3$ & $78$ & $22.8\pm 2.5\pm 0.7$ & $15.2\pm 0.4\pm 0.0$ & $2.2\pm 1.0$ & $9.0\pm 0.4$ \\
$128$ & $\text{2005na}$ & $\text{N}$ & $-18.80\pm 0.02$ & $7.1$ & $3.50\pm 0.8$ & $74$ & $5.4\pm 1.6\pm 1.2$ & $16.5\pm 0.3\pm 0.1$ & $3.4\pm 1.3$ & $9.6\pm 0.2$ \\
$129$ & $\text{2005W}$ & $\text{N}$ & $-18.20\pm 0.02$ & $55.4$ & $10.13\pm 1.6$ & $15$ & $10.6\pm 1.9\pm 0.7$ & $15.8\pm 0.4\pm 0.0$ & $2.4\pm 1.2$ & $9.0\pm 0.2$ \\
$130$ & $\text{2006ac}$ & $\text{N}$ & $-18.60\pm 0.04$ & $21.6$ & $9.37\pm 1.2$ & $70$ & $9.4\pm 1.2\pm 1.0$ & $16.2\pm 0.2\pm 0.1$ & $3.8\pm 1.3$ & $9.4\pm 0.3$ \\
$131$ & $\text{2006ax}$ & $\text{N}$ & $-19.00\pm 0.01$ & $54.2$ & $18.01\pm 1.7$ & $6$ & $18.0\pm 1.7\pm 0.6$ & $15.5\pm 0.4\pm 0.0$ & $2.8\pm 1.3$ & $9.0\pm 0.3$ \\
$132$ & $\text{2006az}$ & $\text{N}$ & $-19.20\pm 0.02$ & $7.3$ & $4.18\pm 0.9$ & $3$ & $4.2\pm 0.9\pm 1.4$ & $16.7\pm 0.2\pm 0.2$ & $4.1\pm 1.4$ & $9.7\pm 0.3$ \\
$133$ & $\text{2006bh}$ & $\text{N}$ & $-18.70\pm 0.01$ & $52.9$ & $11.48\pm 1.0$ & $52$ & $14.1\pm 1.9\pm 0.7$ & $15.7\pm 0.3\pm 0.0$ & $2.9\pm 1.1$ & $9.1\pm 0.4$ \\
$134$ & $\text{2006bq}$ & $\text{N}$ & $-18.80\pm 0.04$ & $3.3$ & $1.36\pm 0.5$ & $14$ & $1.9\pm 0.6\pm 1.3$ & $16.8\pm 0.3\pm 0.4$ & $1.8\pm 0.7$ & $9.9\pm 0.1$ \\
$135$ & $\text{2006br}$ & $\text{N}$ & $-15.70\pm 0.04$ & $5.5$ & $3.17\pm 0.8$ & $21$ & $5.1\pm 2.2\pm 1.2$ & $16.5\pm 0.4\pm 0.1$ & $3.7\pm 1.6$ & $9.5\pm 0.2$ \\
$136$ & $\text{2006bt}$ & $\text{02es}$ & $-18.50\pm 0.01$ & $50.0$ & $30.25\pm 3.0$ & $32$ & $31.4\pm 3.5\pm 0.7$ & $15.5\pm 0.4\pm 0.0$ & $4.6\pm 2.0$ & $9.1\pm 0.2$ \\
$137$ & $\text{2006cp}$ & $\text{N}$ & $-18.80\pm 0.02$ & $25.3$ & $10.38\pm 1.2$ & $60$ & $10.7\pm 1.4\pm 0.5$ & $15.4\pm 0.3\pm 0.0$ & $1.4\pm 0.4$ & $8.8\pm 0.3$ \\
$138$ & $\text{2006D}$ & $\text{N}$ & $-18.60\pm 0.01$ & $13.4$ & $2.34\pm 0.4$ & $4$ & $2.3\pm 0.4\pm 1.2$ & $16.4\pm 0.2\pm 0.3$ & $1.2\pm 0.3$ & $9.6\pm 0.3$ \\
$139$ & $\text{2006ef}$ & $\text{N}$ & $-19.10\pm 0.01$ & $26.2$ & $9.21\pm 1.1$ & $23$ & $9.9\pm 1.4\pm 1.1$ & $15.7\pm 0.3\pm 0.1$ & $2.0\pm 0.7$ & $9.5\pm 0.2$ \\
$140$ & $\text{2006em}$ & $\text{91bg}$ & $-16.50\pm 0.10$ & $54.8$ & $20.23\pm 4.7$ & $45$ & $57.1\pm 25.4\pm 0.8$ & $14.8\pm 0.9\pm 0.0$ & $3.6\pm 2.6$ & $9.2\pm 0.1$ \\
$141$ & $\text{2006en}$ & $\text{N}$ & $-18.80\pm 0.03$ & $5.2$ & $3.18\pm 0.9$ & $36$ & $3.2\pm 0.9\pm 0.7$ & $16.7\pm 0.3\pm 0.1$ & $2.8\pm 1.0$ & $9.1\pm 0.7$ \\
$142$ & $\text{2006et}$ & $\text{N}$ & $-18.80\pm 0.02$ & $10.7$ & $4.19\pm 0.7$ & $47$ & $5.3\pm 1.2\pm 1.2$ & $16.4\pm 0.3\pm 0.1$ & $2.9\pm 1.1$ & $9.6\pm 0.2$ \\
$143$ & $\text{2006gr}$ & $\text{N}$ & $-18.60\pm 0.01$ & $34.7$ & $22.92\pm 2.5$ & $4$ & $23.0\pm 2.5\pm 0.8$ & $16.1\pm 0.3\pm 0.0$ & $8.9\pm 4.1$ & $9.2\pm 0.3$ \\
$144$ & $\text{2006gz}$ & $\text{03fg}$ & $-19.40\pm 0.03$ & $28.5$ & $16.55\pm 1.5$ & $15$ & $17.2\pm 1.9\pm 0.4$ & $15.3\pm 0.4\pm 0.0$ & $1.9\pm 0.8$ & $8.7\pm 0.3$ \\
$145$ & $\text{2006hb}$ & $\text{91bg}$ & $-18.30\pm 0.05$ & $20.0$ & $5.24\pm 0.7$ & $27$ & $6.2\pm 1.1\pm 1.4$ & $16.0\pm 0.3\pm 0.1$ & $1.7\pm 0.5$ & $9.7\pm 0.1$ \\
$146$ & $\text{2006mr}$ & $\text{pec}$ & $-15.10\pm 0.02$ & $9.9$ & $0.83\pm 0.1$ & $18$ & $1.5\pm 0.2\pm 1.3$ & $17.2\pm 0.2\pm 0.5$ & $2.5\pm 0.8$ & $9.9\pm 0.1$ \\
$147$ & $\text{2006ob}$ & $\text{N}$ & $-18.60\pm 0.02$ & $6.5$ & $6.80\pm 1.6$ & $70$ & $9.2\pm 2.6\pm 1.1$ & $16.3\pm 0.3\pm 0.1$ & $5.0\pm 2.0$ & $9.5\pm 0.3$ \\
$148$ & $\text{2006ot}$ & $\text{pec}$ & $-18.70\pm 0.02$ & $6.4$ & $5.90\pm 1.3$ & $40$ & $10.0\pm 3.5\pm 1.2$ & $16.3\pm 0.4\pm 0.1$ & $5.6\pm 2.4$ & $9.6\pm 0.3$ \\
$149$ & $\text{2006S}$ & $\text{N}$ & $-19.00\pm 0.01$ & $11.6$ & $7.52\pm 1.2$ & $60$ & $18.6\pm 5.8\pm 0.9$ & $15.3\pm 0.5\pm 0.0$ & $2.1\pm 0.9$ & $9.3\pm 0.5$ \\
$150$ & $\text{2006X}$ & $\text{N}$ & $-15.60\pm 0.01$ & $48.8$ & $3.26\pm 0.2$ & $48$ & $3.4\pm 0.3\pm 1.0$ & $16.4\pm 0.2\pm 0.2$ & $1.9\pm 0.7$ & $9.4\pm 0.3$ \\
$151$ & $\text{2007A}$ & $\text{N}$ & $-17.90\pm 0.02$ & $10.3$ & $3.19\pm 0.6$ & $4$ & $3.2\pm 0.6\pm 1.3$ & $16.4\pm 0.2\pm 0.2$ & $1.7\pm 0.5$ & $9.6\pm 0.2$ \\
$152$ & $\text{2007af}$ & $\text{N}$ & $-18.10\pm 0.01$ & $46.0$ & $5.02\pm 0.2$ & $87$ & $6.6\pm 0.6\pm 0.5$ & $15.5\pm 0.3\pm 0.1$ & $0.9\pm 0.3$ & $8.7\pm 0.3$ \\
$153$ & $\text{2007ax}$ & $\text{91bg}$ & $-15.90\pm 0.20$ & $7.0$ & $1.04\pm 0.4$ & $52$ & $3.1\pm 1.8\pm 1.4$ & $16.3\pm 0.5\pm 0.2$ & $1.5\pm 0.7$ & $9.7\pm 0.1$ \\
$154$ & $\text{2007ba}$ & $\text{91bg}$ & $-17.80\pm 0.05$ & $13.1$ & $11.15\pm 1.9$ & $68$ & $30.5\pm 9.8\pm 0.6$ & $15.7\pm 0.4\pm 0.0$ & $7.1\pm 3.2$ & $8.9\pm 0.1$ \\
$155$ & $\text{2007bc}$ & $\text{N}$ & $-18.90\pm 0.01$ & $33.0$ & $13.18\pm 1.4$ & $34$ & $13.5\pm 1.6\pm 1.0$ & $15.8\pm 0.3\pm 0.0$ & $3.0\pm 1.3$ & $9.4\pm 0.2$ \\
$156$ & $\text{2007bd}$ & $\text{N}$ & $-19.00\pm 0.02$ & $9.8$ & $5.72\pm 1.0$ & $5$ & $5.7\pm 1.0\pm 1.2$ & $16.3\pm 0.3\pm 0.1$ & $2.7\pm 1.1$ & $9.6\pm 0.2$ \\
$157$ & $\text{2007bj}$ & $\text{N}$ & $-18.90\pm 0.10$ & $4.7$ & $1.87\pm 0.5$ & $1$ & $2.1\pm 0.5\pm 1.6$ & $16.8\pm 0.3\pm 0.4$ & $2.3\pm 0.8$ & $9.9\pm 0.1$ \\
$158$ & $\text{2007bm}$ & $\text{N}$ & $-17.10\pm 0.02$ & $10.6$ & $1.37\pm 0.2$ & $5$ & $1.4\pm 0.3\pm 0.8$ & $16.8\pm 0.2\pm 0.3$ & $1.4\pm 0.4$ & $9.2\pm 0.4$ \\
$159$ & $\text{2007ci}$ & $\text{N}$ & $-17.80\pm 0.04$ & $12.5$ & $4.42\pm 0.7$ & $29$ & $4.6\pm 0.8\pm 1.5$ & $16.4\pm 0.3\pm 0.2$ & $2.7\pm 1.0$ & $9.7\pm 0.1$ \\
$160$ & $\text{2007F}$ & $\text{N}$ & $-19.30\pm 0.01$ & $12.0$ & $5.66\pm 0.9$ & $2$ & $5.7\pm 0.9\pm 0.5$ & $15.9\pm 0.3\pm 0.1$ & $1.4\pm 0.4$ & $8.7\pm 0.3$ \\
$161$ & $\text{2007hj}$ & $\text{N}$ & $-18.10\pm 0.03$ & $15.1$ & $4.10\pm 0.4$ & $25$ & $4.3\pm 0.4\pm 1.3$ & $16.2\pm 0.2\pm 0.2$ & $1.7\pm 0.4$ & $9.6\pm 0.1$ \\
$162$ & $\text{2007kk}$ & $\text{N}$ & $-18.60\pm 0.03$ & $13.3$ & $9.81\pm 1.6$ & $23$ & $10.1\pm 1.8\pm 1.0$ & $16.3\pm 0.3\pm 0.1$ & $5.1\pm 1.8$ & $9.4\pm 0.3$ \\
$163$ & $\text{2007le}$ & $\text{N}$ & $-17.30\pm 0.01$ & $16.2$ & $1.57\pm 0.3$ & $6$ & $1.9\pm 0.8\pm 0.8$ & $16.6\pm 0.3\pm 0.2$ & $1.2\pm 0.4$ & $9.2\pm 0.4$ \\
$164$ & $\text{2007N}$ & $\text{pec}$ & $-15.80\pm 0.07$ & $16.1$ & $5.03\pm 0.8$ & $19$ & $7.1\pm 2.1\pm 1.1$ & $15.8\pm 0.4\pm 0.1$ & $1.6\pm 0.6$ & $9.5\pm 0.3$ \\
$165$ & $\text{2007nq}$ & $\text{N}$ & $-18.90\pm 0.02$ & $24.2$ & $18.27\pm 2.4$ & $20$ & $18.5\pm 2.5\pm 1.5$ & $15.8\pm 0.5\pm 0.0$ & $4.4\pm 2.9$ & $9.7\pm 0.1$ \\
$166$ & $\text{2007O}$ & $\text{N}$ & $-19.20\pm 0.10$ & $10.0$ & $6.52\pm 1.2$ & $44$ & $6.6\pm 1.2\pm 0.6$ & $16.2\pm 0.3\pm 0.1$ & $2.8\pm 1.1$ & $9.0\pm 0.3$ \\
$167$ & $\text{2007on}$ & $\text{N}$ & $-18.20\pm 0.01$ & $70.3$ & $6.33\pm 0.7$ & $3$ & $6.3\pm 0.7\pm 1.4$ & $16.2\pm 0.3\pm 0.1$ & $2.8\pm 1.0$ & $9.7\pm 0.1$ \\
$168$ & $\text{2007S}$ & $\text{N}$ & $-17.80\pm 0.01$ & $11.7$ & $3.31\pm 0.5$ & $9$ & $3.4\pm 0.6\pm 1.1$ & $16.2\pm 0.2\pm 0.2$ & $1.3\pm 0.3$ & $9.5\pm 0.3$ \\
$169$ & $\text{2007st}$ & $\text{N}$ & $18.40\pm 0.20$ & $5.3$ & $2.16\pm 0.6$ & $55$ & $2.8\pm 0.9\pm 1.1$ & $16.8\pm 0.3\pm 0.2$ & $3.3\pm 1.2$ & $9.5\pm 0.4$ \\
$170$ & $\text{2008ae}$ & $\text{02cx}$ & $-17.00\pm 0.03$ & $13.2$ & $7.95\pm 1.0$ & $31$ & $8.0\pm 1.1\pm 0.6$ & $15.9\pm 0.3\pm 0.0$ & $2.2\pm 0.8$ & $9.0\pm 0.6$ \\
$171$ & $\text{2008ar}$ & $\text{N}$ & $-19.00\pm 0.02$ & $5.8$ & $3.10\pm 0.8$ & $20$ & $3.2\pm 0.9\pm 1.4$ & $16.3\pm 0.3\pm 0.2$ & $1.5\pm 0.5$ & $9.7\pm 0.3$ \\
$172$ & $\text{2008bc}$ & $\text{N}$ & $-18.50\pm 0.01$ & $19.6$ & $5.52\pm 0.8$ & $41$ & $6.5\pm 1.2\pm 0.9$ & $15.8\pm 0.3\pm 0.1$ & $1.4\pm 0.4$ & $9.3\pm 0.5$ \\
$173$ & $\text{2008bf}$ & $\text{N}$ & $-19.20\pm 0.02$ & $52.3$ & $28.04\pm 0.9$ & $25$ & $30.4\pm 1.8\pm 1.3$ & $15.7\pm 0.3\pm 0.0$ & $6.1\pm 2.6$ & $9.6\pm 0.1$ \\
$174$ & $\text{2008bi}$ & $\text{N}$ & $-18.00\pm 0.50$ & $4.6$ & $1.24\pm 0.4$ & $76$ & $1.9\pm 0.6\pm 1.2$ & $16.8\pm 0.4\pm 0.4$ & $2.0\pm 1.0$ & $9.7\pm 0.2$ \\
$175$ & $\text{2008cc}$ & $\text{N}$ & $17.40\pm 0.50$ & $8.2$ & $0.97\pm 0.3$ & $57$ & $1.8\pm 0.5\pm 1.3$ & $16.6\pm 0.3\pm 0.4$ & $1.1\pm 0.4$ & $9.9\pm 0.1$ \\
$176$ & $\text{2008fp}$ & $\text{N}$ & $-17.60\pm 0.02$ & $18.5$ & $1.88\pm 0.8$ & $38$ & $2.9\pm 1.4\pm 1.4$ & $16.3\pm 0.5\pm 0.3$ & $1.3\pm 0.6$ & $9.7\pm 0.2$ \\
$177$ & $\text{2008fu}$ & $\text{N}$ & $-17.40\pm 0.50$ & $2.5$ & $2.44\pm 1.1$ & $19$ & $2.7\pm 1.3\pm 0.7$ & $16.7\pm 0.4\pm 0.1$ & $2.3\pm 1.1$ & $9.1\pm 0.7$ \\
$178$ & $\text{2008fw}$ & $\text{N}$ & $19.80\pm 0.50$ & $66.8$ & $10.84\pm 2.5$ & $10$ & $11.0\pm 2.6\pm 0.8$ & $15.9\pm 0.5\pm 0.0$ & $2.9\pm 1.7$ & $9.1\pm 0.4$ \\
$179$ & $\text{2008gl}$ & $\text{N}$ & $-19.00\pm 0.02$ & $23.7$ & $14.45\pm 1.9$ & $13$ & $15.9\pm 2.9\pm 1.3$ & $15.8\pm 0.4\pm 0.0$ & $4.0\pm 1.8$ & $9.6\pm 0.1$ \\
$180$ & $\text{2008gp}$ & $\text{N}$ & $-18.90\pm 0.02$ & $17.4$ & $10.29\pm 1.5$ & $13$ & $10.5\pm 1.6\pm 1.2$ & $15.9\pm 0.3\pm 0.1$ & $3.0\pm 1.3$ & $9.5\pm 0.4$ \\
$181$ & $\text{2008ha}$ & $\text{02cx}$ & $-12.70\pm 0.10$ & $9.4$ & $0.99\pm 0.4$ & $6$ & $1.0\pm 0.5\pm 0.4$ & $16.8\pm 0.4\pm 0.2$ & $0.8\pm 0.3$ & $8.6\pm 0.2$ \\
$182$ & $\text{2008hj}$ & $\text{N}$ & $-20.00\pm 0.50$ & $21.0$ & $14.60\pm 2.0$ & $68$ & $18.0\pm 3.3\pm 1.0$ & $15.2\pm 0.5\pm 0.0$ & $1.8\pm 0.8$ & $9.4\pm 0.3$ \\
$183$ & $\text{2008J}$ & $\text{CSM}$ & $-20.00\pm 0.10$ & $4.8$ & $1.51\pm 0.5$ & $5$ & $1.7\pm 0.6\pm 1.1$ & $16.8\pm 0.3\pm 0.4$ & $1.5\pm 0.6$ & $9.6\pm 0.3$ \\
$184$ & $\text{2008L}$ & $\text{N}$ & $-18.60\pm 0.04$ & $10.6$ & $3.47\pm 0.6$ & $61$ & $5.8\pm 1.5\pm 1.3$ & $16.1\pm 0.3\pm 0.1$ & $1.8\pm 0.6$ & $9.6\pm 0.2$ \\
$185$ & $\text{2008R}$ & $\text{91bg}$ & $-18.30\pm 0.02$ & $14.3$ & $3.42\pm 0.5$ & $20$ & $4.2\pm 1.0\pm 1.4$ & $16.6\pm 0.3\pm 0.2$ & $3.1\pm 1.1$ & $9.7\pm 0.1$ \\
$186$ & $\text{2008s1}$ & $\text{N}$ & $-18.90\pm 0.02$ & $18.4$ & $9.36\pm 1.1$ & $10$ & $10.3\pm 1.8\pm 1.1$ & $15.9\pm 0.3\pm 0.1$ & $2.9\pm 1.2$ & $9.5\pm 0.2$ \\
$187$ & $\text{2009al}$ & $\text{N}$ & $-18.60\pm 0.03$ & $65.5$ & $28.43\pm 3.1$ & $74$ & $33.9\pm 4.9\pm 0.7$ & $15.1\pm 0.5\pm 0.0$ & $2.9\pm 1.4$ & $9.0\pm 0.2$ \\
$188$ & $\text{2009an}$ & $\text{N}$ & $-18.50\pm 0.02$ & $26.8$ & $5.60\pm 1.1$ & $46$ & $6.6\pm 1.6\pm 1.1$ & $16.0\pm 0.4\pm 0.1$ & $2.1\pm 0.9$ & $9.5\pm 0.2$ \\
$189$ & $\text{2009cz}$ & $\text{N}$ & $-19.00\pm 0.01$ & $18.2$ & $7.92\pm 1.1$ & $40$ & $8.5\pm 1.3\pm 1.3$ & $16.2\pm 0.3\pm 0.1$ & $3.7\pm 1.3$ & $9.6\pm 0.2$ \\
$190$ & $\text{2009D}$ & $\text{N}$ & $-18.10\pm 0.02$ & $40.7$ & $18.37\pm 2.0$ & $62$ & $36.7\pm 8.0\pm 0.5$ & $14.7\pm 0.6\pm 0.0$ & $1.9\pm 0.9$ & $8.7\pm 0.3$ \\
$191$ & $\text{2009dc}$ & $\text{03fg}$ & $-19.60\pm 0.02$ & $25.4$ & $11.47\pm 1.3$ & $46$ & $44.0\pm 17.9\pm 0.4$ & $14.8\pm 0.7\pm 0.0$ & $2.8\pm 1.3$ & $8.6\pm 0.5$ \\
$192$ & $\text{2009ds}$ & $\text{N}$ & $-19.80\pm 0.50$ & $12.6$ & $4.99\pm 0.9$ & $11$ & $5.1\pm 0.9\pm 0.8$ & $16.4\pm 0.3\pm 0.1$ & $2.8\pm 1.2$ & $9.2\pm 0.4$ \\
$193$ & $\text{2009F}$ & $\text{pec}$ & $-16.40\pm 0.02$ & $12.2$ & $3.09\pm 0.7$ & $28$ & $3.2\pm 0.7\pm 1.5$ & $16.5\pm 0.3\pm 0.3$ & $2.0\pm 0.8$ & $9.7\pm 0.1$ \\
$194$ & $\text{2009fv}$ & $\text{N}$ & $-16.90\pm 0.10$ & $8.1$ & $3.50\pm 1.1$ & $52$ & $12.4\pm 7.1\pm 1.6$ & $16.2\pm 0.6\pm 0.1$ & $5.3\pm 3.2$ & $9.8\pm 0.1$ \\
$195$ & $\text{2009I}$ & $\text{N}$ & $-16.80\pm 0.20$ & $12.3$ & $6.38\pm 0.9$ & $40$ & $6.9\pm 1.2\pm 0.7$ & $16.2\pm 0.3\pm 0.1$ & $2.6\pm 0.9$ & $9.1\pm 0.3$ \\
$196$ & $\text{2009ig}$ & $\text{N}$ & $-19.00\pm 0.02$ & $22.4$ & $3.37\pm 0.3$ & $10$ & $3.4\pm 0.3\pm 1.3$ & $16.3\pm 0.2\pm 0.2$ & $1.5\pm 0.4$ & $9.6\pm 0.2$ \\
$197$ & $\text{2009le}$ & $\text{N}$ & $19.00\pm 0.50$ & $16.8$ & $5.31\pm 0.8$ & $25$ & $6.3\pm 1.3\pm 0.9$ & $16.2\pm 0.3\pm 0.1$ & $2.8\pm 1.1$ & $9.3\pm 0.4$ \\
$198$ & $\text{2009Y}$ & $\text{N}$ & $-17.50\pm 0.02$ & $26.3$ & $4.69\pm 0.6$ & $16$ & $5.0\pm 0.7\pm 1.3$ & $16.4\pm 0.2\pm 0.1$ & $2.8\pm 1.0$ & $9.6\pm 0.2$ \\
$199$ & $\text{2010B}$ & $\text{N}$ & $-18.70\pm 0.20$ & $8.7$ & $1.98\pm 0.5$ & $23$ & $2.0\pm 0.5\pm 1.5$ & $16.6\pm 0.3\pm 0.5$ & $1.4\pm 0.5$ & $9.7\pm 0.1$ \\
$200$ & $\text{2010kg}$ & $\text{N}$ & $-18.30\pm 0.02$ & $10.7$ & $3.53\pm 0.7$ & $47$ & $4.0\pm 0.9\pm 1.2$ & $16.3\pm 0.3\pm 0.2$ & $1.9\pm 0.7$ & $9.6\pm 0.3$ \\
$201$ & $\text{2011by}$ & $\text{N}$ & $-17.60\pm 0.01$ & $19.9$ & $1.99\pm 0.2$ & $80$ & $7.8\pm 2.3\pm 0.6$ & $15.6\pm 0.4\pm 0.1$ & $1.3\pm 0.3$ & $9.0\pm 0.2$ \\
$202$ & $\text{2011de}$ & $\text{03fg}$ & $-19.70\pm 0.05$ & $78.2$ & $47.49\pm 3.2$ & $9$ & $47.5\pm 3.3\pm 0.3$ & $14.9\pm 0.5\pm 0.0$ & $3.5\pm 1.8$ & $8.5\pm 0.1$ \\
$203$ & $\text{2011fe}$ & $\text{N}$ & $-17.80\pm 0.01$ & $277.9$ & $9.03\pm 0.2$ & $35$ & $9.2\pm 0.3\pm 0.5$ & $15.7\pm 0.2\pm 0.0$ & $1.8\pm 0.6$ & $8.7\pm 0.3$ \\
$204$ & $\text{2011hb}$ & $\text{N}$ & $-19.00\pm 0.01$ & $19.4$ & $11.38\pm 1.2$ & $5$ & $11.4\pm 1.2\pm 0.8$ & $16.3\pm 0.2\pm 0.0$ & $5.7\pm 1.8$ & $9.2\pm 0.3$ \\
$205$ & $\text{2011im}$ & $\text{N}$ & $-18.30\pm 0.01$ & $22.2$ & $7.10\pm 1.6$ & $80$ & $11.4\pm 3.4\pm 1.2$ & $16.0\pm 0.5\pm 0.1$ & $3.3\pm 1.8$ & $9.6\pm 0.2$ \\
$206$ & $\text{2012cg}$ & $\text{N}$ & $-16.90\pm 0.02$ & $15.0$ & $1.19\pm 0.2$ & $40$ & $2.0\pm 0.6\pm 0.9$ & $16.5\pm 0.3\pm 0.2$ & $1.1\pm 0.3$ & $9.3\pm 0.4$ \\
$207$ & $\text{2012dn}$ & $\text{03fg}$ & $-18.90\pm 0.03$ & $36.7$ & $6.94\pm 1.5$ & $65$ & $9.4\pm 2.5\pm 0.4$ & $15.5\pm 0.5\pm 0.0$ & $1.3\pm 0.5$ & $8.7\pm 0.3$ \\
$208$ & $\text{2012fr}$ & $\text{N}$ & $-19.90\pm 0.02$ & $48.1$ & $4.21\pm 0.2$ & $49$ & $6.8\pm 0.9\pm 1.0$ & $16.3\pm 0.2\pm 0.1$ & $3.4\pm 1.1$ & $9.4\pm 0.3$ \\
$209$ & $\text{2012hr}$ & $\text{N}$ & $-19.00\pm 0.01$ & $93.1$ & $17.93\pm 3.5$ & $76$ & $24.2\pm 6.0\pm 0.5$ & $15.2\pm 0.6\pm 0.0$ & $2.4\pm 1.4$ & $8.8\pm 0.2$ \\
$210$ & $\text{2012ht}$ & $\text{N}$ & $-17.90\pm 0.01$ & $22.1$ & $2.57\pm 0.2$ & $59$ & $4.1\pm 0.6\pm 0.3$ & $15.8\pm 0.3\pm 0.1$ & $0.9\pm 0.2$ & $8.4\pm 0.2$ \\
$211$ & $\text{2012Z}$ & $\text{02cx}$ & $-17.80\pm 0.01$ & $47.3$ & $7.04\pm 0.6$ & $16$ & $7.1\pm 0.6\pm 0.8$ & $15.9\pm 0.2\pm 0.1$ & $1.7\pm 0.5$ & $9.2\pm 0.3$ \\
$212$ & $\text{2013aa}$ & $\text{N}$ & $-19.50\pm 0.07$ & $194.6$ & $15.04\pm 7.3$ & $3$ & $15.0\pm 7.3\pm 0.5$ & $15.5\pm 0.9\pm 0.0$ & $2.1\pm 2.1$ & $8.8\pm 0.3$ \\
$213$ & $\text{2013cv}$ & $\text{91T}$ & $-19.50\pm 0.05$ & $3.3$ & $2.36\pm 1.8$ & $30$ & $2.4\pm 1.9\pm 0.3$ & $16.2\pm 0.9\pm 0.1$ & $0.9\pm 0.5$ & $8.5\pm 0.2$ \\
$214$ & $\text{2013dy}$ & $\text{N}$ & $-17.90\pm 0.01$ & $25.0$ & $2.40\pm 0.2$ & $45$ & $9.0\pm 3.3\pm 0.3$ & $15.4\pm 0.5\pm 0.0$ & $1.1\pm 0.3$ & $8.4\pm 0.4$ \\
$215$ & $\text{2013gs}$ & $\text{N}$ & $-18.90\pm 0.01$ & $10.7$ & $3.79\pm 0.7$ & $25$ & $5.6\pm 1.8\pm 0.6$ & $15.8\pm 0.4\pm 0.1$ & $1.2\pm 0.4$ & $9.0\pm 0.3$ \\
$216$ & $\text{2013gy}$ & $\text{N}$ & $-19.10\pm 0.01$ & $33.4$ & $7.04\pm 1.5$ & $4$ & $7.1\pm 1.6\pm 0.8$ & $15.7\pm 0.4\pm 0.1$ & $1.4\pm 0.5$ & $9.2\pm 0.3$ \\
$217$ & $\text{2013Q}$ & $\text{N}$ & $-18.60\pm 0.01$ & $41.5$ & $14.90\pm 1.8$ & $27$ & $15.5\pm 2.1\pm 0.7$ & $16.1\pm 0.2\pm 0.0$ & $6.5\pm 2.4$ & $9.1\pm 0.3$ \\
$218$ & $\text{2014dg}$ & $\text{N}$ & $-16.30\pm 0.01$ & $6.2$ & $0.43\pm 0.1$ & $54$ & $1.1\pm 0.4\pm 0.6$ & $16.9\pm 0.4\pm 0.3$ & $1.1\pm 0.5$ & $9.2\pm 0.4$ \\
$219$ & $\text{2015F}$ & $\text{N}$ & $-18.40\pm 0.01$ & $97.3$ & $9.36\pm 0.3$ & $50$ & $10.2\pm 0.6\pm 0.7$ & $16.0\pm 0.2\pm 0.0$ & $3.2\pm 1.0$ & $9.1\pm 0.3$ \\
$220$ & $\text{2016bln}$ & $\text{91T}$ & $-19.20\pm 0.02$ & $170.1$ & $82.20\pm 6.2$ & $22$ & $101.0\pm 14.7\pm 0.1$ & $14.6\pm 0.5\pm 0.0$ & $4.8\pm 2.4$ & $7.5\pm 0.1$ \\
$221$ & $\text{2016dwu}$ & $\text{03fg}$ & $-19.70\pm 0.12$ & $23.3$ & $6.70\pm 1.1$ & $45$ & $7.4\pm 1.4\pm 0.3$ & $15.6\pm 0.4\pm 0.0$ & $1.2\pm 0.5$ & $8.4\pm 0.3$ \\
$222$ & $\text{2018oh}$ & $\text{N}$ & $-19.50\pm 0.01$ & $5.3$ & $1.21\pm 0.4$ & $29$ & $1.3\pm 0.5\pm 0.4$ & $16.6\pm 0.3\pm 0.2$ & $0.9\pm 0.3$ & $8.6\pm 0.2$ \\
$223$ & $\text{LSQ12gdj}$ & $\text{91T}$ & $-20.10\pm 0.01$ & $27.9$ & $13.40\pm 1.7$ & $17$ & $13.7\pm 1.9\pm 0.5$ & $15.1\pm 0.4\pm 0.0$ & $1.1\pm 0.3$ & $8.8\pm 0.6$ \\
$224$ & $\text{PTF11kx}$ & $\text{CSM}$ & $-19.30\pm 0.01$ & $6.9$ & $6.35\pm 1.1$ & $14$ & $7.2\pm 1.7\pm 0.6$ & $15.8\pm 0.3\pm 0.1$ & $1.5\pm 0.4$ & $9.0\pm 0.6$ \\\\[-3mm]
\hline\\[-3mm]
\caption{List of SNe used in this work. Ia-types: N = normal, 91bg = SN Ia like 1991bg, 91T = SN Ia like 1991T, 02cx = SN Ia like 2002cx, 02es = SN Ia like 2002es, 03fg = SN Ia like 2003fg, CSM = SN Ia of interaction with circum stellar medium, pec = peculiar SN Ia; $\theta$ = angular offset (error estimation globally $\delta \theta = 1"$); $r_{\!\perp}$ = perpendicular offset from host; $\phi$ = SN position angle (error estimation globally $\delta \phi = 2^{\circ}$); $r$ = 3d-offset from host; $\rho_{\rm DM}$ = local DM density in the vicinity of the SN; $v_{\rm DM}$ = velocity dispersion in the vicinity of the SN explosion but outside the progenitor WD potential well; $t$ = local stellar age; stat = statistical error; sys = systematic error (due to stellar migration).
}
\label{tab:data-sn}
\end{longtable}
\end{footnotesize}

\bsp	
\label{lastpage}

\end{document}